\documentclass{article}

\usepackage{arxiv}

\usepackage{amssymb}

\usepackage{tabularx}
\usepackage{graphicx}
\usepackage[export]{adjustbox}
\usepackage{subcaption}
\usepackage{mwe}
\usepackage{hyperref}
\usepackage{xcolor}
\usepackage{amsmath}
\usepackage{booktabs}
\usepackage{cuted}
\usepackage{comment}

\newcommand{\squareone}{\textcolor[HTML]{DE89FF}{\rule{2mm}{2mm}}}
\newcommand{\squaretwo}{\textcolor[HTML]{73BF02}{\rule{2mm}{2mm}}}
\newcommand{\squarethree}{\textcolor[HTML]{00C4FF}{\rule{2mm}{2mm}}}
\newcommand{\squarefour}{\textcolor[HTML]{4E453C}{\rule{2mm}{2mm}}}
\newcommand{\squarefive}{\textcolor[HTML]{FF8905}{\rule{2mm}{2mm}}}
\newcommand{\squaresix}{\textcolor[HTML]{FE5585}{\rule{2mm}{2mm}}}
\newcommand{\squareseven}{\textcolor[HTML]{02BC93}{\rule{2mm}{2mm}}}
\newcommand{\squareeight}{\textcolor[HTML]{D4B3AF}{\rule{2mm}{2mm}}}
\newcommand{\squarenine}{\textcolor[HTML]{999999}{\rule{2mm}{2mm}}}

\title{Studying Fake News Spreading, Polarisation Dynamics, and Manipulation by Bots: a Tale of Networks and Language}

\author{
Giancarlo Ruffo \\
  DISIT,
  Universit\`a degli Studi del\\ Piemonte Orientale ``A. Avogadro''\\
  Alessandria, Italy \\
  \texttt{giancarlo.ruffo@uniupo.it} \\
   \And
Alfonso Semeraro \\
  Dipartimento di Informatica\\
  Universit\`a degli Studi di Torino\\
  Torino, Italy \\
  \texttt{alfonso.semeraro@unito.it} \\
  \And
Anastasia Giachanou \\
  Utrecht University\\
  Utrecht, The Netherlands\\
  \texttt{a.giachanou@uu.nl} \\
  \And
Paolo Rosso \\
  PRHLT Research Center\\ 
  Universitat Politècnica de València\\
  València, Spain\\
  \texttt{prosso@dsic.upv.es} \\
}

\begin{document}

\maketitle

\begin{abstract}
With the explosive growth of online social media, the ancient problem of information disorders interfering with news diffusion has surfaced with a renewed intensity threatening our democracies, public health, and news outlets' credibility. Therefore, thousands of scientific papers have been published in a relatively short period, making researchers of different disciplines struggle with an information overload problem. 
The aim of this survey is threefold: (1) we present the results of a network-based analysis of the existing multidisciplinary literature to support the search for relevant trends and central publications; (2) we describe the main results and necessary background to attack the problem under a computational perspective; (3) we review selected contributions using network science as a unifying framework and computational linguistics as the tool to make sense of the shared content. Despite scholars working on computational linguistics and networks traditionally belong to different scientific communities, we expect that those interested in the area of fake news should be aware of crucial aspects of both disciplines. 
\end{abstract}


\section{Introduction}
\label{sec:introduction}

``Fake News'' is not a recent threat, nor even a recent word, since its origin roots at least one century ago, when in 1925 the Harper's Magazine~\cite{Harpers} raised the concern on how the new technologies would have disrupted classic journalism; in fact, the history of journalism and news diffusion is tightly coupled with the effort to dispel hoaxes, misinformation, propaganda, unverified rumours, poor reporting, and messages containing hate and divisions. However, the recent explosive interest in this topic is probably mainly due to the advent of Social Media platforms that are progressively used by many as a source of information, and also as a tool for the diffusion of the information itself. Along with the mainstream public's attention to this problem, scientists' interest also spiked around 2016, when worldwide media appraised the possible interference of misinformation, disinformation, and other forms of information disorders\footnote{There are subtle but significant differences between misinformation, disinformation, and malinformation, that are all referred as instances of information disorders. Terminology will be properly defined in Sec.~\ref{sec:terminology}.} during two historical political events: the U.S. Presidential campaign and Brexit. Since 2016, the yearly academic production about information disorders issues has increased enormously, and thousands of researchers have published theoretical and experimental findings contributing to this new area of study, not to mention the even more recent interest to infodemic, caused by the COVID-19 threat and how the information about it has been managed by mainstream and alternative news outlet.

This enormous production is certainly beneficial for science, but it has its drawbacks. It is quite difficult to have a clear picture of what is happening in the field, when researchers are stacking findings after findings in such a short time. This trend creates two problems. First, sometimes it is hard to keep track of the latest prints out, to properly evaluate each contribution, and to get meaningful information from this massive amount of papers. Second, it is likely that this newborn scientific field did not find its ultimate structure yet. Many works differ in the definitions, there is often a confusing terminology, and such issues have consequences on the methodology and the results. It is not rare that new evidence is in open or apparent contradiction with previous findings; quite similarly, we have models designed to explain the emergence of spreading phenomena or how opinion dynamics can lead to echo-chambers that yield different results (e.g., Del Vicario et al.~\cite{delvicario2016modeling}, and Perra and Rocha~\cite{perra2018modelling}). Plus, the social media landscape is continuously mutating, and even two observations of the same phenomenon over years may differ substantially. 

As a survey, this work is an attempt to look back at thousands of papers published so far focusing on fake news and other kinds of information disorders. In this massive production, that spans across different disciplines and scientific communities, surveys should have the key role of finding the most relevant sub-narratives and highlighting the most influential contributions. This is not the first survey about this topic; in particular, there are many solid works about the latest methods to automatically detect fake news (e.g.,~\cite{conroyetal2015,Shuetal2017,ZhouZafarani2020}), a sub-field which is evolving at an extraordinary pace. However, this survey contributes to the scientific debate because of its attention to networks and language, and to how they can be mixed together. 
The formalism of nodes linked by edges mimics perfectly the relationship between individuals connected to each other on social media platforms, information flows through links in a way that depends also on the topological structure of the physical network. Thus, networks are a great tool for explaining why and how users engage with fake news, how these spread online, and how we can try to stop them. Network structure may foster segregation and radicalisation of users, as well as fuel or curb down any (mis)information outbreak. Language, on the other hand, is crucial to understand how fake news works on users' minds. Deception is often conveyed through rhetorical tricks. Apparently, the language of fake news is inherently different from the language of mainstream journalism, and it works on a subconscious, emotional layer, exploiting the readers' biases~\cite{FerraraYang2015,stella2018bots}. Hints of deceptive language may be found in the style of low-quality articles, and in turn, the style of a text can be leveraged for detection purposes. Together, language and network provide a unifying framework to study how fake news is crafted and what is the environment they spread in. While the problem of disinformation has been framed either as a network problem or as a stylistic and semantic problem until some years ago, it is from the mix of the two disciplines that we see the most relevant contributions of recent and, very likely, next years.

Last, but not least, it is very likely that in the future many academic courses, also at bachelor's and master's degree level, will be delivered to introduce the scientific and methodological basis to study how social media manipulation is perpetrated and how to counter it, from both technical and behavioural points of view (e.g., ``Social Media Manipulation 101'' course was launched at Indiana University in Fall 2022\footnote{ \url{https://osome.iu.edu/education/social-media-manipulation-101}}). We think that a unifying framework allowing instructors to introduce coherently the different perspectives to look at the issues behind online disinformation is needed. Also with this purpose in mind, we give an overview of the problem, from its psychological motivations to the dynamics of opinion fragmentation and polarisation, from the detection of false news by their text to the activity of automated accounts, using network science and computational linguistics as methodological and computational tools to better understand the emerging ``Science of Fake-News''~\cite{Lazer1094}.

This survey is partially guided by data, too. Traditionally, surveys are written by experts based on their knowledge and their background, and this work is not an exception. However, we decided to rely also on a wide repository of scientific publications about fake news, that we collected and analysed to have a deeper understanding of sub-narratives and also to minimise our own biases and limited knowledge of existing literature, performing a citation analysis (See Section~\ref{sec:network}). As a by-product of our research, we created a library of relevant papers and we developed a search engine accessible via the Web at \url{http://fakenewsresearch.net/} to allow others to explore the given library with personalised queries.

\section{The problem with terminology}
\label{sec:terminology}

The term ``fake news'' has become extremely popular in the last few years, trespassing the domain of journalism and social media users and entering the everyday language. Intuitively, it reminds of some kind of false information broadcasted by one or more news outlets, and generally speaking to the manipulation of the information systems in order to deceive someone's opinion. This pseudo-definition is however broad and unclear, as it contains several shades of disinformation practices, while at the same time it excludes bad practices that indeed contribute to the diffusion of misleading news. In the previous sections, we mentioned terms such as ``fake news'', ``misinformation'', ``disinformation'': the purpose of this section is to report the state of the art on the debate on the rigorous definitions of these terms.

\subsection{What is fake news?}
\label{subsec:what_is_fake_news}
Especially after the nomination as 2017 ``Word of the year'' by the Collin's Dictionary, ``fake news'' has been widely used for several, often very different, bad practices. This kind of confusion is so common that some scholars tried to put order. Already in 2015, Rubin et al.~\cite{Rubinetal2015} discussed three kinds of fakes: a) serious fabrications (uncovered in mainstream or participant media, yellow press or tabloids); b) large-scale hoaxes; c) humorous fakes (news satire, parody, game shows).  Later, Tandoc et al.~\cite{Tandoc2018}, collected 34 previous works about ``fake news'', only to find out that they were targeting six distinct practices: satire, news parody, fabricated news, manipulated pictures, advertising, and political propaganda. When different kinds of manipulation conflate in the term ``fake news'' without a global agreement on what it actually means, and when researchers target and study as ``fake news'' quite different phenomena, it is hard to agree on findings, and to verify the results with scientific rigour. 
Such a buzzword needs to be replaced by proper terminology, a terminology that solves two problems. First, false information online may come in different fashions, and one term could not include them all. Second, definitions must be exact. If we want to compare the answers, we must ask the same questions. For instance, the Merriam-Webster Dictionary, while explaining the decision not to include ``fake news'' in the Dictionary, gives it a short definition~\cite{Merriam-Webster}:
\begin{small}
\begin{quote}
    Fake news is, quite simply, news (``material reported in a newspaper or news periodical or on a newscast'') that is fake (``false, counterfeit'').
\end{quote}
\end{small}

However, this simple definition raises many questions, as argued by Mould~\cite{Mould2018}:
\begin{small}
\begin{quote}
    What is considered a news source? The New York Times is a news source, but how about mock news outlets such as the Daily Show or the Onion? How about blogs, comment sections, [...]?
\end{quote}
\end{small}

And to what extent a piece of news must be ``fake'' to be considered ``fake news''? Fabricated, non-factual news is certainly fake news, but what about slight manipulations of partially true news? Does the omission of a crucial detail make the news fake? And what about verified news framed in biased narratives?
Tandoc et al.~\cite{Tandoc2018} is also an important entry point for getting to the core of the ``fake news'' problem, as it introduces a possible taxonomy of problems distinct by two dimensions: \emph{facticity}, i.e., adherence to factuality of news, and \emph{intentionality}, i.e., if the news is written with the intent of deceiving the reader. The accent on the intention of the writer seems to be an agreed factor, when defining ``fake news''. Allcott et al.~\cite{Allcott2017}, also insist on the publisher's intentions: 
\begin{small}
\begin{quote}
    We define ``fake news'' to be news articles that are intentionally and verifiably false, and could mislead readers.
\end{quote}
\end{small}
Another similar definition is given by Gelfert~\cite{Gelfert2018-GELFNA}, with the specification that ``fake news'' have the goal to mislead ``by design'':
\begin{small}

\begin{quote}
     This paper argues that it should be reserved for cases of deliberate presentation of (typically) false or misleading claims as news, where these are misleading by design. The phrase `by design' here refers to systemic features of the design of the sources and channels by which fake news propagates and, thereby, manipulates the audience’s cognitive processes. 
\end{quote}
\end{small}

The previous considerations narrow the domain of ``fake news'' to false information shared with harmful purpose, which to the best of our knowledge is accepted as a good definition, because it includes both the veracity of the news, and the malicious intention of the author. However, there is no doubt that this is only part of a bigger problem. Some of the above questions remain unanswered, as partially false or re-framed pieces of information may have the same impact in the end that entirely fabricated contents: how to consider them? In addition, in the last few years we have observed the rise of many issues related to the dissemination of fake news online, like coordinated disinformation campaigns held by automated accounts, constellations of false information supporting conspiracies, and propagation of unverified rumours.

\subsection{Beyond fake news}
\label{subsec:taxonomy}
For the reasons stated before, the popular term ``fake news'' is now considered insufficient to represent a broader range of malicious activities. A 2017 report by the Council of Europe~\cite{CouncilofEurope} first, and a 2018 UNESCO handbook~\cite{UnescoHandbook} later, found a more comprehensive definition in \emph{information disorder}, a term which in turn encompasses misinformation, disinformation, malinformation as three distinct problems.

The rationale behind this taxonomy is a two-dimensional evaluation of a piece of information, similar to the one in Tandoc et al.~\cite{Tandoc2018}: the veracity of the news and the malicious intentions of the author. \emph{Misinformation} is information false or misleading, but shared without harmful purposes. All the false news published with \emph{bona fide} mistakes are examples of misinformation, as well as news in which the title or a captivating picture are not adequately supported by the body of the article, like click-baiting~\cite{BLOM201587,Chen2015}. \emph{Malinformation} is (partially) true, verifiable information, but shared with malicious intent. Articles characterised by hate speech~\cite{Paz2020} belong to this category, like publishers cherry-picking~\cite{Asudeh2020} real news in order to foster hate against some minorities. \emph{Disinformation} is, finally, an intersection of the two above: false information shared with harmful intent. Impostor content, like information falsely credited to authoritative newspapers, falls in this category, as well as: manipulated content, such as true facts distorted to serve a false narration~\cite{benkler2018network,guess2020misinformation}; fabricated content, i.e., proper fake news in the definition of Allcott et al.~\cite{Allcott2017}; and true news framed in a narrative they do not belong (in this case it is the context to be false, rather than the content). Table~\ref{table:taxonomy} shows an attempt to define some key terms as instances of disinformation/misinformation/malinformation, to give a general overview of the information disorder categories and main practices, including the key terms proposed in~\cite{Liangetal2019}. Of course, the list of terms in Table~\ref{table:taxonomy} is by no means exhaustive. We also included a peculiar form of deceptive news (satire) which is part of the spectrum of the possible categories of news, also according to~\cite{Tandoc2018}.

\begin{table*}[ht!]
\begin{small}
\centering
\begin{tabularx}{\textwidth}{ l X } 
 \hline
 misinformation & unintentionally spread false information deceiving its recipients~\cite{UnescoHandbook,CouncilofEurope} \\ 
 \hline
 malinformation & information that is partially or totally true, but spread with malicious intent~\cite{UnescoHandbook,CouncilofEurope}\\ 
 \hline
 disinformation & intentionally spread and/or fabricated misinformation~\cite{UnescoHandbook,CouncilofEurope}  \\ 
 \hline
 fake news & disinformation in the format of news~\cite{UnescoHandbook,CouncilofEurope}\\  
 \hline
 hoax &  disinformation that can have also humorous purposes~\cite{situngkir2011spread} \\ 
 \hline 
 rumour & information that can be true and accurate, but still unverified; if it is falsified, it becomes misinformation~\cite{qazvinian2011rumor,MEEL2020112986}\\ 
 \hline
 conspiracy theory &  explanation of an event that assume a conspiracy by powerful group; a theory can make use of fake news, rumours as well as true information~\cite{Douglas2019,SUTTON2020118} \\ 
 \hline 
 urban legends & kind of folklore made of rumours characterised by supernatural, horrifying or humorous elements~\cite{blank2018slender} \\ 
 \hline 
 infodemic & mixture of misinformation and true information about the origins and alternative cures of a disease; especially observed during COVID-19 pandemic~\cite{world2021infodemic,Solomon2020} \\  
 \hline
 propaganda & malinformation that aims to influence  an audience and a political agenda~\cite{benkler2018network,guess2020misinformation} \\ 
 \hline
 click-bait & misinformation based strategy to deceive Web users and enticing them to follow a link~\cite{BLOM201587,Chen2015}  \\ 
 \hline
 cherry-picking & malinformation practice that selects only the most beneficial information to the author's argument from what is available~\cite{Asudeh2020}. \\  
 \hline
 hate speech &  abusive malinformation that targets certain groups of people, expressing prejudice and threatening~\cite{Paz2020} \\ 
 \hline 
 cyberbullying & form of bullying that uses electronic communication, usually social media, that can contain misinformation, rumour and hate speech~\cite{8947876,MAFTEI2022107032}  \\ 
 \hline 
 troll & social media user that uses disinformation to increase the tension between different ideas~\cite{7752428} \\ 
 \hline 
 astroturf & disinformation practice of orchestrating a campaign masking its supporters and sponsors as grass-roots participants~\cite{Kovic_Rauchfleisch_Sele_Caspar_2018,Zerback2021} \\  
 \hline 
 crowdturf & crowdsourced astroturf \\  
 \hline 
 spam &  unsolicited information that overloads its recipients~\cite{Jindal2008,FEI2017141} \\ 
 \hline
 social bot  & a social media user controlled by a software that mimic human behaviour; often used a tool for spamming, spreading misinformation, and astroturfing~\cite{Stieglitz2018} \\  
 \hline 
 satire &  false information but intentionally harmless in the majority of cases, even if often it has strong political references~\cite{Lamarre09theirony} and can be misused as a propaganda practice \\ 
 \hline
\end{tabularx}
\caption{A list of terms referring to different key aspects characterising information disorder.}
\label{table:taxonomy}
\end{small}
\end{table*}

Other important contributions to the debate on terminology can be found in Lewandowsky et al.~\cite{BeyondMisinformation}, Fallis~\cite{whatIsDisinformation}, Floridi~\cite{Floridi}, Stahl~\cite{Stahl}, Fetzer~\cite{Fetzer}, Berghel~\cite{Berghel}. The interested reader can also check out  \url{fakenewsresearch.net} to explore the paper's dataset presented in Sec.~\ref{sec:network}.

\section{An information overload problem: bibliometric analysis}
\label{sec:network}

This survey is partially guided by data. Although we have a general understanding of the state of the art, that we have also contributed to on our own (e.g., ~\cite{Giachanou2019,GiachanouNLDB2020,Fakenewslab}), we are also aware that in the last years the size of related literature has grown considerably, and that it is difficult to overview the state of the art objectively and exhaustively. Moreover, results have been published and discussed in journals and conferences that belong to different scientific communities, often distant from our areas of expertise. In this section, we describe the methodology we followed to discover many of the contributions we review in this survey, and how a data-driven analysis supported us to fill some gaps and spot underlying narratives. To be noticed that the analysis presented here is based on methodologies from network science and, partially, computational linguistics. General methods, definitions and basic concepts are reviewed in the Supplementary Material. Some well-consolidated guidelines to set up a bibliometric analysis on a generic topic can be found in Donthu et al.~\cite{DONTHU2021285}.

\subsection{Data collection and library creation}
The principal objective of our analysis was to create a corpus of all the recent contributions on fake news, and related topics, plus the main references. The purpose of such a dataset is to eventually maintain a comprehensive and continuously evolving library of the existing literature on this topic. On top of this library on fake news research, we implemented some tools to explore the papers, follow citations, and execute some network analysis to identify clusters of the related graph that could suggest the spontaneous emergence of underlying sub-narratives. Moreover, we wish to easily and objectively identify published contributions that should not be missed in a survey; in fact, such a dataset can be represented by a directed graph, whose nodes are papers, links are citations, and network centrality measures can be exploited to find publications that are likely to be relevant, and not to be skipped.

To create the ``fake news research'' library, we started from a set of seed papers we already knew. Then, we download all the papers, along with their metadata, whose titles contain the following keywords: \emph{fake news}, \emph{disinformation}, \emph{misinformation}, \emph{hoax}, \emph{false news}, \emph{social bot}, \emph{fact-checking}, \emph{infodemic}, \emph{information disorder}. To build the dataset we initially queried the Microsoft Academic Graph (MAG)~\cite{MAG}, then we adopted OpenAlex~\cite{OpenAlex} since the beginning of 2022. 
Some stemming and lemmatization was applied to include the most common variants: e.g., ``hoax'' and ``hoaxes'', as well as ``fact checkers'', ``fact checking'', ``fact-checkers'' and ``fact-checking''. This search yielded 19,294 papers, for any of which we kept this information: paper id, title, authors, authors' institution, publication date, abstract, list of the so called \emph{Concepts} (i.e., tags that the OpenAlex engine extracts from the paper), list of references (only to papers in OpenAlex databases), and abstract. We also saved an estimate of the number of global citations the paper received.

\begin{figure}[ht!]
    \centering
    \begin{subfigure}[b]{0.47\textwidth}
        \centering
        \includegraphics[width=0.93\textwidth]{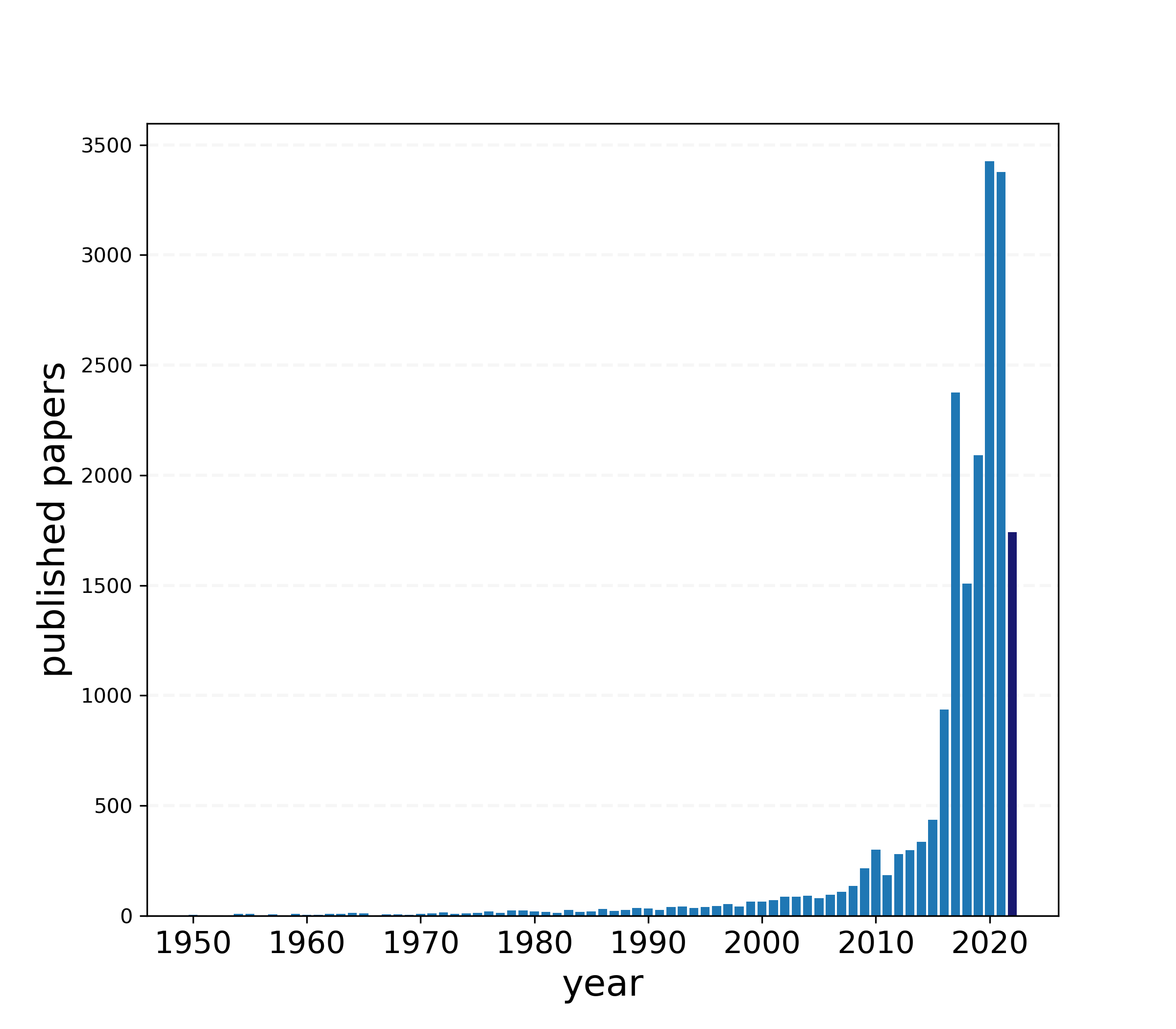}
        \caption{{\small Yearly percentages of papers w.r.t. the size of the collection itself. Although the trend is growing since 2000s, it is after 2016 that the scientific production really spiked. 2022's bar is in progress.}}   
        \label{fig:papers_year}
    \end{subfigure}
    \hfill
    \begin{subfigure}[b]{0.47\textwidth}  
        \centering 
        \includegraphics[width=0.8\textwidth]{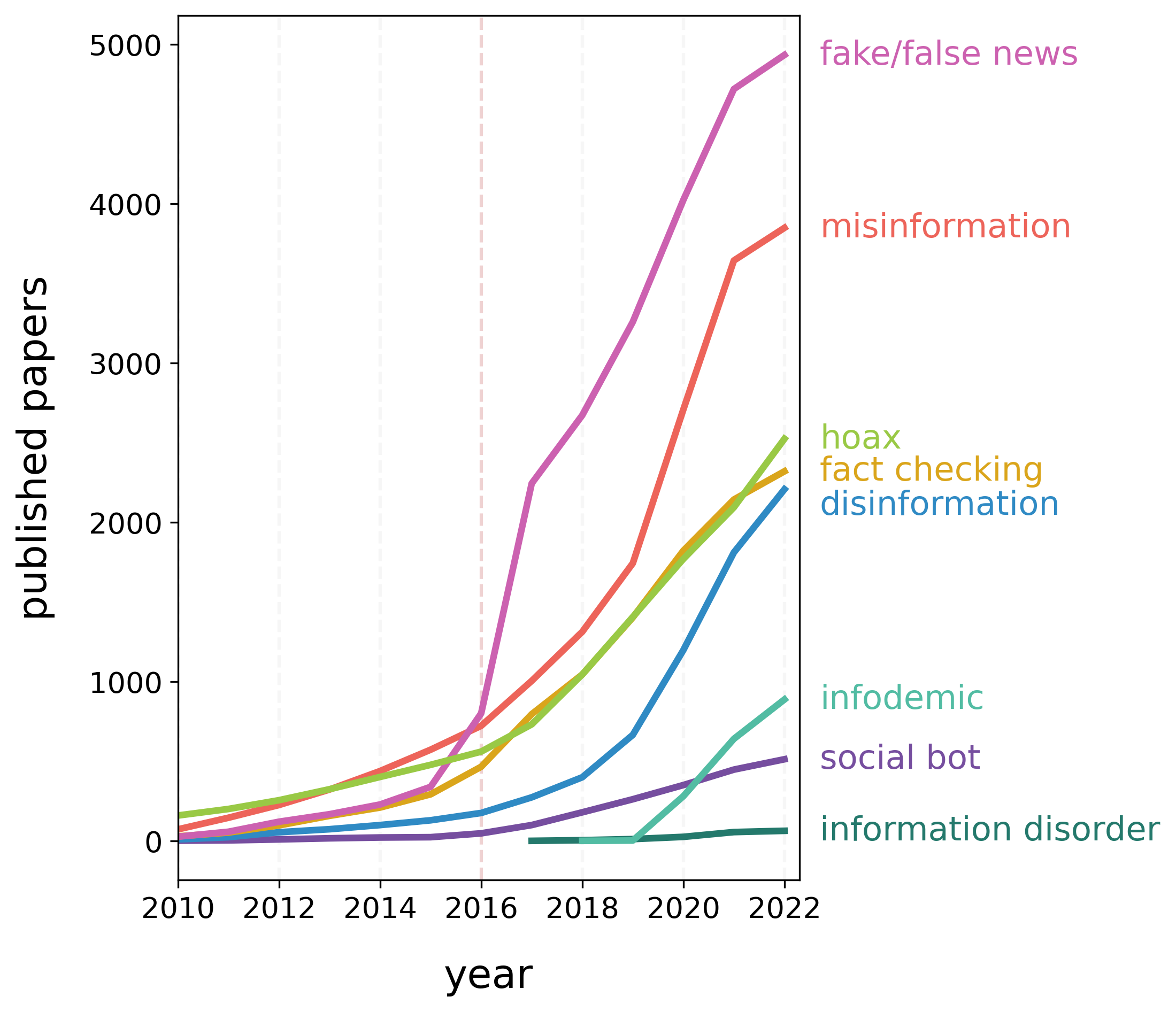}
        \caption[]%
        {{\small Cumulative number of papers by year and by keyword. ``Misinformation'' and ``hoax'', related not only to fake news, have a longer history of research. All the keywords have come to huge scientific attention mostly after 2016.}}   
        \label{fig:flag_by_year}
    \end{subfigure}
    \vskip\baselineskip
    \caption{Preliminary observations in the ``fake news research'' library's core.} 
    \label{fig:corelib}
\end{figure}

We filtered out all the contributions in languages other than English, and we kept only papers published in journals or conference proceedings. We call this set of papers the \emph{library's core}, since it contains contributions that focus explicitly on the above mentioned key terms. Fig.~\ref{fig:papers_year} shows the number of these papers by year. This is a very trendy field, yet it is newborn: most of the thousands of papers published - most of the knowledge we have - is just 6 years old, or even less.

Fig.~\ref{fig:flag_by_year} shows the cumulative number of papers published with one of the query keywords in the title. To properly comment on the plot, there is an important caveat to address: the keyword ``misinformation'' is often related to the ``misinformation effect'' in psychology, i.e., how patients (mis)remember facts and how memories can be manipulated years later. Given that, although ``misinformation'' and ``hoax'' have a long history of scientific publications, it is in the last years that they became popular, along with other keywords, ``fake news'' on top of all. Also, this research field is an extremely multidisciplinary one: OpenAlex automatically extracts from the papers a list of tags, namely \emph{Concepts}; Concepts are organised in a 5-level hierarchical acyclic directed graph, where each Concept may have parents Concepts and children Concepts. Level-5 Concepts are specific topics, techniques or keywords related to some specific problem; level-0 Concepts are the most general Concepts, and they roughly define disciplines more than topics. 
\begin{figure}[ht!]
        \centering 
        \includegraphics[width=0.47\textwidth]{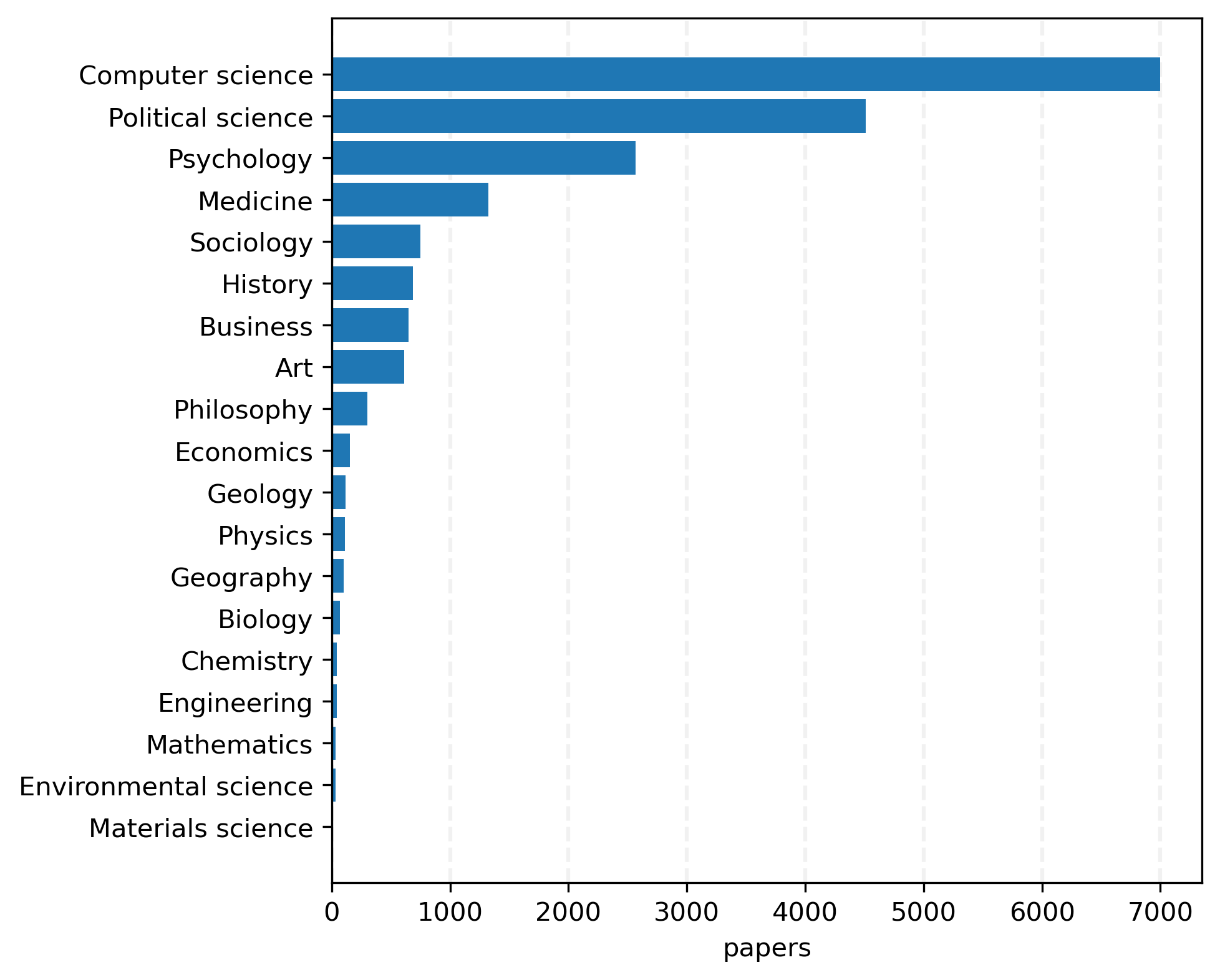}
        \caption[]%
        {{\small Percentages of papers per level-0 Concepts (disciplines) to the total library's size. Computer, political sciences, and psychology are the most interested disciplines in the fake news problem.}}  
        \label{fig:fos_distribution}  
\end{figure}
In Fig.~\ref{fig:fos_distribution}, we reported the distribution of level-0 Concepts. This field of research is dominated by contributions that suggest how the spread of misinformation in social media can have a big impact on democracies, elections, and government agendas, so it is not surprising that computer and political science are the predominant disciplines. Psycho and social scholars are heavily engaged in this field of research, even if we have to consider that this is largely due to the polysemy of the word ``misinformation'', which is also related to a cognitive effect that leads to inaccurate memories. Of course, many psychological explanation of ``post-truth'' era we are living in can be traced back to the misinformation effect~\cite{BeyondMisinformation}, but we will review cognitive biases later in Sec.~\ref{sec:psychological}. The other most involved disciplines are medicine, sociology, history, art and business, and so on, as a demonstration that disinformation is a complex problem, that can be studied from several perspectives. 

Assuming the presence of topicality in the underlying citation graph, as it happens in other information networks~\cite{wengetal2015}, we also downloaded all the new papers that were not retrieved after the first round of queries to OpenAlex APIs. This is equivalent to running one step of a snowball sampling from references in the core, and it has two immediate consequences: we have a greater chance to find other relevant papers that we skipped by searching only by means of keywords, and then to include in our library also some fundamental papers of interest for whom is planning to do research on fake news, even if the presented contribution is not directly related to this topic. For example, this can happen for methodological papers, or for other ``classics'' that can be only loosely connected to disinformation, or that pioneered this field when it was not a trendy topic at all. We call this new set of 71,272 papers the ``library's periphery''. The entire library counts 90,566 papers, with a total of 606,574 references between them. Of course, this collection may include some miscast paper, but we decided to keep any reference. 

\begin{figure*}[ht!]
    \begin{subfigure}[b]{0.47\textwidth}
    \centering
    \includegraphics[width=\textwidth]{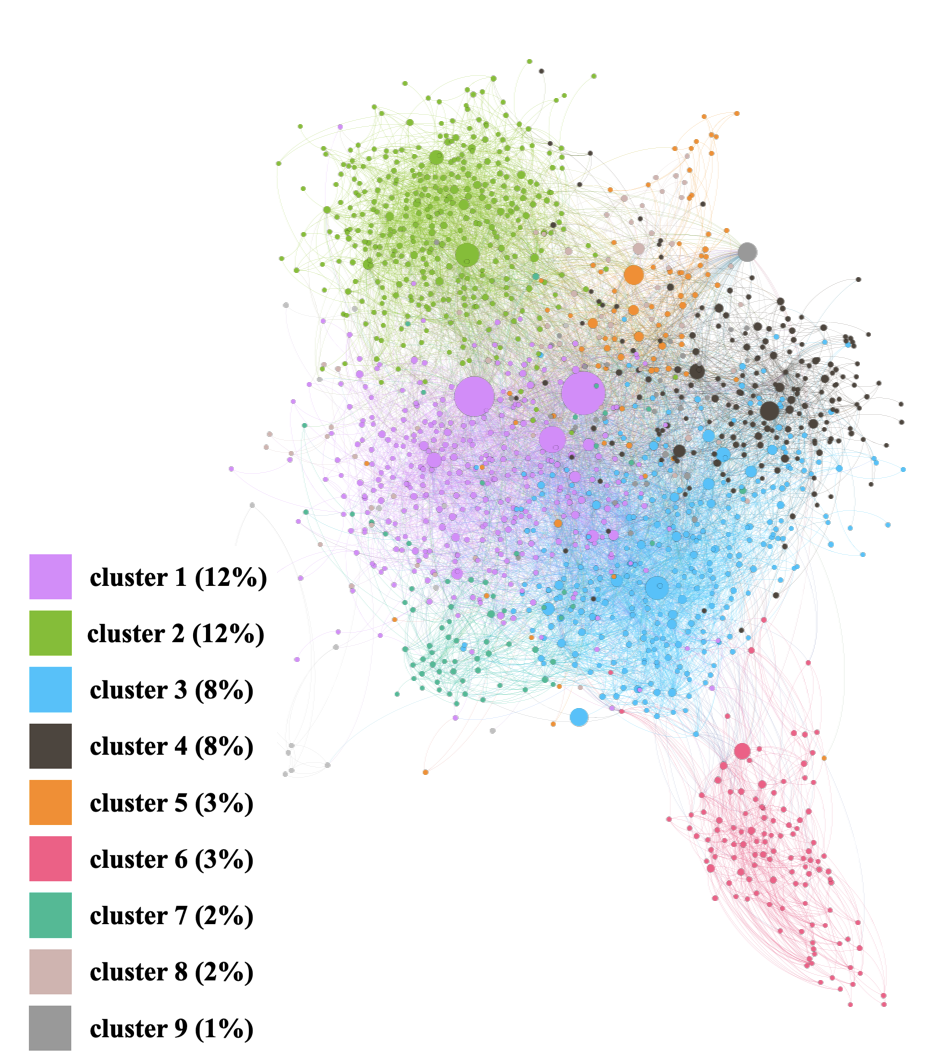}
    \caption{Nodes size is proportional to the number of papers' global citations; edges directions to be read clock-wise. Smaller clusters are shown in grey. Nodes whose in-degree is smaller than 3 have been filtered out to reduce overlapping noise: the most cited 1,530 papers out of the 19,294 in the core are represented here.}
    \label{fig:core_lib_network_a}
    \end{subfigure}
     \begin{subfigure}[b]{0.47\textwidth}
     \centering
    \includegraphics[width=\textwidth]{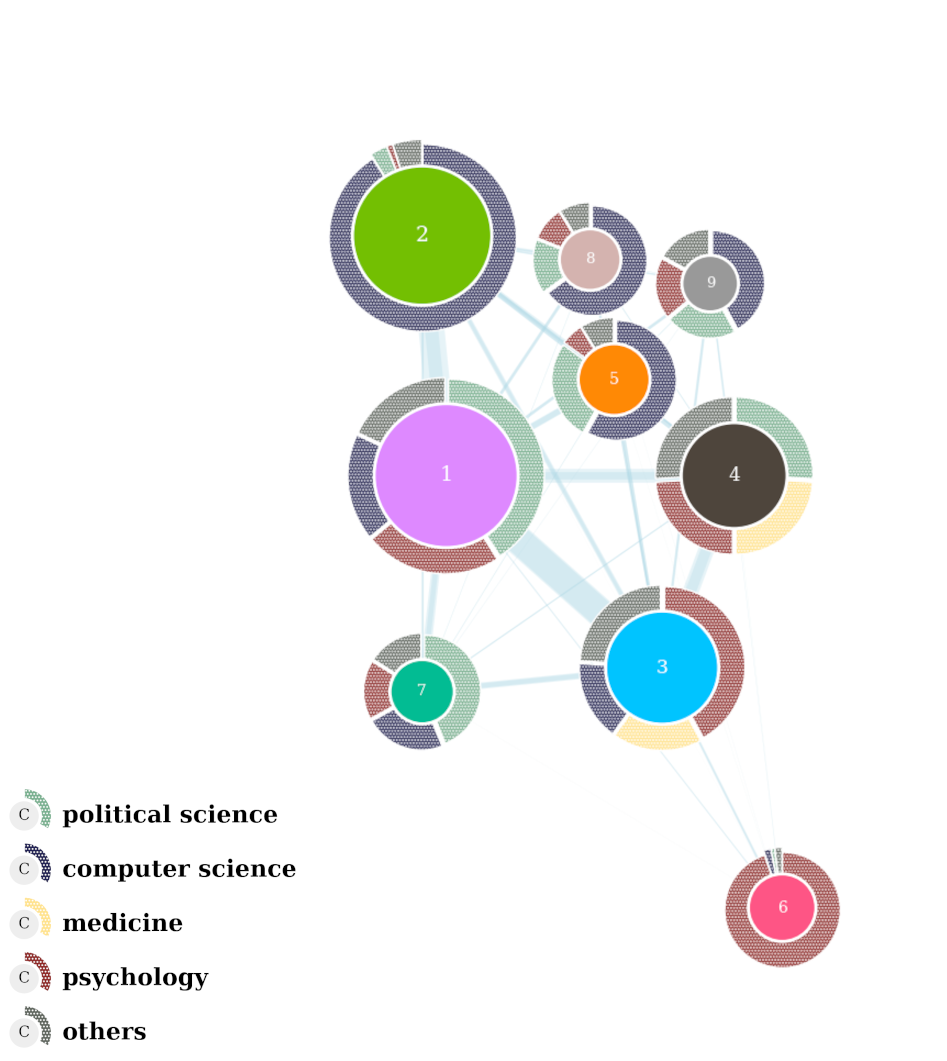}
    \caption{Links between clusters are weighted accordingly the references between papers in different clusters. Node's size is proportional to each cluster's number of nodes. The ring around each cluster node is a donut chart representing disciplines representativeness. Values are in Table~\ref{tab:communities}}
    \label{fig:core_lib_network_b}
    \end{subfigure}
     \caption{Visualisations of the core network's giant component at two different resolutions: nodes are papers to the left, and clusters to the right. A force based layout is applied to the left, and clusters are positioned accordingly to the right. Nodes and clusters to which they belong share the same colour code.}
    \label{fig:core_lib_network}
\end{figure*}

\subsection{Citation graph}
Our aim is is to make sense of this huge and multidisciplinary corpus, and to this purpose we want to better understand the structure of the network representing the library, to find clusters, and to identify central nodes. To create a citation (directed) graph, we represent every paper in the library with a node, and a citation between two papers with a directed edge. In the rest of our survey, we will call the ``core network'' the subgraph that contains all the papers in the library's core, and the ``periphery network'' the subgraph that contains the papers in the library's periphery. In this section we mainly focus on sub narratives that we detected from the graph's clustering analysis, and we refer the reader to the Supplementary Material for the in-depth discussion on central nodes of the graph (i.e., the likely most influential papers in the field).

First of all, we observe that the core network is made of hundreds of very small disconnected components, plus the giant component whose size is $\approx 53.03\%$ of the core network. The remaining papers in the core do not have any citations: most of them are very new at the time of writing this survey, as $20\%$ of them have been written from 2021 on. Others have been ignored by the rest of the scientific community, at least so far. 

\subsection{Cluster's analysis}
In Fig.~\ref{fig:core_lib_network}, we visualise the giant component that emerges from our core library only. We run the Louvain community detection algorithm~\cite{Blondel_2008} to identify tightly-knit groups of papers, to let emerge a coarse-grained characterisation of the underlying sub narratives. In Table~\ref{tab:communities}, we describe high-level differences between these clusters, in terms of their papers' main disciplines. We deduced the nodes' disciplines downloading the ordered lists of level-0 concepts for each corresponding papers, and assuming that the first Concept in these lists represents the main paper's discipline. Then, for each cluster, we extract the three most frequent disciplines.

\begin{table*}[]
\begin{footnotesize}
    \centering
    \begin{tabular}{|c|c|c|c|c|}
    \hline
    \multicolumn{2}{|c|}{\textbf{Cluster}} &  \multicolumn{3}{c|}{\textbf{Most frequent disciplines}}\\
   \hline
    \textbf{id} & \textbf{num of papers}  & \textbf{1st} & \textbf{2nd} & \textbf{3rd}\\
    \hline
    {\squareone} 1   & $1158$   &  political science ($41\%$)  &  psychology ($23\%$)& computer science ($18\%$)\\
    \hline
    {\squaretwo} 2   & $1003$   &  computer science ($91\%$)  &  political science ($3\%$) & psychology ($1\%$)  \\
    \hline
    {\squarethree} 3  & $778$  & psychology ($42\%$)  & medicine ($18\%$) & computer science ($16\%$) \\
    \hline
    {\squarefour} 4  & $700$   & political science  ($26\%$) & medicine ($24\%$) & psychology ($24\%$) \\
    \hline
    {\squarefive} 5   & $289$  &  computer science ($58\%$) & political science ($27\%$) &  psychology ($6\%$)   \\
    \hline
    {\squaresix} 6  &  $237$  & psychology ($95\%$)  & computer science ($2\%$)  & political science ($1\%$) \\
    \hline
    {\squareseven} 7  & $199$  & political science ($44\%$)  & computer science ($23\%$) &  psychology ($17\%$)\\
    \hline
    {\squareeight} 8  &  $164$   & computer science ($65\%$) &  political science ($16\%$) & psychology ($1\%$) \\
    \hline
    {\squarenine} 9  &  $104$  & computer science ($42\%$)  &  political science ($22\%$) & psychology ($19\%$) \\
    \hline
    \end{tabular}
    \caption{The three most frequent disciplines in each of the nine biggest clusters found in the core network. Disciplines are deduced from OpenAlex level-0 Concepts. Colour code as in Fig.~\ref{fig:core_lib_network}.}
    \label{tab:communities}
\end{footnotesize}
\end{table*}

The structure of the core networks (Fig.~\ref{fig:core_lib_network}) and the papers' disciplines (Table~\ref{tab:communities} suggest that we have a central sub-graph of highly multidisciplinary contributions, that is composed of clusters $1$, $3$, $4$, and $7$ whose predominant disciplines are psychology, political science, and medicine. Quite interestingly, we also have at least two clusters that are much more mono-disciplinary than others: cluster $2$, at the top of the giant component, that is distinctly focused on computer science, and cluster $6$, at the opposite side, that is definitely specialised on psychology. At a closer look, other smaller clusters, like $5, $$8$ and $9$, that are in the top right of the Fig.~\ref{fig:core_lib_network}, have a clear lean towards computer science, while the already mentioned cluster $3$, at the centre-bottom, is more characterised by psychological contributions. 

Let us observe that the bottom ``psychology'' cluster $6$ is more loosely connected to the rest of the giant component, w.r.t. the other clusters. We think that this is mainly due to the ambiguity on the ``misinformation'' term that we mentioned above. However, it is worth noticing that cluster $1$, $3$, and $4$, that are more likely to contain contributions actually related to fake news research problems, have a significant psychological characterisation, with relatively more links to cluster $6$ than others. This observation helps us to visually explore the backbone Fig.~\ref{fig:core_lib_network_b} from a top-down perspective: it looks like that applications and technological solutions on the top are layered on problems that arise in particular in health and political sciences, and that are probably explained in terms of psychological effects and cognitive biases. 

\subsection{Most influential papers}
Fig.~\ref{fig:core_lib_network_a} shows also some information on bibliographic references that emerge clearly in the core network: node's size is proportional to the number of citations received by the corresponding paper. Let us recall here that a reference is usually an endorsement of someone else's work, hence, highly cited papers are publications to which many other researchers recognise value. The multidisciplinary contributions in cluster $1$ that attracted more attention corresponds to the fundamental contributions from Vosoughi et al.~\cite{Vosoughi1146}, Allcott and Gentzkow~\cite{Allcott2017}, and Lazer et al.~\cite{Lazer1094}: the first and the second paper presents two strong empirical arguments on the prevalence of fake news in social media, the third is the already mentioned ``the science of fake news''  manifesto. These studies contributed significantly to the adoption of the general terminology (see Sec.~\ref{sec:terminology}), to the general understanding of the psychological appeal of misinformation using computational linguistics tools (see Sec.~\ref{subsec:appeal}, and to the acceptance of modelling the misinformation spread with networks (see Sec.~\ref{sec:models}).

Hubs in clusters $2$, $6$, and $8$, are also quite representative of the major challenges from a computer science perspective: the role of echo-chambers in the spread of misinformation online, introduced by Del Vicario et al.~\cite{DelVicario554}, the problem of ``fake news'' detection, as presented by Shu et al.~\cite{Shuetal2017}, and the rise of social bots according to Ferrara et al.~\cite{ferrara2016}. We developed further these problems respectively in Sec.~\ref{sec:echochambers}, \ref{sec:linguistic}, and~\ref{sec:bots}.

Furthermore, other significant contributions have important implications on public health, political agendas and fact-checking strategies. Some of them are also likely to be found in cluster $3$, and $4$, e.g., the psychological difficulties of correcting misinformation studied by Lewandowsky et al.~\cite{lewandowskyetal2012}, and the early warning launched by Zarocostas though The Lancet in February 2020 on the forthcoming COVID-19 infodemic~\cite{zarocostas2020}. The psychology biases that make mis-/dis-information difficult to eradicate are reviewed in Sec.~\ref{sec:psychological}. Instead, we will cover some of the most recent contributions on the infodemic emergency in Sec.~\ref{sec:network_language}.

\subsection{Discussion and limitations}
As already discussed in the introduction, we focus on a dual network science and computational linguistics perspective to provide the computer scientist that is approaching this multidisciplinary field with a general and comprehensive framework. There are other surveys focused on the many existing fake news detection methods (e.g., \cite{DONG201857}), or written for a political science~\cite{tucker18} or psychological audience~\cite{WANG2019}. However, from the perspective of a computer scientist, a challenge in this research field is rarely reduced to training a machine learning classification model with data: there are many different methodological and conceptual approaches that could serve to design better solutions. It is worth noticing again that an indisputable fraction of papers that attracted high attention so far, are characterised by results of multidisciplinary interest, and by the adoption of both empirical and theoretical methodologies that represent data and systems as networks. Also, advanced computational linguistics techniques are often used to make sense of textual content, complementing what we can understand observing the structure and the dynamics of information and social networks. To better illustrate why a mixed network/linguistic based approach is necessary to make sense of the most significant contributions published so far, we will outline some state-of-the-art case study in Sec.~\ref{sec:network_language}.

Finally, we must acknowledge the limitations of a data-driven approach to find ``hidden gems'' in scientific literature, and to spot predominant narratives; in fact, even if OpenAlex has good coverage of scientific disciplines, it does not include the 100\% of published papers (an estimate of the coverage of OpenAlex predecessor, MAG, can be found in~\cite{hug2017,soton408647}). Many relevant papers may have been excluded and many references may be missing, and both impact not only the papers' repository coverage, but also the network structure and therefore its centrality measures. Also, we have our own biases, and although a data-driven approach supported us to give relevance also to contributions that can be far from our scientific interest and expertise, the selection of papers to overview is still the result of a personal, subjective choice of topics and publications. Therefore, in order to allow everyone to perform their own queries to the ``Fake News Research'' library, we decided to make our search tool accessible through the Web at \url{http://fakenewsresearch.net}. We meant not only to release the dataset we employed, as in this ever-changing field this dataset will age-old soon, but a long-lasting service, an always up-to-date publication repository about academic research on disinformation, misinformation and other related problems. The user can query the database searching for papers ranked by citation count, date, page rank, betweenness, hub or authority scores, and also by author, institution or Concept (as provided now by OpenAlex~\cite{OpenAlex}, which replaced the now dismissed MAG). Plus, we provide a basic recommendation system that suggests, for each paper, the top 10 most similar publications, where similarity is simply computed on the abstracts with Doc2Vec~\cite{Le2014}.

\section{Psychology of Fake-News: the problem of cognitive biases}
\label{sec:psychological}

In Sec.~\ref{sec:network} we presented the results of the network analysis of the citation graph underlying our ``Fake News Research'' library, a collection of scientific papers on fake news, disinformation, and related topics. We found that psychology is one of the disciplines that studied misinformation the most, even if this is partially due to the polysemy of the word 'misinformation' itself, which also refers to a well studied cognitive effect that occurs to persons that alter the memory of an event with information acquired successively. The so called \emph{post-event misinformation effect} has been studied for decades, also due to its implications on the accuracy of eyewitness testimony (e.g., see Loftus et al. in 1978~\cite{Loftusetal1978}, McCloskey and Zaragoza in 1985~\cite{McCloskeyZaragoza1985}), and should not be confused with the recent interest to the spread of fake news on social media. However, we also observed that there are clusters of papers that focus on the psychological reasons behind our tendency to share misinformation, even if the fake content has already been debunked; in fact, there are many cognitive biases that usually drive our judgement for the best, but that can mislead us when exposed to deceptive news. Therefore, in this section we mention the most important psychological effects that are used to give tentative explanations to reported empirical observations on how and why fake news spread on social networks. 

\subsection{The appeal of disinformation}
\label{subsec:appeal}
Why do users fall into fake news in the first place? Researchers have enquired about the relationship between users and false news online, which keeps being attractive even when blatantly false. Disinformation works through a series of cognitive hooks that make the information appealing, triggering a psychological reaction in the reader. First of all, fake news is attractive. They usually contain more novelty than regular news~\cite{Vosoughi1146}, something that is captivating for a distracted reader. Plus, they deliver information simpler to process~\cite{Horne2017} compared to regular news, and a simple message that can be easily digested even when a person is paying low attention. Also, fake news often triggers an emotional response~\cite{emotionsPartisanship, Bakir2018, stella2018bots, Vosoughi1146}, proven to be one driver for online virality. Readers prefer high arousal news~\cite{Bakir2018}, which catches more attention than regular news and sticks easily on mind.

\subsection{Exposure to disinformation}
\label{subsec:exposure}
Another factor that determines the success of disinformation is the repeated exposure to the same fake news, as argued by Pennycook et al.~\cite{PriorExposure}, because repetition increases the content perceived accuracy. Repetition is a well known propaganda mechanism, and it is exploited by social media filtering algorithms that decide what content deliver to whom, by indulging the users' tastes, eventually building a reinforcement feedback loop on users' walls. This makes appealing content even more visible, as long as the user interacts with them, and conversely it may hinder content that the user dislikes. When applied to political issues, for instance, this algorithmic bias easily produces a filter bubble~\cite{filterbubble}, a content selection mechanism that the social media platforms may apply to one user's online feed. Additionally, when algorithms highlight the content shared by a user's friend according to their opinions, this effect may grow stronger, triggering the emergence of echo chambers~\cite{Sunstein2002}.
Disinformation may grow on the user's mind, as repetition has been proven as one of the most effective techniques of persuasion for online news~\cite{Unkelbach2019}, a phenomenon also known as \emph{validity effect}, or \emph{illusory truth effect}~\cite{Hasher1977}. 

Repeated exposure to others' behaviours is also a key element of the so called `complex contagion' effect (see the Supplementary Material). In other words, if a user is exposed repeatedly to an opinion (or a behaviour, an innovation, and so on), the probability of adopting it is higher. The hypothesis that information and diseases spread as `simple contagions', and that behaviours require multiple interactions with the source was proposed by Centola and Macy in~\cite{centolamacy2007}. Also, a behaviour spreads faster in clustered networks because of social reinforcement, as proven by Centola with his experiment on a controlled social system~\cite{Centola1194}.

\subsection{Social biases}
\label{subsec:social_biases}
Other reasons to engage with low-quality, partisan content often root in social behaviour: the \emph{bandwagon effect}~\cite{Nadeau1993}, for instance, is the tendency to align to the opinion of what is perceived as a big mass of other individuals; our perception of what is right and what is wrong may depend on the social group we think we belong: in this mindset, the group's opinion is assumed to be the public opinion, whereas discordant news are perceived as hostile attacks towards the public. The \emph{normative social influence theory}~\cite{C.NathanDeWall2011SAaR} hypothesises that people conforms to the behaviours and ideas of the social group they belong to, while the influence of social identity has been showed to play an important role in news perception~\cite{Schulz2020}. These misperceptions are then backed up by other cognitive biases inherited by our social behaviour in case of us-against-them dynamics. It is the case of the \emph{third-person effect}~\cite{Davison1983}, the tendency to believe that disinformation affects more the other group than the one the individual belongs to. These effects are often associated with what is called \emph{naive realism}~\cite{NaiveRealism}, the tendency of an individual to assume that the reality they perceive is objective and factual: if applied on news and the perception of the world, it creates a \emph{false consensus effect}~\cite{ROSS1977279}, a tendency for individuals to think their beliefs are widely accepted; under this perspective, adverse news by hostile news outlets can be classified as a malicious attempt to undermine a widespread acknowledged fact.

All of these biases are reflected and also amplified in the structure of the social network the users belong to; in fact, ego-networks, i.e., the persons connected to single individual (the ``ego''), are not unbiased, and they do not cover the full spectrum of available beliefs, ideas, opinions. Homophily in networks (for a review on this topic, see McPherson et al.~\cite{McPherson2001}) is a well documented tendency of individuals and their on-line personas to get along with similar others: socio-demographic indicators are usually very homogeneous in the list of acquaintances of an individual, and this may favour an homogeneity of opinions toward economics, politics and society. This is a key factor for the formation of clusters and echo-chambers in social networks, as we will discuss in Sec.~\ref{sec:echochambers}.

\subsection{Individual biases and the challenges of correction}
\label{subsec:individual_biases}
Finally, there are many biases working as reinforcement feedback for one individual's beliefs, in a loop that may strengthen their misperception. In fact, reinforcement and repetition of news are not only the result of algorithmic biases, echo chambers, and clustered social networks, but also a self-inflicted deception. People tend to notice one content if they are already thinking of it, a mechanism known as \emph{attentional bias}~\cite{williams1997cognitive}: if we convince ourselves of a shocking, disturbing fact, we may start noticing that fact often. The highly mentioned \emph{confirmation bias}~\cite{ConfirmationBias} also plays an important role in news selection, as users tend to focus on information that confirms prior beliefs, instead of news that challenges them. Confirmation bias is often found in literature after the name of \emph{selective exposure}~\cite{FREEDMAN196557}; in fact, people firmly rely on news that agrees with them, but rarely seek for disproofs of their beliefs, a bias known as \emph{congruence bias}~\cite{Wason1960}. Some headlines can be perceived as stronger despite the weak arguments that support the conclusion in an article, if the conclusion resonates with one's expectations, as the effect of the \emph{belief bias}~\cite{leighton2004nature}. Also, as mentioned in Sec.~\ref{subsec:appeal}, disinformation often is conveyed through articles with a high emotional valence. The \emph{emotional bias}~\cite{Barrett2002} may distort the perception of argumentation, for instance bringing one person to refute to acknowledge facts that upset them. Ferrara and Yang~\cite{FerraraYang2015} proposed an analysis of ``emotional contagion'' with Twitter data to find that individuals overexposed to negative content are likely to share negative messages, and vice-versa. In fact, emotional content can be exploited by malicious agents to provoke inflammatory responses and exacerbate polarisation in online debates, as observed by Stella et al.~\cite{stella2018bots} on Twitter during the Catalan independence referendum campaign. 

Misinformation correction and debunking are difficult for many different reasons. Generally speaking, individuals tend to overstate their knowledge about some topics, and to perceive their partial information as highly accurate, two biases named \emph{illusion of asymmetric insight}~\cite{Pronin2001} and \emph{overconfidence effect}~\cite{Pallier2002}, which may lead one person to think they are undoubtedly right. Correction of unsubstantiated beliefs about politics often fails when individuals with strong ideological opinions are corrected, as experimentally analysed by Nyhan and Reifler in~\cite{NyhanReifler2010}. They observed also the so called \emph{backfire effect}, i.e., correction may actually increase misperception and ideological biases. Quite interestingly, this publication is one of the most cited in our ``Fake News Research'' library periphery (see Sec.~\ref{sec:network} and the Supplementary Material), suggesting its importance as a pioneering contribution in this field.

Apparently debunking could be supported by the \emph{hyper-correction effect}, i.e., when an individual commits a mistake with high confidence and receives corrective feedback, the same individual is more likely to remember the right information. However, Butler et al.~\cite{butler2011} observed that the hyper-correction effect seems to vanish after one week, and that misperceptions with higher confidence are more likely to return after a given time.

Lewandosky et al.~\cite{lewandowskyetal2012} review many cognitive biases and effects that make misinformation correction so difficult, suggesting a list of guidelines for debunkers, educators, journalists, science communicators, and health professionals. However, the \emph{continued influence of misinformation}~\cite{ecker2010}, that makes people rely on information that is already retracted or falsified, is still a crucial and controversial point for designing proper correction narratives: a common strategy is to avoid repeating misinformation during the correction, because it may strengthen the effectiveness of false information. However, Ecker et al. in 2017~\cite{Ecker2017RemindersAR} found that retractions that explicitly mentioned the misinformation were more effective than corrections, because they avoided repeating the false information. The analogies and differences between the continued influence effect and the post-event misinformation effect (mentioned at the beginning of this section) have been studied by Ecker et al. in 2015~\cite{Ecker2015HeDI}, establishing a link between two important traditions of causal explanations.

\section{The problem of the interplay between network, opinions, and misinformation}
\label{sec:echochambers}

The relationship between fake news spreading and echo-chambers formation is one of the most studied aspects of this newborn scientific sub-field; in fact, misinformation's prevalence in social media is often attributed to the so called ``echo chambers'', tightly knit clusters of users that keep interacting with one-sided content until they get radicalised, as a result of a reinforcing feedback loop. Misinformation has been observed to circulate consistently within echo chambers~\cite{Bessi2015ViralMisinfo,DelVicario554}; research has showed that polarisation ``plays a key role in misinformation spreading''~\cite{DelVicarioFakePol}, while for Osmundsen and al.~\cite{osmundsen2021} it is the ``primary motivation''. Studying the relationship between users and their one-sided neighbourhood is particularly interesting because of a non-linear characterisation of the corresponding causality. Trivially speaking, homophily triggers social network clustering, that amplifies opinion polarisation, leading to the emergence of echo chambers. However, we can also observe a process following a reversed direction: echo chambers amplify polarisation, which increases network clustering, which leads to higher homophily. Co-evolution of network structure and dynamics is also exacerbated by filter bubbles and social media business models.
The process of radicalisation of users has been explained by several dynamics that can nudge an idea to a user and to their neighbours, slowly building a cluster of like-minded social media users. How opinions are adopted by individuals and spread across a network of relationships is a research problem on its own, with many related academic publications that have been published over the last decades; we briefly review the most relevant contributions to opinion dynamics in the Supplementary Material.
In this section, we focus on those models and empirical results of particular interests for phenomena that could be observed in social media, also to better contextualise the interplay between network structures, the formation of echo chambers and the spread of misinformation.

\subsection{The emergence of polarisation: some definitions}

Since echo chambers are well connected to the concept of polarisation, we need a definition. As stated in Dimaggio et al.~\cite{Dimaggio1996HaveAS}:

\begin{quote}
    Polarization is both a \emph{state} and a \emph{process}. Polarization as a state refers to the extent to which opinions on an issue are opposed in relation to some theoretical maximum. Polarization as a process refers to the increase in such opposition over time.
\end{quote}

\begin{figure}[ht!]
    \centering
    \begin{subfigure}[b]{0.475\textwidth}
        \centering
        \includegraphics[width=0.3\textwidth]{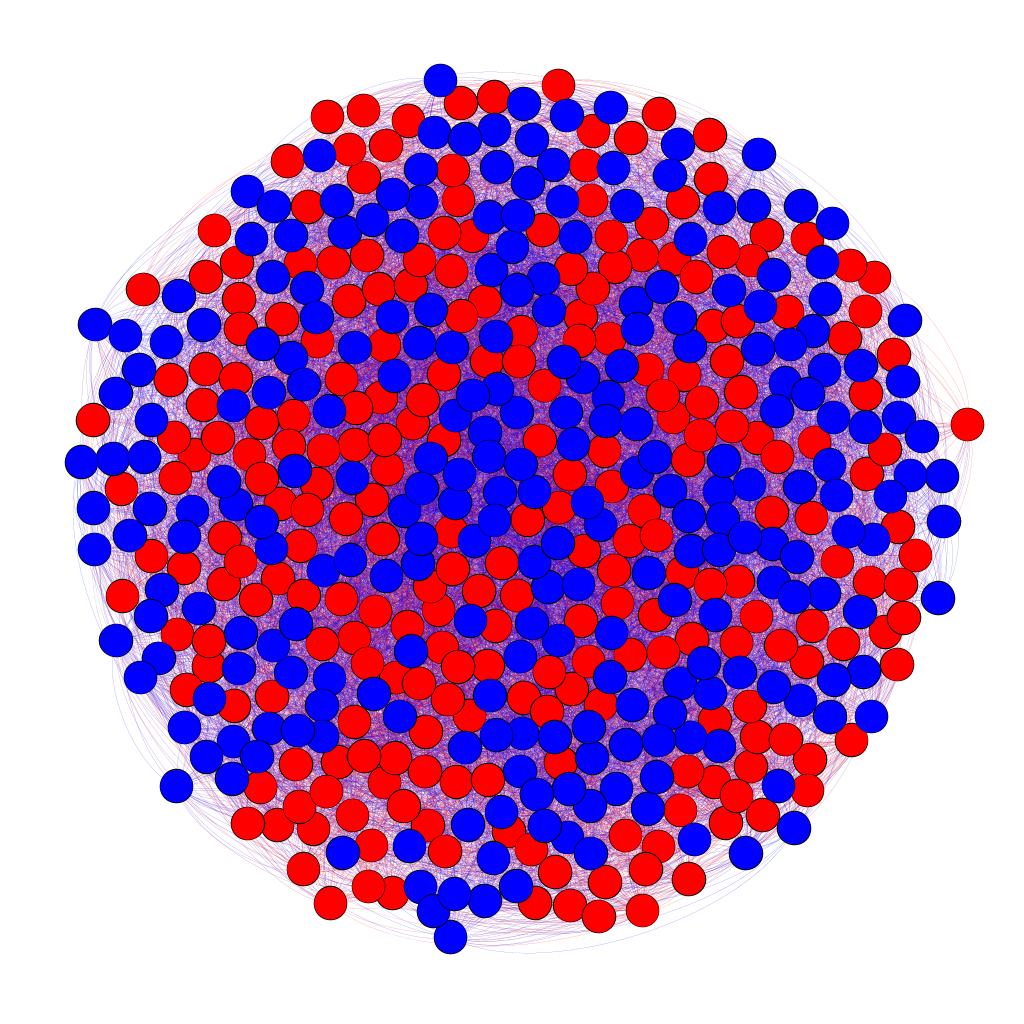}
        \caption{{\small well mixed network, with no clustering}} 
        \label{subfig:mixednet}
    \end{subfigure}
    \hfill
    \begin{subfigure}[b]{0.475\textwidth}  
        \centering 
        \includegraphics[width=0.6\textwidth]{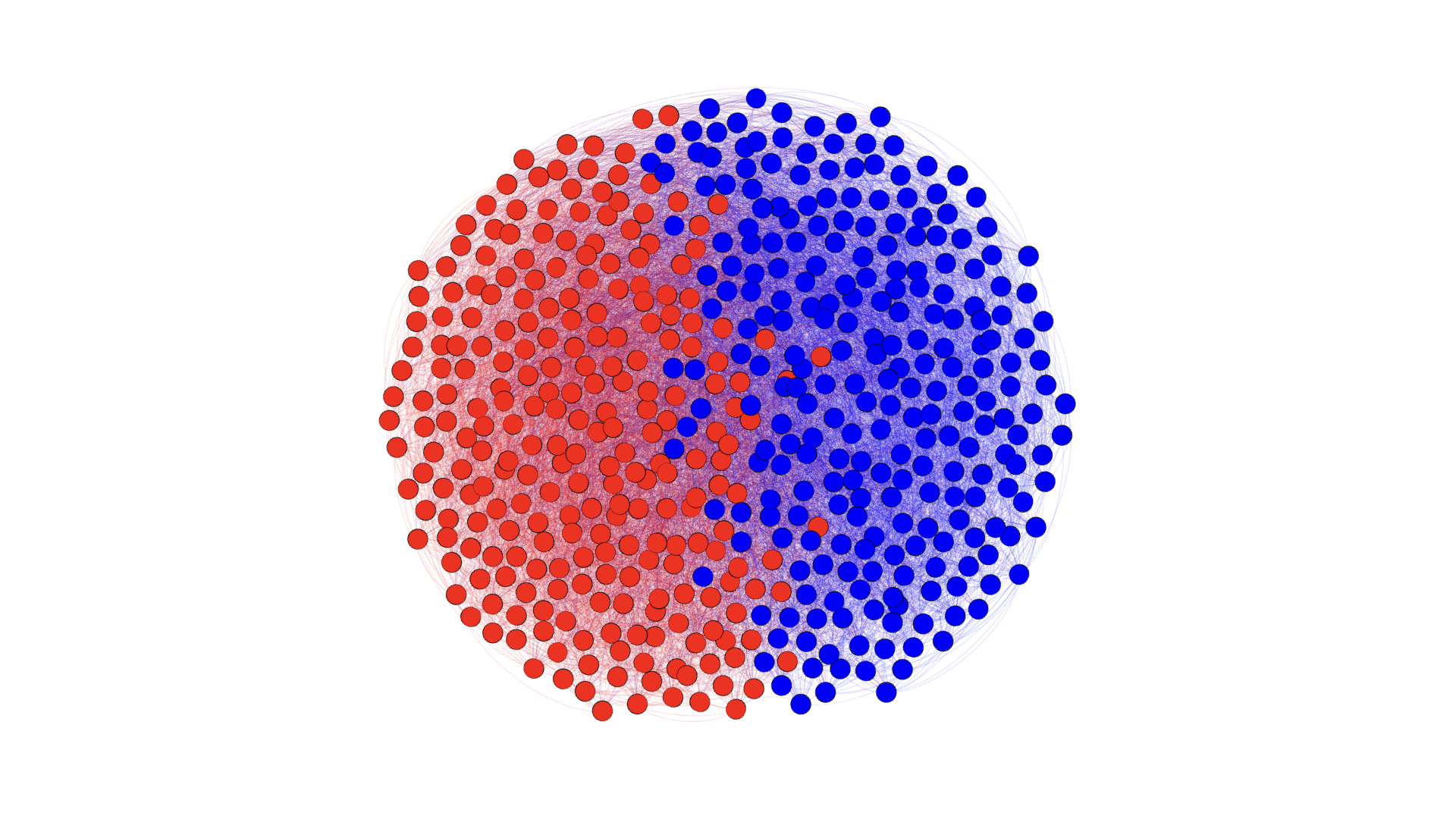}
        \caption[]%
        {{\small high homophily may coexist with a cohesive society}}   
        \label{subfig:coexist} 
    \end{subfigure}
    \vskip\baselineskip
    \begin{subfigure}[b]{0.475\textwidth}   
        \centering 
        \includegraphics[width=0.5\textwidth]{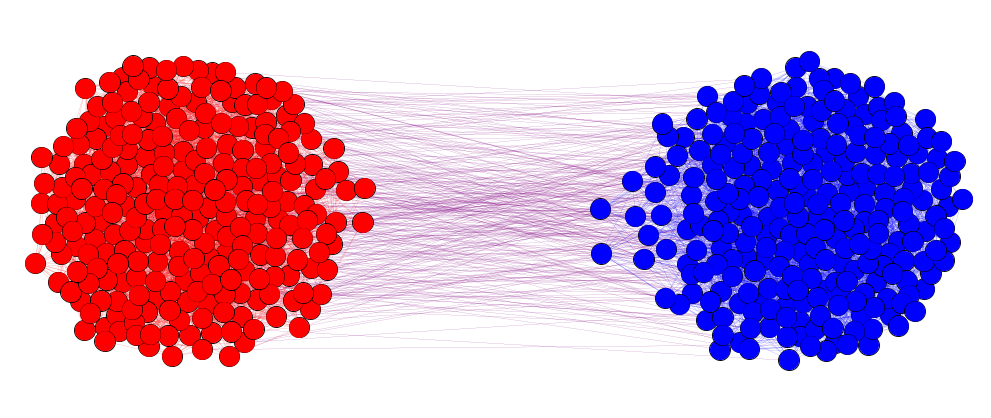}
        \caption[]%
        {{\small higher separation leads to polarisation and echo-chambers}}  
        \label{subfig:polarisation}  
    \end{subfigure}
    \hfill
    \begin{subfigure}[b]{0.475\textwidth}   
        \centering 
        \includegraphics[width=0.5\textwidth]{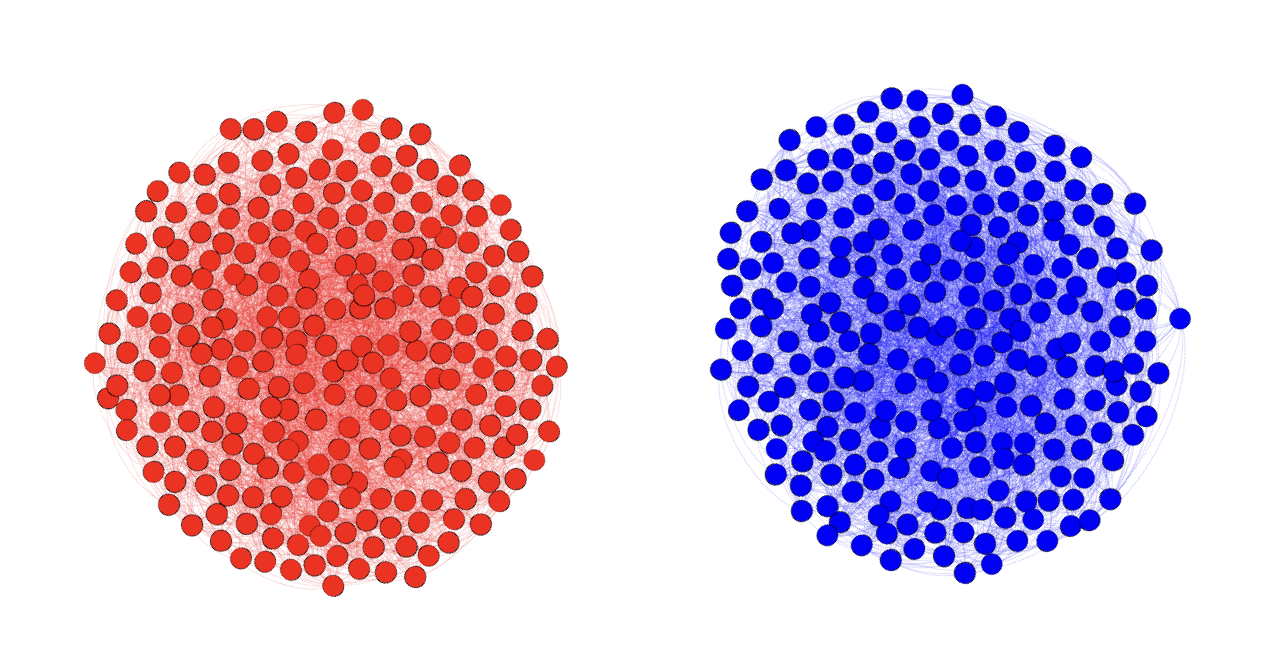}
        \caption[]%
        {{\small extreme polarisation can be irremediably divisive}}    
        \label{subfig:division}
    \end{subfigure}
    \caption{Illustrative example of the structural evolution of an artificially generated network of 1000 nodes, each belonging to a different class (e.g., exposing two different opinions, or adopting not compatible innovations). The network, artificially generated with a stochastic block model, keeps the same internal groups cohesion, but with a progressively higher separation between the two classes of nodes, until no cross links are left, and the graph is finally divided in two components.} 
    \label{fig:polarization1}
\end{figure}

Henceforth, we will refer to opinions as states of the users (or the nodes of a social network), and polarisation as mainly the process that leads to an irredeemable divide, i.e., an equilibrium point or a stationary state of the system, between two groups, each representing a one sided view over a topic. An illustrative example of such a process is given in Fig.~\ref{fig:polarization1}, where we show four possible states of a social network, the first representing a well-mixed society where different opinions do not correspond to a structural division, to an extreme polarisation whose outcome is a graph made of two disconnected components. Echo chambers emerge when polarisation is close or equal to its theoretical maximum, such as in Fig.~\ref{subfig:polarisation} and~\ref{subfig:division}.
Models explaining the emergence of consensus or polarisation in social media by means of opinion dynamics are reviewed in Sec.~\ref{subsec:opinion_dynamics}. 

As stated above, there are documented empirical observations suggesting that the spreading of misinformation is amplified by echo chambers; however, fake news circulates widely even in networks without a clear structural polarisation. In the subsections below, we will give references to the debate that is still going on, pointing out some research questions that remain unanswered. A review of the many empirical findings on well-known information networks is accounted for in Sec.~\ref{subsect:magnitude_polarisation}.

\subsection{Modelling opinion dynamics on online social networks}
\label{subsec:opinion_dynamics}

Analogously to the literature on opinion dynamics on social networks, we define opinions as continuous values $x$: an agent $i$ has a given opinion $x_i(t)$ at time $t$. Then, we study the equilibrium eventually reached by an agent-based system where each agent $i$ has been initiated (at time $t=0$) to a state $x_i(0)$. If a steady state is reached at time $t^*$, it can represent a \emph{consensus}, where every node shares the same opinion, a \emph{polarisation}, where a group of nodes shares one opinion, and the rest share the opposite opinion. Another typical stationary state is \emph{fragmentation}, when opinions are concentrated around more than two values. 

Many classic opinion spreading models define a set of rules according to which agents may keep their state, or change opinions. Some cognitive effects, such as confirmation and overconfidence biases (see Sec~\ref{sec:psychological}), are traditionally taken as inspiration to define such a set of opinion update rules. For instance, in the Bounded Confidence Model (BCM) by Deffuant et al.~\cite{deffuantetal2000} agents evaluate others' opinions only if they are close enough, and that may result in polarisation and fragmentation scenarios depending on the definition of the acceptance interval~\cite{Hegselmann2009}; other models embedded a \emph{resistance} for an agent to change their mind and agree with others~\cite{FriedkinJohnsen1990,FriedkinJohnsen1999}. However, these effects are not the only possible biases that could explain the emergence of polarisation in current social media. In particular, \emph{limited attention} can have an impact on the users' perception of online news~\cite{PhysRevLett.112.048701}. All the traditional models assume that each user would evaluate every neighbour's ideas at each time $t$. This assumption may not hold in real social networks: each user may evaluate only a fraction of the available ideas among their friends, as the attention that online users offer to news content is actually limited~\cite{Weng2012}, with repercussions on what information is eventually absorbed~\cite{Leskovec2009} (see Sec.~\ref{subsec:appeal}). Hence, this theoretical framework can be adapted to get even closer to a realistic online social network, in order to account not only for users' biases, but also for automatically selected information or other platform-induced biases. 
In fact, there are externalities that can impact opinions diffusion in social media, such as broken or rewired links, or content sorted and filtered according to the strategies of underlying personalisation algorithms. Friending and unfriending users, i.e., establishing or revoking a connection with another user, is a factor that can contribute to polarisation, too. Sasahara et al.~\cite{Sasahara_2020} define a network of agents, each of them with a limited screen that shows the most recent messages posted (or reposted) by their neighbours. Each of these messages conveys an opinion with value $x \in [-1,1]$. Opinions are defined as concordant or discordant as in the \emph{Bounded Confidence Model (BCM)}. At each time step $t$, an agent $i$ sees only the first $l$ messages on the screen. The neighbours that posted or re-posted these messages are listed in set $N^l_i$ The opinion of user $i$ then changes based on the concordant messages on the screen:

\begin{equation}
    x_i(t+1) = x_i(t) + \mu\frac{\sum_{j\in N_i^l} \delta_{ij}(t) (x_i(t) - x_j(t))}{\sum_{j\in N_i^l} \delta_{ij}(t) (x_i(t))}
\end{equation}

where $\mu$ is an \emph{influence strength} parameter (a measure similar to $1 - g_i$, where $g_i$ is a resistance factor, see the Supplementary Material), and the Kronecker delta $\delta_{ij}$ checks concordance w.r.t.  $\epsilon$, that represents mind broadness of the agent, or also a quantification of the agent's confirmation bias:

\begin{equation}
    \delta_{ij}(t) =
    \begin{cases}
        1, &  \text{if $|x_i(t) - x_j(t)| \leq \epsilon $} \\
        0, & \text{otherwise}
    \end{cases}
\end{equation} 
Also, with probability $q$, each agent unfollows a discordant neighbour $j$ and establishes a new connection with a new user. Authors try different strategies, like friending random users, concordant users not currently among friends, or a random user among the originator of a re-post on $i$'s screen: opinions are modelled as explicit messages to be posted or re-posted. Eventually, this system converges to a polarised setting, independently of the rewiring strategy employed. Also, Sasahara et al. found that unfollowing friends accelerates the convergence of the polarisation process, and the formation of two distinct clusters. 

Polarisation is not necessarily a consequence of structural clustering: in Fig.~\ref{subfig:mixednet} we showed an illustrative example of a not clustered network where nodes have two opposite views. However, we know that the presence of clusters makes the diffusion of new ideas very difficult~\cite{morris2000}. Hence, if a fake-news is very common within a very cohesive cluster, then its debunking may be difficult, and sometimes it could radicalise beliefs due to the back-fire effect. This is why echo chambers, as a particular instance of polarisation, are considered particularly dangerous by many. Studying the co-evolution of networks and opinions is one of the most important points in the scholars' agendas. The first models considering co-evolution are due to Holme and Newman~\cite{Holme2006} and Gil and Zanette~\cite{GIL200689}, Gross et al.~\cite{Gross2006} and I\~niguez et al.~\cite{Iniguez2009}. Effects of rewiring in a network may be heavy, even when the tendency to break links with discordant opinions is mild. Schelling's classic model of spatial segregation~\cite{Schelling} shows that even when an agent allows one third of its neighbours to be discordant, still the system converges on a segregated population, i.e., homogeneous neighbourhoods. 

Del Vicario et al.~\cite{delvicario2016modeling} proposed a mix of a particular bounded confidence model and a rewire strategy. They define an \emph{Unbounded Confidence Model (UCM)}, i.e., a BCM where discordant opinions are also taken into account but with opposite values. Agent $i$ updates its status following $j$'s opinion as:


\begin{equation}
    x_i(t+1) =  
    \begin{cases}
		x_i(t) + \mu(x_j(t) - x_i(t)) \\ \text{if $j$ is concordant,} \\ \\
        x_i - \mu(x_j(t) - x_i(t) - \rho(x_j(t) - x_i(t))) \\ \text{otherwise}
	\end{cases} 
\end{equation}

with $\mu$ being the influence factor as defined above, and $\rho$ being a \emph{correction factor} for extreme opinions. Then, both strategies are paired with a rewiring mechanism, yielding other two variants (RBCM and RUCM), where discordant links are broken and random new links are established. Given non-zero $\epsilon$ and $\mu$ parameters, all the four variants of this model lead eventually to a bi-modal distribution of opinions, with two distinct chambers in the network.

Finally, users' choices about which opinion to acquire and neighbours to unfriend are ultimately determined by the algorithms that populate their news feeds. Simple models that mimic a feed algorithm that maximises user interaction can also successfully explain the rise of echo chambers. In Perra and Rocha~\cite{perra2018modelling}, authors propose a model of opinion dynamics that takes into account, among other factors, an algorithmic personalisation of each user's feed. This model shows some differences w.r.t models discussed above, like a binary A-B opinion, and a network of users that are not active at each step $t$, meaning that they share ideas and change their opinions only occasionally. Authors present several scenarios where opinions from neighbours to user $i$ are arranged in a queue, sorted with different strategies (LIFO, FIFO, random, according to the similarity to $i$'s ideas, or with an opinion preferentially nudged by the system) and limited to the first $N$ entries, simulating an attention limit from user $i$. Simulations clearly show that algorithmic personalisation of content can create a fragmentation where each group of neighbours has a dominant idea, isolating the subordinate ones, when the underlying network is clustered enough and the initial distribution of ideas uneven.

\subsection{Empirical findings on polarisation and echo-chambers}
\label{subsect:magnitude_polarisation}

Social media are also an excellent source of data that can be analysed to check hypotheses, and to validate proposed models. It is not surprising that they have been extensively used by scholars as a platform for empirical studies. Network analysis and computational linguistics provide a wide spectrum of practical tools that allow the researcher to find, on large scale networks, signals of homophily,  sentiments and emotions, users' stances and opinions, agreement and disagreement, and so on. 

The tendency of online users to only engage with some other users and select news that are in agreement with prior beliefs, while disregarding information that challenges them, is well documented.
Conservatives versus liberals~\cite{Conover2011,Colleoni2014}, pro-Euro versus anti-Euro~\cite{Gorodnichenko2018SocialMS}, or pro-vax versus anti-vax~\cite{cossard2020}, are a few examples of the many divisive topics that fuel heated debates on online social media. It happens on Facebook~\cite{Bakshy1130}, Twitter~\cite{Colleoni2014}, Youtube~\cite{Bessi2016Users}, Reddit and Gab~\cite{Cinellie2023301118}. It happens also on blogs~\cite{Adamic2005}, and even email recipients appear divided into two stances~\cite{Nikolov2015MeasuringOS} about political topics. 

Some topics are divisive by nature and ultimately require each observer to take one of two possible sides. Politics is the main apple of discord~\cite{Gorodnichenko2018SocialMS}: it is the case, for instance, of the polarisation in the American politics in which both voters~\cite{Conover2011,Fiorina2008} and candidates~\cite{Andris2015,Theriault2006} are getting more and more polarised. In Bakshy et al.~\cite{Bakshy1130} authors determined the political alignment of 10 million Facebook users, and found that neighbours of liberal users are way more likely to be liberals than neutral or conservative, and the opposite holds for conservatives. Similar findings are displayed in Colleoni et al.~\cite{Colleoni2014}, where authors report that Democrats more than Republicans are more likely to belong to an ideologically enclosed network, on average, but the highest homophily is due to highly partisan Republicans. The same findings are also supported by Boutyline et al.~\cite{Boutyline2017SocialStructure}. Adamic et al.~\cite{Adamic2005} shows that even conservative blogs are more likely to reference other conservative blogs, compared to liberal blogs. However, opposite communities have been found between science and conspiracies supporters~\cite{Bessi2015ViralMisinfo,Bessi2016Users}, or between users debating about climate policies~\cite{WILLIAMS2015126}.

Polarisation is not only measured in terms of homophily of users' neighbours, but also in terms of interaction with other users online. Barberà et al.~\cite{Barbera2015} estimated the political leaning of 3.8 million Twitter users that produced over 150 million tweets, confirming that information about politics circulated mostly in closed loops between ideologically lenient users. Conover et al.~\cite{Conover2011} looked at 250,000 tweets before the 2010 U.S. Presidential campaign, confirming that the network of retweets shows high partisanship, as most of the retweets (intended as a form of endorsement) are in-group for both conservatives and liberals. In-group and out-group dynamics are predominant in Del Vicario et al.~\cite{DelVicario554}, Bessi et al.~\cite{Bessi2015ViralMisinfo}, Zollo et al.~\cite{Zollo2017Debunking} and Bessi et al.~\cite{Bessi2016Users}, where pro-science users are opposed to pro-conspiracies users, and the two communities rarely overlap. Homophily and ideological segregation, however, only trigger partisanship on flammable content, like politics. Barber\'a et al.~\cite{Barbera2015} measured polarisation over 12 diverse topics, and found that mainly political issues caused high partisanship of users, while other events did not (such as Super Bowl, Oscar 2014, Winter Olympics). Garimella et al.~\cite{Garimella2018PoliticalDiscourse} collected 10 datasets of tweets, divided into political (i.e., online debates over political topics) and non-political. They also found that political topics are more divisive, while non-political did not trigger the same partisan response. 

Users' individual and social biases play a relevant role in polarisation dynamics. For instance, Bessi~\cite{Bessi2016PersonalityTraits} investigates personality traits of conspiracy followers and pro-science users. Boutyline et al.~\cite{Boutyline2017SocialStructure} instead establishes a link between a preference for certainty, stability and familiarity of conservative voters and political extremists, and higher levels of homophily and in-group sense of belonging. Such conformity to in-group ideas brings social advantages to the individual. In Garimella et al.~\cite{Garimella2018PoliticalDiscourse} authors noted that bi-partisan users, i.e., users that bridge opposite echo chambers, show lower Pagerank centrality in the network, while also receiving less explicit appreciation (e.g., with likes), retweets and mentions than partisan users. Politicians on Twitter with an extremist, polarising stance have more followers than politicians with same activity and same presence on other media~\cite{hong2016PoliticalPolariz}. On the contrary, out-group dynamics can be openly adversarial. Conover et al.~\cite{Conover2011} noted high homophily between in-group users in retweets, but also a big, mixed cluster with low homophily between in- and out-group mentions, that authors verified to be made of provocative interactions. Williams et al.~\cite{WILLIAMS2015126} show similar findings on the two communities of skeptics and activists about climate change on Twitter. Lai et al.~\cite{LAI2019101738,Laietal2020} confirmed that also during Brexit and the 2016 Italian constitutional referendums, retweets were mainly used as an endorsement, as opposed to mentions and ``reply to'' messages, that were adopted by users to engage a debate with someone with an opposite view. They also noticed that very active users tend to mild their own stances on Twitter after referendums results. 
In Bail et al.~\cite{Bail9216} authors surveyed Democrats and Republicans users before, during and after a month of observation, during which they exposed users to a bot spamming content from the opposite party. Republicans that followed the liberal bot resulted to be substantially more radicalised than before. Munson et al.~\cite{Munson2013EncouragingRO} suggested that educating users about how much they are exposed to a single-minded environment may lead to better results in correcting polarisation.

Polarisation may be the long-term outcome of a process driven by inner and outer biases on each user's online experience. The interplay between personal biases, social biases and algorithmic filters is addressed in Geschke et al.~\cite{triplefilter}, where agents with limited attention are exposed to information coming from neighbours, personal searches, and algorithms. The effectiveness of algorithms in shielding users from discordant ideas is in fact disputed: along with empirical findings that confirm the existence of filter bubbles~\cite{filterbubble} accountable to algorithms~\cite{Spohr2017}, there is also evidence of the contrary, of a minor or non-existent role of algorithms in driving the polarisation process~\cite{donotblame}. 
Nikolov et al.~\cite{Nikolov2015MeasuringOS} measured the diversity of information consumed by approximately 100,000 users in mails, search logs and social media. They found that social media algorithms lead to a narrower variance of opinions the user could pick, compared to web searches. This view is shared also by Cinelli et al.~\cite{Cinellie2023301118}, where authors compared homophily on several social networks, claiming that substantial differences between Reddit's and Facebook's content consumption patterns are due to the news feed algorithm of the latter. Also Flaxman et al.~\cite{Flaxman2016FilterBubbles} found similar results: authors show that ``articles found via social media or web search engines are indeed associated with higher ideological segregation than those an individual reads by directly visiting news sites''. However, they also highlight how users are more exposed to diverse ideologies through social media than through web engines, addressing to users' biases the lack of variance in news read. The same consideration holds in Bakshy et al.~\cite{Bakshy1130}, where conservative Facebook users' news feed was estimated not far from a random sample of political ideas. In the authors' words, ``Within the population under study here [10.1 million users], individual choices more than algorithms limit exposure to attitude-challenging content in the context of Facebook''. Haim et al~\cite{Haim2018Burst} reject the hypothesis that Google news feed may overindulge with one-sided news.
Although the radicalisation of political views is not entirely to blame on the web~\cite{Iyengar2009,donotblame}, it is still reasonable to believe that online social media may play a role in this process, because they represent a platform for media outlets to reach visibility, and also for user generated content to potentially spread at an unprecedented rate.

It is worth noting that also the existence or the impact of echo chambers on the spreading of misinformation is not universally accepted as facts. The aforementioned contributions from Barber\'a et al.~\cite{Barbera2015} and Flaxman et al.~\cite{Flaxman2016FilterBubbles} claim that the echo-chambers phenomenon is overstated. 
A research from the Pew Research Center  shows how ``only 23\% of U.S. users on Facebook and 17\% on Twitter now say with confidence that most of their contacts’ views are similar to their own. 20\% have changed their minds about a political or social issue because of interactions on social media''~\cite{maeve2016}. Dubois et al.~\cite{Dubois2018EchoChamber} shows the results of a survey conducted in the United Kingdom, and concludes that only a small fraction of people find themselves in an echo chamber. Guess et al.~\cite{Guess2018Avoiding} is also critical of the echo chamber phenomenon, and the authors present an exhaustive list of arguments to support their claim. 

Last, we focused on polarisation mainly, but we have to point out that this binary lineup does not necessarily mean that there are only two surviving opinions on one issue. Individuals are entitled to doubts and soft opinions, they may acknowledge partial grounds to the opposite party, or even change their opinion when confronted with new evidence. For example, the Twitter's debate on immigration in Italy shows that there is an ``almost'' silent majority of users with mild or less polarised opinions between others expressing radicalised views or following political leaders with a clear (sometimes extreme) stance on this topic (see Vilella et al.~\cite{vilellaetal2020}). Fragmentation is likely more common in debates than polarisation, even if an empirical observation of such phenomena is in general more challenging due to the difficulties of assigning scores to multi-faceted opinions that have values in a continuum. However, studying fragmentation in real scenarios instead of polarisation could be one of the forthcoming problems to be addressed to better understand how misinformation spreads even in less radicalised debates.

\section{Fact-checking and the concurrent spread of misinformation and its correction}
\label{sec:models}

\subsection{Paradigms of fact-checking}
\label{sec:paradigms}

Fact-checking is the process of verifying the accuracy of information, and is a critical component of journalism and the online media system. The role of fact-checkers is not just to debunk false information, but also to provide context and clarification for statements that may be partially true or open to interpretation. Fact-checking involves a variety of tasks, including checking sources, verifying claims and statistics, and consulting experts. Professional fact-checkers also use a variety of computational resources, such as databases and search engines, to aid in their research, which in turn ignited the research on computational fact-checking~\cite {Cazalens2018}. Fact-checking has a history as long as mass disinformation: according to Amazeen et al.~\cite{Amazeen2020}, ``fact-checking may be understood as a democracy-building tool that emerges where democratic institutions are perceived to be weak or are under threat'': in the era of fake news and the spread of misinformation, fact-checking has become increasingly important in ensuring that the public has access to accurate information. Professional fact-checkers may work for a news organisation, fact-checking service, or other entity that focuses on ensuring the accuracy of information. Some examples of such organisations, often cited as sources for the research on misinformation, are PolitiFact~\footnote{\url{https://www.politifact.com/}, accessed on 12/12/2022.}, FactCheck~\footnote{\url{https://www.factcheck.org/}, accessed on 12/12/2022.},  Snopes~\footnote{\url{https://www.snopes.com/}, accessed on 12/12/2022.}, TruthorFiction~\footnote{\url{https://www.truthorfiction.com/}, accessed on 12/12/2022.}), GossipCop (dismissed in 2021), just to name a few. There are also international organisations of fact-checkers, such as the International Fact-Checking Network, which share the same code of principles, ensuring transparency and objectivity to the fact-checkers activity~\cite{Poynter}. However, the landscape of fact-checking organisations is complex; Graves~\cite{Graves2018} offers one of the first attempts to map the transnational network of fact-checking outlets.

Despite the importance of professional fact-checking, its effectiveness may be hindered because of the difficulty to apply it on a large scale. Fact-checking is a time-consuming and labor-intensive activity; it is usually carried out by a small number of experts, that must deal with the constant flow of (mis-)information published online. There is a limited number of items a group of fact-checkers may check, while millions of misinformation items flood the Web every day. Professional fact-checking by experts, however, is not the only existing form of fact-checking. ``Crowdsourced fact-checking'' is a form of fact-checking that leverages the contribution of many non-experts, usually social networks' users that flag problematic content as fake or worthy of checking. This approach can lead to demonstrably good results in identifying false news~\cite{Tschiatschek2018,SOUZAFREIRE2021115414}, even in delicate contexts such as the COVID-19 infodemic~\cite{Kou2021}. Crowdsourced fact-checking is based on the assumption that a non-expert fact-checker may be less reliable than an expert journalist trained in fact-checking, but the overall sum of hundreds or thousands of non-experts signals eventually leads to the same ratings of professionals. This is the conclusion of Allen et al.~\cite{Allen2021}, which noted that a small but controlled crowd of non-experts rates news similarly to fact-checkers, but it is a disputed conclusion~\cite{Godel2021}. Saeed et al.~\cite{Saeed2022} observed methodological differences between non-experts users that joined the Twitter's Birdwatch program and professional fact-checkers. Experts and non-experts may also work in synergy, by leveraging the signals produced by the latter to highlight content that needs professional fact-checking by the former~\cite{Pinto2019}.

If crowd-based systems scale better than experts' organisations on facing a large number of items to fact-check, automated fact-checking provides powerful tools that can process large amounts of data with responsiveness. Similarly to the task of fake news detection described in Sec.~\ref{sec:linguistic}, the task of automatic fact-checking leverages techniques borrowed from Natural Language Processing in order to verify one claim. However, while the models proposed in Sec.~\ref{sec:linguistic} are designed to detect disinformation pieces, automated fact-checking models need to disprove a false claim by producing a correction. Guo et al.~\cite{Guo2022Survey} decomposed such a task in (i) extracting one claim from a statement, (ii) retrieving shreds of evidence (checking the claim against a knowledge base), and finally (iii) producing and (iv) justifying a verdict; Zeng et al.~\cite{Zeng2021factchecking} described automated fact-checking in a similar way. A comprehensive overview of the variety of methods and techniques of computational fact-checking can be found in Thorne and Vlachos~\cite{Thorne2018} or Hassan et al.~\cite{hassan2015quest}.

Finally, it is worth mentioning a different paradigm of fact-checking, that combines both human annotators and automated tools. We will refer to this model as ``Human-in-the-loop'' fact-checking, quoting Demartini et al.~\cite{demartini2020human}. The rationale of this system is to balance the trade-off between the accuracy of expert human fact-checkers, the explainability and the control of biases of human (experts and non-experts) annotators, and the high scalability and low cost of automated tools. Nguyen et al.~\cite{Nguyen2018} argue that a fact-checking system must be fast, but also accurate and explainable at the same time, otherwise it will not be easily trusted. Similarly, Hassan et al.~\cite{Hassan2019} argue that the combination of ``crowds, professionals, and computer-assisted analysis could increase efficiency and decrease costs in news organisations that involve fact-checking''. For instance, Kim et al.~\cite{Kim2018Leveraging} designed an automated system that evaluates crowd signals to select news items to be verified by experts. Shabani and Sokhn~\cite{Shabani2018} proposed a model that works the other way around: a machine learning algorithm analyses pieces of news, and determines whether to ask for a human evaluation by a crowd of non-experts or not. In such a scenario, the automation of the software reduces an intractable amount of information to a feasible task, while human annotators are accountable for the accuracy of the overall system. A summary of the above-mentioned fact-checking paradigms is reported in Table~\ref{tab:fact_checking}, along with an assessment of their cost and reliability inspired by Demartini et al.~\cite{demartini2020human}.

\begin{table*}[]
\begin{footnotesize}
    \centering
    \begin{tabular}{|l|c|c|c|c|}
    \hline
     \textbf{} & \textbf{Scale}  & \textbf{Cost} & \textbf{Accuracy} & \textbf{Explainability}\\
    \hline
    Professional fact-checking & Low & High & High & High\\
    \hline
    Crowdsourced fact-checking & Medium & Medium & High & High\\
    \hline
    Automated fact-checking & High & Low & Low & Low\\
    \hline
    Human-in-the-loop & High & Low & High & High\\
    \hline
    \end{tabular}
    \caption{Four different fact-checking paradigms. Human fact-checkers are more accurate and reliable, but also more expensive. Automated fact-checkers can be applied on a mass scale with little cost, but they can be less accurate and provide no meaningful fake news correction. Human-in-the-loop systems try to exploit the perks of both systems.}
    \label{tab:fact_checking}
\end{footnotesize}
\end{table*}

\subsection{Early (mis)information spreading models}
Understanding how misinformation spreads, and which are the characteristic temporal patterns of such dynamics, is an instance of a broader problem of understanding information diffusion on communication networks and social media. Such simple observation is a promising starting point, because we have tools for tracking cascades of ``memes'' propagating in social media, as well as empirical studies on temporal patterns that can be found in the news cycle, and models that characterise information spreading. For example, Leskovec et al.~\cite{Leskovec2009} proposed a methodology to track memes and to study their temporal dynamics, and Matsubara et al.~\cite{Matsubaraetal2012} developed a model to observe and understand the rise and fall of information diffusion patterns; also, the role of the exposure to information through social media platforms and its implications in re-sharing has been empirically studied extensively since the work of Bakshy et al.\cite{Bakshyetal2012}. 

Some of the earliest studies on the peculiarities of rumour and misinformation spreading are due by Moreno et al.~\cite{moreno2004dynamics}, Acemoglu et al.~\cite{acemoglu2010spread}, Chierichetti et al.~\cite{CHIERICHETTI20112602}, Kwon et al.~\cite{kwonetal2013}. Acemoglu et al.~\cite{acemoglu2010spread} proposed a game-theoretic approach to model (mis)information spreading, with some selected agents that can influence the opinions of their neighbours, without changing their own (i.e., a behaviour that is elsewhere referred to as \emph{zealotry}, as in Mobilia et al.\cite{Mobilia_2007}). They also assume that all the agents may receive information from others. Under these assumptions, they observe that all the beliefs reach a stochastic consensus, coherently also with the classic opinion dynamics model by French–Harary–DeGroot~\cite{French_1956,harary1959criterion,DeGroot}. Rumour spreading is also represented by means of diffusion of beliefs and opinions over networks by Chierichetti et al.~\cite{CHIERICHETTI20112602}, Moreno et al.~\cite{moreno2004dynamics}. In search of an empirical characterisation of rumour propagation, Kwon et al.~\cite{kwonetal2013} studied how some stories spread in online social networks, finding that news cycles follow daily fluctuations and are likely to resurface to the attention of social media users from time to time. These temporal patterns can also be found in the analyses that we will discuss in more detail in Sec.~\ref{sec:socialbots} by Vosoughi et al.~\cite{Vosoughi1146}, Shao et al.~\cite{shao2018spread}, and others: even if the fake news has been corrected by some, it can still be shared by many. Another class of models takes into account branching processes, as in Del Vicario et al.~\cite{delvicario2016modeling}, aiming at understanding opinion polarisation and its interplay with misinformation spreading, as we already discussed in Sec.~\ref{sec:echochambers}.  

In previously discussed models, fact-checking is considered a remedy after the misinformation infection, and the underlying mechanisms embedded in these frameworks are related to how individuals can be influenced by others they are in contact with, or how the structure of the network can change diffusion dynamics or vice-versa. However, the competition between a fake-news and its correction, may actually change misinformation spreading patterns in ways that do not affect ``ordinary'' news cascades. As already discussed in Sec.~\ref{sec:network}, debunking does not imply necessarily that the information will be properly corrected or removed from social media: hyper-correction, back-fire effects, and other cognitive biases, can lead people who have been exposed to misinformation, in the long term, to forget the fact-checking, or to remember only the false information, or also to deny facts to continue supporting previous beliefs and opinions. Therefore, even worse than ineffective, fact-checking can be counterproductive: not only fact-checkers contribute, during the publication of the correction, to the spread of the original (false) news (see the discussion on the ``continued influence of misinformation'' debate in Sec.~\ref{subsec:individual_biases}), but they may also play a fundamental role in the emergence of opinion polarisation and echo-chambers (see Sec.~\ref{sec:echochambers}). As a consequence, although indirectly, debunking could be one of the elements that fertilises the soil for further propagation of fake news. Nevertheless, we need stronger arguments and evidence to better understand the role of debunking, its efficacy, and under which conditions, if any, it may help to eradicate misinformation from a network. Moreover, we may wonder if, analogously with compartmental modelling and the definition of the basic reproductive number $R_0$, is there a measure that can be compared with the ``epidemic threshold'', that tells us if and when we can stop worrying because the misinformation will be vanishing without any intervention.

The intuition that there exists a natural analogy between the diffusion of diseases and rumours, can be traced back to 1964 and the seminal work of Daley and Kendall~\cite{daley1964epidemics}. Their approach was based on compartmental models: the population is divided into compartments that indicate the stage of the ``infection'', similarly to SIR and SIS, whose states can be Susceptible, Infected, or Recovered. The evolution of the spreading process is ruled by transition rates between states in differential equations. In the adaptations of the models to rumours and news, a similarity between the latter and infectious pathogens is considered. 

Instances of compartmental models that have been used to study misinformation diffusion include the addition of an \emph{exposed} state by Jin et al.~\cite{jinetal2013}, a forgetting transition and an \emph{hibernator} state by Zhao et al.~\cite{zhao2013rumor}, periodical bursts in time evolution by the already mentioned Matsubara et al.~\cite{Matsubaraetal2012}, super-spreaders and apathy with heterogeneous activation patterns by Borge-Holthoefer et al.~\cite{borge2013emergence}. 

\subsection{The hoax epidemic model}
Fact-checking, as one of the underlying mechanisms that shape the dynamics of the diffusion process, has been introduced with the \emph{hoax epidemic model}\footnote{This model is also referred as the SBF (or SBFC) model, because of the Susceptible-Believer-FactChecker states the single agent can get. However, the same acronyms have been already used for models proposed in biology and complex systems. To prevent any confusion with existing and not related approaches, we prefer to use here the original name chosen by the authors.}
by Tambuscio et al.~\cite{tambuscio2015fact,tambuscio2018network,tambuscio2019fact}. Other mechanisms such as a varying forgetting behaviour and different underlying network structures are ingredients of the model, to embed other cognitive and relational driven characteristics to the analytical framework, and to test different what-if scenarios. 

\begin{figure}[ht!]
\centering
\includegraphics[width=0.5\textwidth]{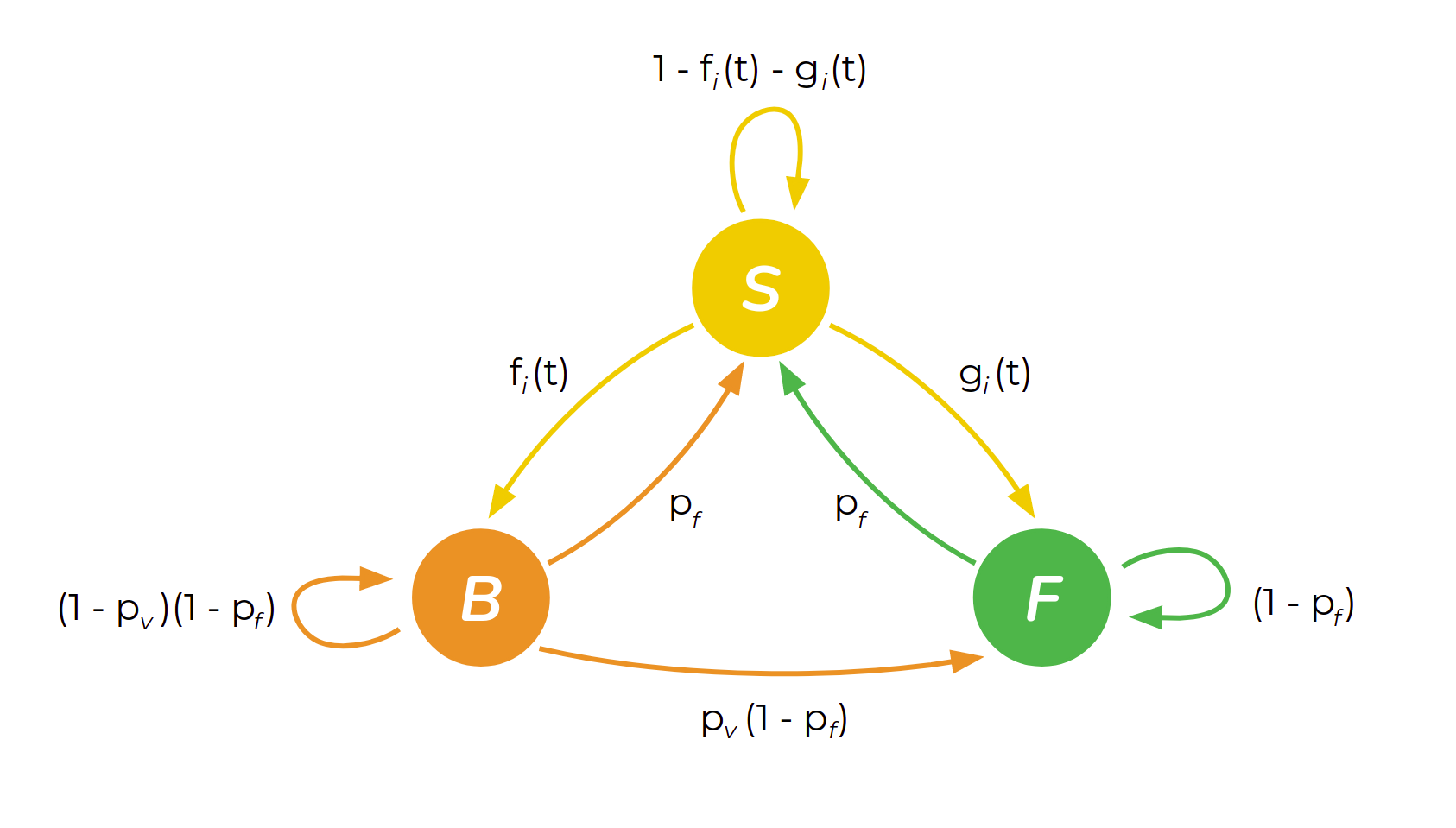}
\caption{The transitions states for the generic $i_\text{th}$ agent of Tambuscio et al.'s ``hoax epidemic model''~\cite{tambuscio2015fact,tambuscio2018network,tambuscio2019fact}.
}
\label{fig:hoaxmodel}
\end{figure}

An agent in the hoax epidemic model can be in any of the following three states: ‘Susceptible’ (S), if they have not been exposed either to the fake news or the fact-checking, or if they have previously forgotten about it; ‘Believer’ (B), if they believe in the news and choose to spread it; and ‘Fact-checker’ (F) if they know the news is actually false — for example after having consulted a trusted debunking site — and choose to spread the fact-checking. The $i\text{th}$ agent at time step $t$ is a node in a social network and it has a number of neighbours in state $X \in \{S,B,F\}$ denoted by $n_i^X(t)$. Also, the fake-news has an intrinsic credibility $\alpha \in [0,1]$, and that the overall spreading rate is $\beta \in [0,1]$. Transition probabilities between states are defined accordingly and displayed in Fig.~\ref{fig:hoaxmodel}.
The model assumes that a believer can verify the news and then turn into a fact-checker, but not the opposite. However, this can happen for an agent that forgets the correction or previous belief, and go back to the susceptible state. Unlikely compartmental models used to study the transmission of biological pathogens, some social contagion aspects are embedded here: the spreading probabilities admit bandwagon effects and group pressures. Hence, they change in function of the number of believers or fact-checkers in the neighbourhood of agent $i$ at time $t$ (mathematical details in~\cite{tambuscio2015fact}).

The hoax epidemic model can be used to study the behaviour of the agents at equilibrium. Starting from a few agents seeded as believers, and by means of agent simulations on different network topologies, and with analytical analysis by mean field equations, it can be shown that the number of susceptible agents at infinity stabilises, independent of the network topology (randomly generated), the value of $p_v$, and $\alpha$.
In particular, and analogously with the SIR and SIS models, it is possible to calculate an \emph{epidemic threshold}, to predict under which conditions the fake news would stop from spreading in homogeneous networks, i.e., when the number of believers at equilibrium converges to zero:

\begin{equation}
    p_v \ge \frac{1-\alpha}{\alpha}p_f \Rightarrow  B_\infty \rightarrow 0
    \label{eq:epithresh1}
\end{equation}

This means that there exists a critical value of the verifying rate $p_v$  above which the fake news is eradicated from the network, and such value depends on the credibility $\alpha$ of the information and the forgetting probability $p_f$, but not the spreading rate $\beta$. 

As investigated in~\cite{tambuscio2018network}, the model allows putting light also to the interpretation of the given epidemic threshold and to polarisation phenomena in segregated networks. Leaving the details to the paper, 
it is possible to assign different values of $\alpha$ to sub-populations: the individuals can be  \emph{sceptic} (with low $alpha$) or \emph{gullible} (with high $alpha$). Then, groups of sceptic and gullible agents can be well-mixed together or highly separated in different segregation scenarios, like the red and blue nodes in Fig.~\ref{fig:polarization1}. Using agent-based simulations with different segregation configurations, it was observed that if the forgetting probability $p_f$ is large (e.g., in the case of mere rumours), segregation has a different effect on the spread of the fake news, and the presence of links among clusters can promote the diffusion of the misinformation also within the sceptic groups.  However, if such a forgetting rate is small (e.g., the case for conspiracy theories) then the fraction of the population that believes the fake news will be large or small depending on whether the network is, respectively, segregated or not.

The results presented so far have implications for understanding the effect of fact-checking on individuals debating conspiracy theories: as already discussed in Sec.~\ref{sec:echochambers}, when individuals have strong opinions toward a topic, the emergence of polarisation and echo chambers is very likely, as observed by Bessi et al.~\cite{Bessi2015ViralMisinfo,Bessi2016Users} between followers of science or conspiracies pages and channels in social media. That means that we can expect that these cases are characterised by large segregation between sceptic and gullible individuals, and low forgetting probability. According to the hoax epidemic model, under these conditions, the fraction of believers in the population is maximised, and we can expect to find believers even among the sceptic individuals, making the eradication of the fake news quite impossible.

It is important to question the efficacy of fact-checking when conspiracy theories are into play
to identify the most promising ``immunisation'' strategies to contain the spread of misinformation as much as possible. Several instances of the hoax epidemic model have been compared in the \emph{what-if analysis} presented in~\cite{tambuscio2019fact}. 
Apparently, under very pessimistic assumption, if every node, even fact-checkers, is allowed to change its status, the fake news will always win the competition against debunking, and the system will reach a consensus toward a population of believers. Vice versa, verification ($p_v > 0$), and the presence of zealot fact-checkers that do not ever forget (see again Mobilia et al.,~\cite{Mobilia_2007}, and Acemoglu et al.,~\cite{acemoglu2010spread}), can give a chance to fake news to vanish out or being confined only to gullible clusters: fact-checking can be ineffective, but a world without debunking would be much worse. In fact, if the system is forced to keep a minority of some strongly committed debunkers, some positive equilibria are possible: for example, if the top $10\%$ of the hubs of the sceptic population (i.e., highest degree sceptic individuals) are zealot fact-checkers, the fake-news will be spreading mainly within the gullible clusters, without ``infecting'' the sceptic groups.

\subsection{Discussion and limitations}
There are many shortcomings to be considered when we try to apply models' outcomes: first, validation of results is difficult because many nodes can be in a hibernator state, as observed by Zhao et al.~\cite{zhao2013rumor}, e.g., individuals believed the fake news, they do not share their belief on social media, but they can still convince someone else off-line. In such cases, it is difficult to estimate the numbers of ``infected'' individuals, and all the subjects that have been exposed should be tested directly, like with epidemics where also asymptomatic patients could transmit the disease. Second, convincing influencers to openly debunk fake news could be hard. Third, calculating values for parameters such as $\alpha, \beta, p_v$ and $p_f$ on real stories can be unrealistic, especially at the earliest moment of the cascades formation. However, this could help a social media company to embed, in their platforms, automatic methods to identify ``influencers'', and label as ``likely false'' some of the stories they have been exposed to (for example, using one of more detection systems, as in Sec.~\ref{sec:linguistic}), to suggest them to use some trusted debunking site before spreading the news.


\section{The Problem of Prevention: A Computational Linguistics Perspective}
\label{sec:linguistic}

The development of systems that can automatically detect fake news and prevent their propagation is very challenging since they can be manipulated in a number of different ways in order to confuse the users. One of the most important source of information for the systems can be extracted from the content of the news article. Theories developed by forensic psychology have implied that deceptive statements are different to the true ones in terms of writing style and sentiments expressed~\cite{zuckerman1981verbal}. In addition, the study by Vosoughi et al.~\cite{Vosoughi1146} showed that false stories inspired fear, disgust, and surprise in replies whereas true stories inspired anticipation, sadness, joy, and trust. Also, they showed that the content of false news was more novel than true news.

Computational linguistics, that is the scientific study of language from a computational perspective, and the tools that provides can be very useful in the development of automatic fake news detection systems. In this Section, we will present related work that focuses on the detection of fake news from the perspective of computational linguistics either applied only on the textual content of the article or in combination with additional information extracted from images or user profiles. 

\subsection{The role of external evidence}

As pointed out in Sec.~\ref{sec:paradigms}, a number of fact-checking websites are available, and they rely on domain-experts that check and verify the news content. Some of those websites, including Politifact, Snopes, and GossipCop, have been extensively used by the researchers as a means to create annotated collections on which the proposed automated methodologies can be evaluated.  

One type of information that has been explored is the credibility of the websites that publish the article. That is because a website that has already published false information in the past, it is likely to do that again in the future. Journalists as well take into account the source of the news as a fundamental information when they want to verify the credibility of an article. One of the works that focused on the prediction of the factuality of a news site was presented by Baly et al.~\cite{baly2018predicting}. For the prediction, Baly et al. used a wide range of features extracted from a sample of articles published by the news site, its Wikipedia page, its Twitter account, its URL structure and web traffic information. The experiments that were performed with a Support Vector Machine (SVM) classifier showed that the textual features extracted from the articles led to the highest performance on factuality prediction.

\begin{figure*}[ht!]
\centering
\includegraphics[width=0.7\textwidth]{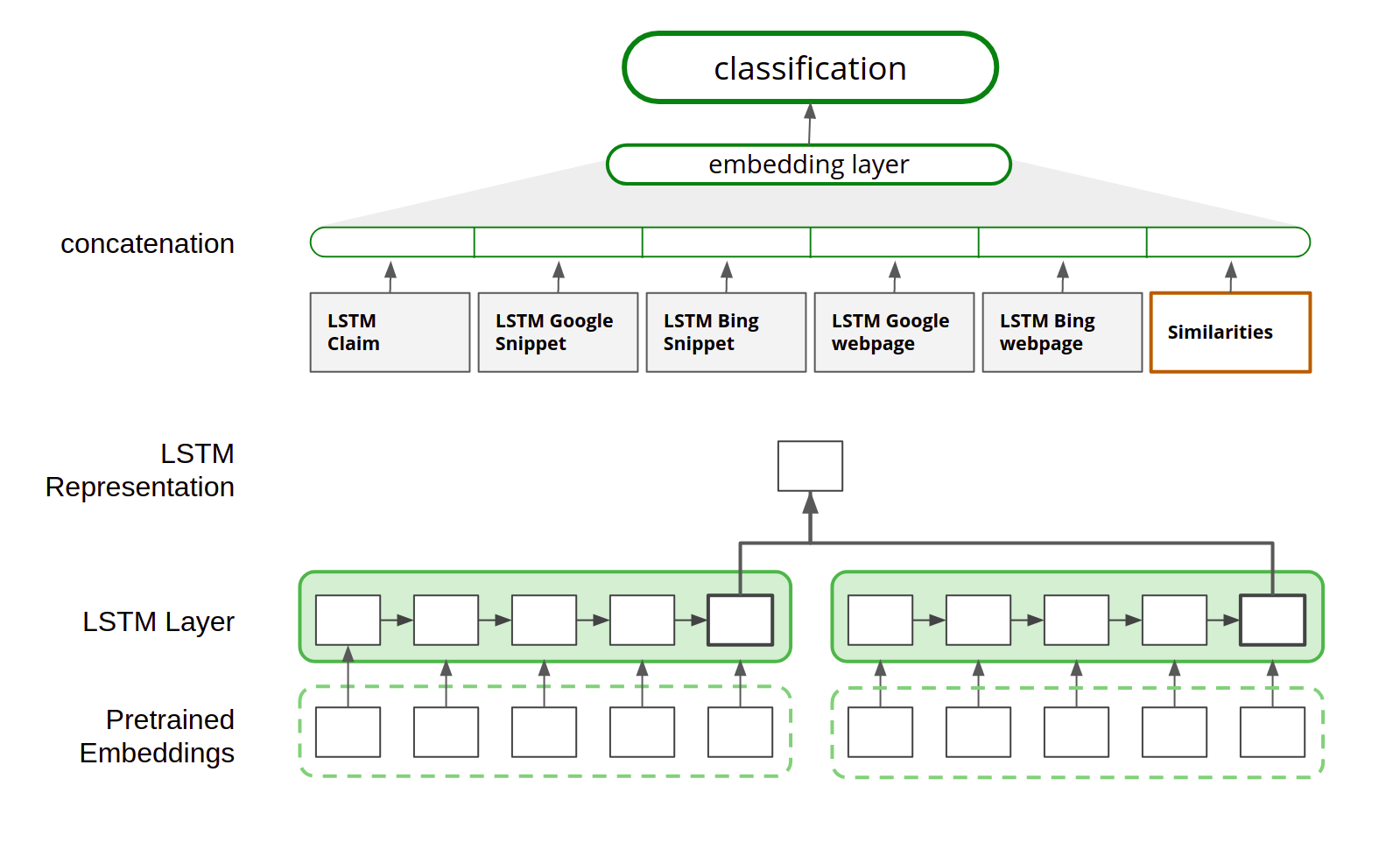}
\caption{Architecture of the system proposed by~\cite{karadzhov2017fully}}
\label{fig:preslav}
\end{figure*}

Other researchers used external knowledge to verify a claim as an additional source of information. Karadzhov et al.~\cite{karadzhov2017fully} proposed a deep neural network with Long Short Term Memory (LSTM) encoding and task-specific embeddings to verify a claim. To address the task, they retrieved documents that were relevant to the claim using a search engine. The snippets and the most relevant sentences were then compared to the claim. The final classification was performed with an SVM classifier. The architecture of the proposed system is depicted in Figure~\ref{fig:preslav}.

\begin{figure*}[ht!]
\centering
\includegraphics[width=0.95\textwidth]{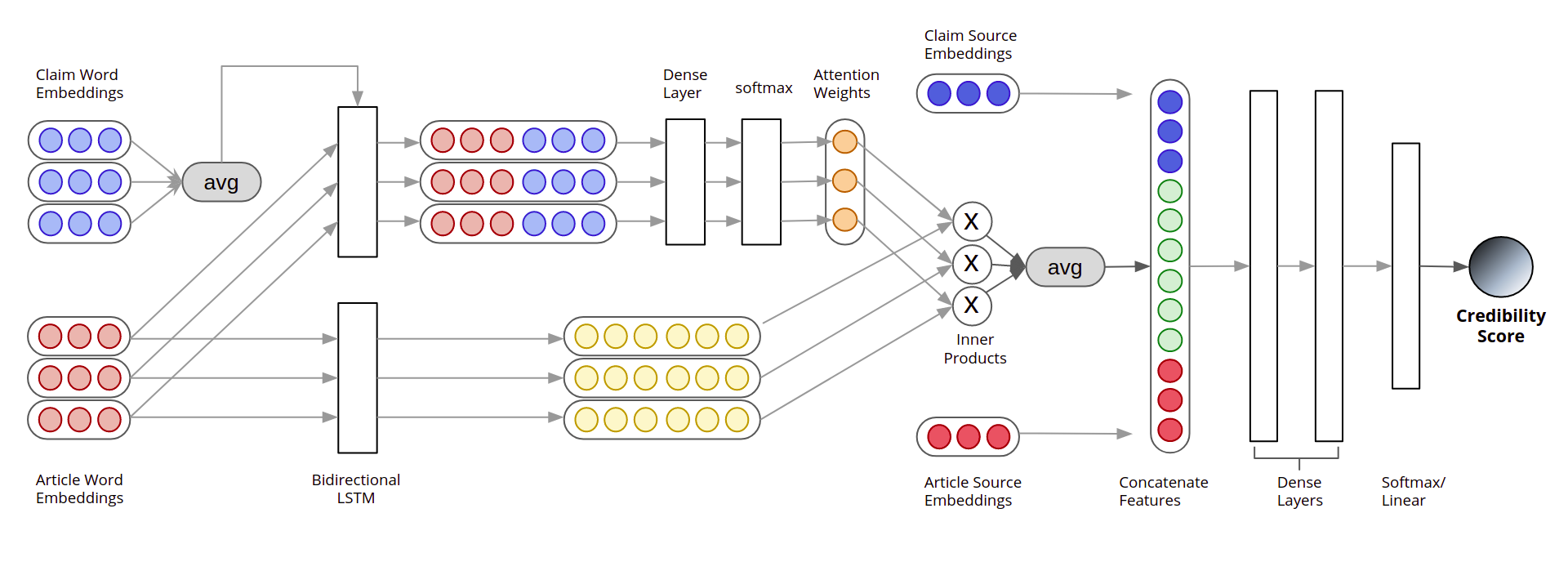}
\caption{Architecture of the DeClarE system proposed by~\cite{popat2018declare}}
\label{fig:declare}
\end{figure*}

Neural networks have been also used to model the external evidence. For example, Popat et al.~\cite{popat2018declare} proposed DeClarE, a neural network model that aggregated signals from external evidence articles, the language of these articles and the trustworthiness of their sources. The system first used the claim as input to retrieve relevant documents from the Web. A bidirectional LSTM (biLSTM) was then applied to capture the language of the web articles and an attention layer to focus on the parts of the articles that were relevant to the claim. Information about the claim source web article contexts, attention weights, and trustworthiness of the underlying sources was aggregated for the final assessment of the claim. The architecture of the proposed system is depicted in Figure~\ref{fig:declare}.

\subsubsection{The Role of Textual Content}

A large number of studies focused on information extracted from the textual content of the news articles that is one of the most fundamental information for the detection of fake news. In particular, several works have studied the impact of linguistic and semantic information on fake news detection~\cite{Castillo2011,Giachanou2019}. The textual information proposed in the different studies ranges from simple features such as word frequencies and text length~\cite{Castillo2011,Zhou2020} to more complicated such as latent representations extracted using deep learning methodologies~\cite{jwa2019exbake}. 

Early works were based on manual features and traditional representations of text such as Bag-of-Words and term frequency–inverse document frequency (TF-IDF)~\cite{Castillo2011,Perez2017}. These representations lead to high dimensional feature vectors and do not capture the co-occurrence statistics between words. To overcome the limitations of the traditional representations, researchers explored the effectiveness of low text representations. These representations, known as embeddings, can be at word~\cite{mikolov2013efficient}, sentence or document level~\cite{Le2014}, and can be used as input to traditional classifiers~\cite{Zhou2020} or neural network architectures~\cite{Giachanou2019,ghanem2019emotional, rashkin2017truth, william2017}. Some studies have used pre-trained embeddings~\cite{popat2018declare, GiachanouNLDB2020} such as GloVe~\cite{pennington2014} and ELMo~\cite{peters2018dissecting} whereas others built word embeddings from the input text~\cite{ma2016detecting}. 

Recently, some researchers have proposed methods that are based on the Bidirectional Encoder Representations from Transformers (BERT) model~\cite{devlin2018bert} with the aim to capture the context representation of a sentence~\cite{jwa2019exbake,Kaliyar2021}. BERT is based on Transformers, a deep learning model in which each output element is connected to every input element, and the weightings between them are dynamically calculated based on their connection. Jwa et al.~\cite{jwa2019exbake} used BERT to analyse the relationship between the headline and the body text and its impact on fake news detection, whereas Kaliyar et al.~\cite{Kaliyar2021} proposed FakeBERT by combining different parallel blocks of the single-layer Convolutional Neural Networks (CNN) with BERT.

\begin{figure*}[ht!]
\centering
\includegraphics[width=0.6\textwidth]{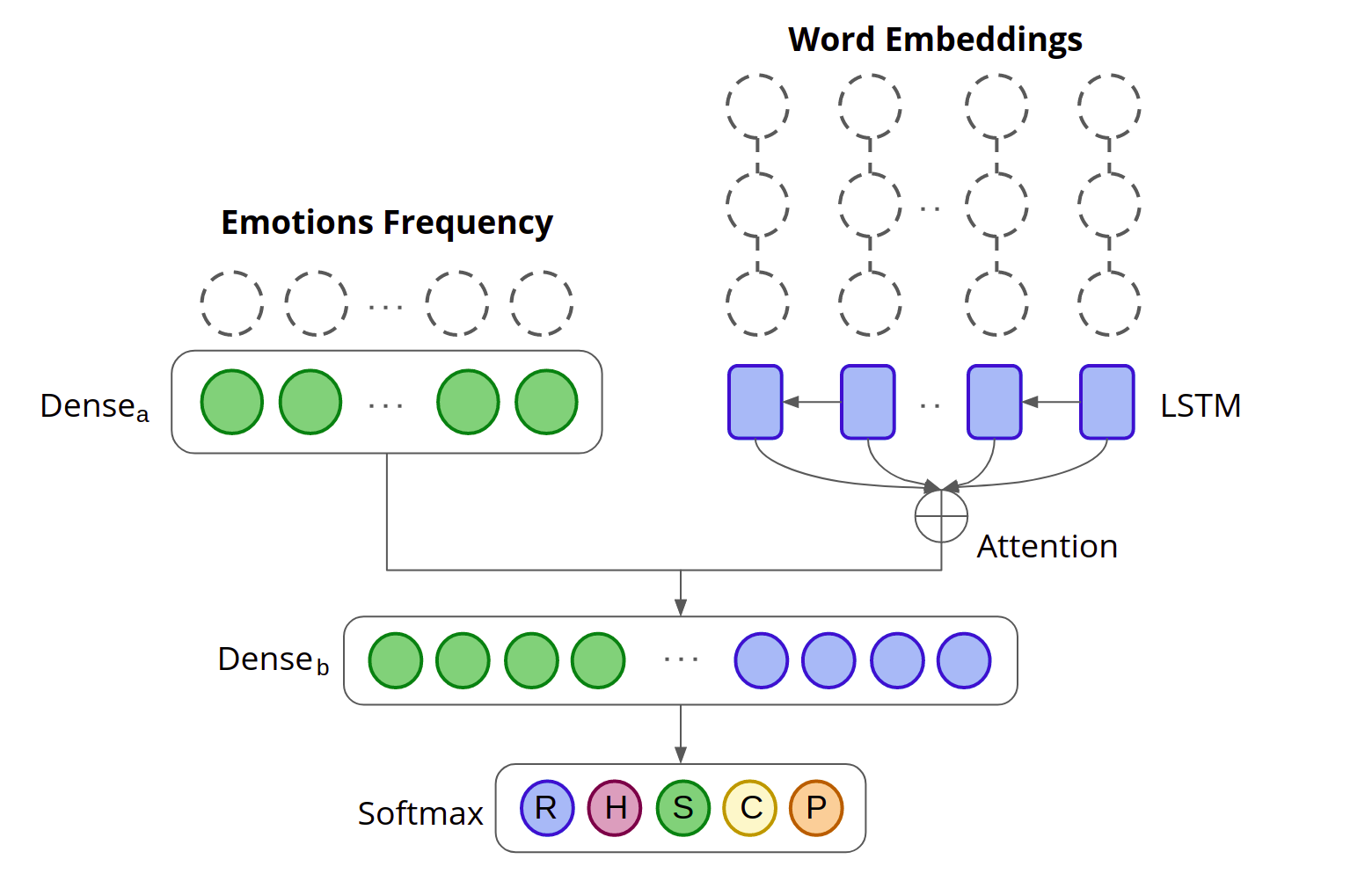}
\caption{Emotionally-infused neural network architecture for false information detection proposed by~\cite{ghanem2019emotional}}
\label{fig:GhanemTOIT}
\end{figure*}

Together with the general textual representation, some researchers were also interested in exploring the effectiveness of semantic information. To this end, the role of emotions on fake news detection has been extensively analysed due to the fact that fake news tend to produce different emotions and of different intensity to the users compared to the real news~\cite{Vosoughi1146}. Giachanou et al.~\cite{Giachanou2019} proposed EmoCred, a system for credibility assessment based on emotional signals extracted from the text of the claims. The architecture of EmoCred was based on an LSTM model and concatenated textual and emotional features for credibility detection. Three different approaches were explored for estimating the emotional signals based on emotional lexicons and a CNN network. The results showed that incorporating emotional signals can increase the effectiveness of fake news detection. Ghanem et al.~\cite{ghanem2019emotional} explored the role of emotions in the detection of the different types of fake news (i.e., satire, hoax, propaganda, clickbaits and real news). The proposed network consisted of two branches, the first one was based on word embeddings and the second on emotion frequencies. 
Figure~\ref{fig:GhanemTOIT} shows the architecture of the system proposed by Ghanem et al~\cite{ghanem2019emotional}. 
In a follow-up work, the FakeFlow system was proposed to model the flow of affective information in fake news articles using a neural architecture~\cite{Ghanem2021EACL}. Recently, Guo et al.~\cite{guo2021dean} proposed a dual emotion-based fake news detection model to learn content and comment emotion features and their relationship for publishers and users respectively.

\subsubsection{Multimodal Approaches}

\begin{figure*}[ht!]
\centering
\includegraphics[width=0.7\textwidth]{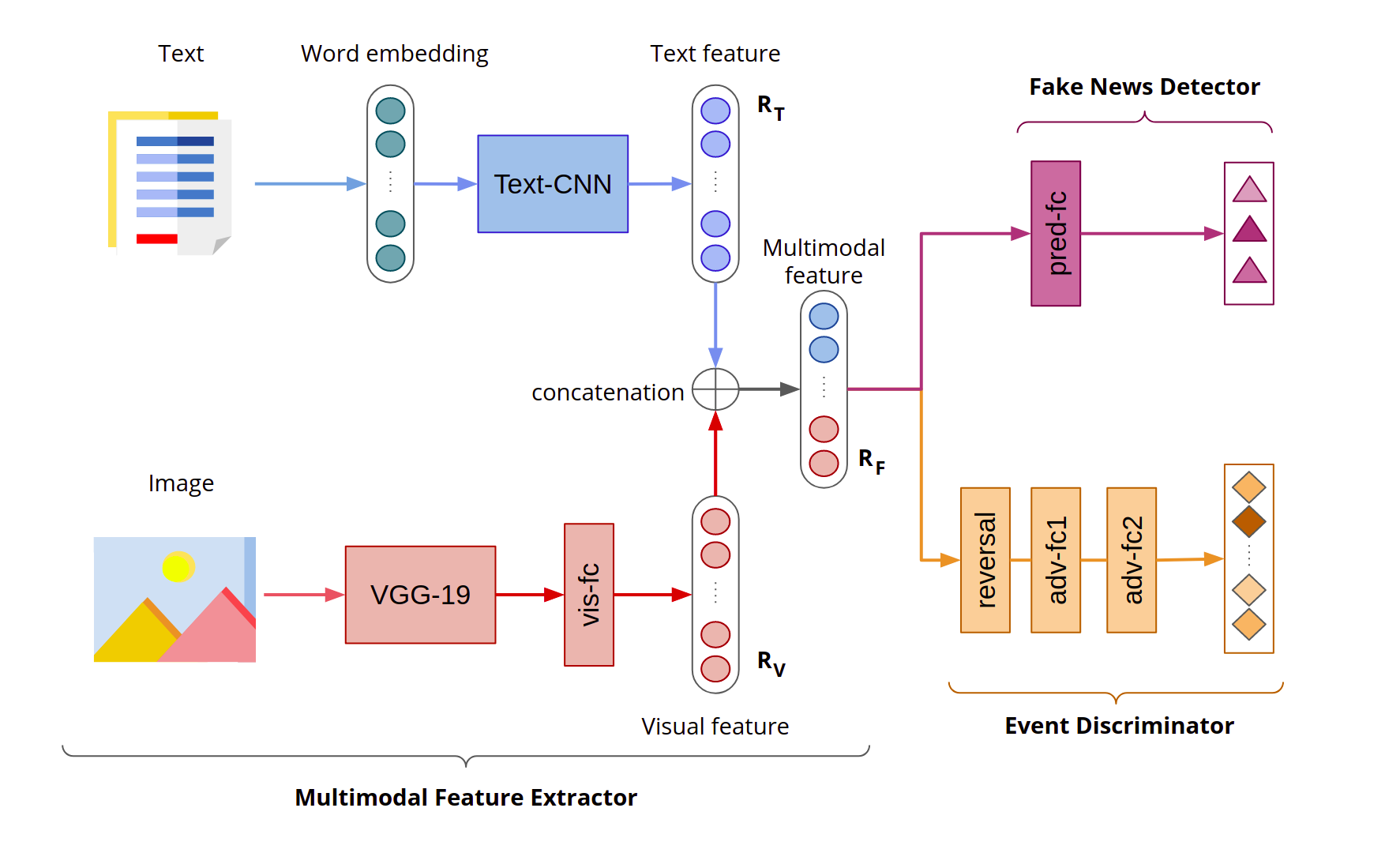}
\caption{The architecture of Event Adversarial Neural Networks proposed by~\cite{wang2018eann}}
\label{fig:EANN}
\end{figure*}

It is very common that fake news also contain images that have been manipulated in a number of different ways. Visual information can be useful for the systems to detect whether an article is fake or not. Several researchers proposed methodologies that incorporated visual information and analysed its impact on the performance of fake news detection. Visual information can be captured by hand-crafted features~\cite{Jin2017} or latent representations extracted by neural networks~\cite{wang2018eann,khattar2019mvae}. Wang et al.~\cite{wang2018eann} proposed the Event Adversarial Neural Networks (EANN) model that consisted of a textual component represented by word embeddings and a visual one that was extracted using the VGG-19 model pre-trained on ImageNet. 
Figure~\ref{fig:EANN} shows the architecture of EANN and its different components. 
Khattar et al.~\cite{khattar2019mvae} proposed the Multimodal Variational Autoencoder model based on bi-directional LSTMs for the textual representation and VGG-19 for the image representation. Singhal et al.~\cite{singhal2019spotfake} proposed SpotFake, a multimodal system for fake news detection based on text and image modalities. Singhal et al. applied BERT to learn text features and to incorporate contextual information, and pre-trained VGG19 to extract the image features. 

Some researchers incorporated the similarity between the image and the text in their models since this information can be valuable during the learning process. Giachanou et al.~\cite{Giachanou2020TSD} proposed a multimodal system that combined textual, visual and semantic information. For the textual information they used word embeddings and sentiment. For the visual information Giachanou et al. combined image tags extracted with 5 different pre-trained models and local binary patterns, whereas for the semantic information they used the cosine similarity between the embeddings of the image tags and text. Different to the majority of the systems which were based on visual features extracted from one image, Giachanou et al.~\cite{Giachanou2020DSAAA} proposed to use visual features extracted from more than one image. In particular, they combined visual features extracted from three images, textual features and image-text similarity. The textual representation was based on pre-trained BERT, whereas for the image representation they leveraged the VGG-16 model to extract image tags that were then passed to an LSTM model to capture the sequence of the images. Finally, the image-text similarity was calculated using the cosine similarity between title and image tags embeddings. Another multimodal approach was proposed by Zhang et al.~\cite{Zhang2022} who extracted the place, weather  
and season scenes from each image and showed that there are statistically significant differences regarding the frequency of those scenes in fake and real news.

Zlatkova et al.~\cite{zlatkova2019fact} focused on the relevant problem of detecting fake claims about images. To address this task, Zlatkova et al. explored a variety of features extracted from the claim (e.g., claim text), the image (e.g., google tags, URL domains) and the relationship between the two (e.g., cosine similarity, embedding similarity). In their study, they found that URL domains play a very important role in predicting the factuality of a claim with respect to an image. 

\subsubsection{The Role of Users}

Although the textual and visual information extracted from the content of the articles is essential for the effective detection of fake news, the role of users is also important. Users are the ones that decide whether to share a piece of fake news or not when this arrives on their screen. Contrary to what one may expect, Vosoughi et al.~\cite{Vosoughi1146} showed that bots accelerated the spread of true and false news at the same rate, implying that humans have a significant contribution to the spread of false news. This result has consequences on computational social science research and the interpretation of the outcomes of many real data analyses, as we will discuss further in Sec.~\ref{sec:socialbots}.  

Given the role of users in the propagation of fake news, profiling users that are involved either in the propagation or prevention of fake news based on their profile information or the language they use is very important. Shu et al.~\cite{Shu2018UserProfiles} analysed different characteristics of users that were sharing fake news and those who were sharing real news. They explored the impact of explicit profile features such as number of statuses and implicit profile features such as age, location and personality traits on fake news detection. An analysis on the importance of the features showed the registration time and whether the user is verified or not are the two most important features for the detection of fake news.

With the emergence of debunking sites, a new group of social media users appeared who aim on raising awareness regarding the factuality of news. Those users are known as fact-checkers, and their impact on correcting the news has already been discussed in Sec.~\ref{sec:models}. Giachanou et al.~\cite{GiachanouNLDB2020} analysed and explored the impact of different linguistic features and personality traits on differentiating between fact checkers and fake news spreaders. Vo and Lee~\cite{Vo2019} analysed linguistic characteristics of fact-checking tweets and found that the fact-checkers tend to use formal language and use few swear words and slang. In addition, they proposed a deep learning framework to generate responses with fact checking intention. The Factchecking Response Generator framework is based on Seq2Seq with attention mechanisms.

In another study, Giachanou et al.~\cite{giachanouIS2020} focused on users that tend to propagate conspiracy theories, known as conspiracy propagators. In their study, Giachanou et al. performed a comparative analysis over various profiles, psychological and linguistic characteristics of conspiracy and anti-conspiracy propagators. In addition, they proposed ConspiDetector, a model based on a CNN and which combined word embeddings with psycho-linguistic characteristics extracted from the tweets of users with the aim to differentiate between conspiracy and anti-conspiracy propagators. Ghanem et al.~\cite{ghanem2020factweet} proposed an approach to detect non-factual Twitter accounts such as propaganda, hoax and clickbait accounts. They treated post streams as a sequence of tweets' chunks and explored a range of different semantic and dictionary based features including word embeddings, emotion, sentiment, morality and style. Their methodology was based on a LSTM neural network.

In order to facilitate research on profiling fake news spreaders, Rangel et al.~\cite{fnspreaders2020} organised an evaluation task on profiling users who have shared some fake news in the past. In order to create the collection Rangel et al. collected tweets that were relevant to news articles that were annotated as fake or real on different fact-checking websites. After a manual inspection of the tweets, they collected the timelines of the users that have shared a high number of either fake or real news. Users that had shared at least one fake news was labelled a potential fake news spreader. Otherwise, the user was labelled as real news spreader. A wide range of features (e.g., word embeddings, n-gram, emotions, personality traits, punctuation marks) and learning approaches (e.g., SVM, Logistic Regression, CNN, LSTM) were used by the participants. The highest performances were achieved by the approaches that used n-grams~\cite{Pizarro2020,Buda2020}. Finally, Sakketou et al.~\cite{Sakketou2022}, created the FACTOID dataset derived from the monitoring of the discussion about political topics within certain subreddits in the period between January 2020 and June 2021. The data collection represents roughly 4000 users, counting a total of 3.4 million posts, each of which was annotated in three ways: binary (fake or real news), by means of a fine-grained credibility scale (from very low to very high), and according to the degree of political bias (from extreme right to extreme left).

\section{The problem of manipulation by Bots and Trolls}
\label{sec:bots}

Bots (i.e., automated accounts) aim at influencing users for commercial, political or ideological purposes. They often spread disinformation trying, for instance, to amplify a specific political opinion or to support a political candidate during elections. For example, Stella et al.~\cite{stella2018bots} showed that $23.5\%$ of 3.6 million tweets about the 1st October 2017 referendum for the Catalan independence were generated by bots. About $19\%$ of the interactions were from bots to humans, in form of retweets and mentions, as a way to support them. Regarding the former US presidential election, Bessi and Ferrara~\cite{bessi2016social} showed that, in the week before the election day, around 19 million bots tweeted to support Trump or Clinton\footnote{\url{http://comprop.oii.ox.ac.uk/2016/11/18/resource-for-understanding-political-bots/}}. Paul and Matthews~\cite{paul2016russian} stated that in Russia fake accounts and bots have been created to spread disinformation\footnote{\url{http://time.com/4783932/inside-russia-social-media-war-america/}}, and around 1,000 Russian trolls would have been paid to spread fake news about Hillary Clinton\footnote{\url{http://www.huffingtonpost.com/entry/russian-trolls-fake-news_us_58dde6bae4b08194e3b8d5c4}}. 

But how effective are bots in propagating disinformation? Can they effectively nudge disinformation content in the users’ feed, convincing them of false news? Aiello et al.~\cite{aiello2012} reported the result of an experiment where one bot that reached unexpected popularity in a popular social media platform for book lovers, was found very effective for leading users to accept link recommendations and also for determining polarisation in the underlying social and communication networks. Although many bots can have benign or also harmless purposes, they can be designed with the goals of persuading, manipulating, or deceiving humans, as brought to attention by Ferrara et al.~\cite{ferrara2016}. and Stella et al.~\cite{stella2018bots}. 

Disentangling the effects of bots propaganda from the individual responsibility of real users, however, is not a trivial question. Bots may repeatedly share manipulative content, but real users propagate it. Who is ultimately to blame for the spread of fake news? The answer to this question remained controversial for some time, also because of two very influential papers published in high impact journals (respectively Science and Nature Communications), reporting apparently contradictory results: Vosoughi et al.~\cite{Vosoughi1146}, after the analysis of $\approx 126,000$ stories tweeted by $\approx 3$ million people more than $4.5$ million times, concluded that humans, and not bots, contributed more to the acceleration of the spread of fake-news;  Shao et al.~\cite{shao2018spread}, after the analysis of $14$ million messages spreading $400$ thousand articles on Twitter during ten months in 2016 and 2017, observed instead a ``disproportionate'' role of bots in spreading low-credibility articles. However, this is an insubstantial contradiction, because Shao et al. found that bots adopt the following strategy: they amplify the diffusion of articles before they go viral, also exploiting the structure of the network and the role of single accounts. For example, bots target users with followers through replies and mentions. We know that humans are vulnerable to manipulations, and they tend to re-share content posted by others if their messages are aligned with their personal beliefs (see Sec.~\ref{sec:psychological}). When an account controlled by an (influential) human shares an article, the probability that the message will become viral is much higher. If bots are responsible for injecting fake news into social media, humans should be in charge of checking the veracity of the message before re-sharing it. When they are fooled, and they are also followed by many users, the strategy reached bots' ultimate goal: misleading a much larger portion of humans, as summarised by Aral in~\cite{aral2020hype}. 

In his book ``The Hype Machine''~\cite{aral2020hype}, S. Aral points out that social media platforms must take into account that humans are vulnerable to deception, and change their business plans to embed new countermeasures to protect their users against bots' manipulative actions. For example, shutting down accounts that are likely controlled by software~\footnote{Twitter 'shuts down millions of fake accounts', BBC news, 9 July 2018, \url{https://www.bbc.com/news/technology-44682354}} has been presented by social media companies as one signal in that direction. However, this means that the research problem of automatically detecting bots (and trolls) will be in many scholars' agendas for years to come. It should also be noticed that targeting users with many followers, is not the only successfully implemented strategy by malicious agents; in fact, it is possible to exploit natural fragmentation of the network. As observed by Stella et al.~\cite{stella2018bots} during the 2017 Catalan independence referendum campaign, bots started mentioning incessantly accounts controlled by humans in peripheral areas of independentists and costitutionalists groups. A similar bots' strategy was observed more recently by Vilella et al.~\cite{vilella2021impact} during the Italian debate on immigration on Twitter: in this fragmented but not necessarily polarised context, social bots tend to inject in the network more content from unreliable sources w.r.t. humans. However, such content often remains confined to a limited number of clusters. Quite interestingly, bots share also a huge amount of content from reliable (or mainstream) sources in order to diversify their reach, and to provoke wider and deeper cascades through the underlying, highly fragmented, information network.

\subsection{Detection of Bots and Trolls}

Pioneer researchers such as Lee et al.~\cite{lee2010uncovering,lee2011seven} proposed the use of honeypots to identify the main characteristics of online spammers. To this end, they deployed social honeypots in MySpace and Twitter as fake websites that act as traps to spammers. They found that the collected spam data contained signals strongly correlated with observable profile features such as contents, friend information or posting patterns. They used these observable features to feed a machine learning classifier with, in order to identify spammers with high precision and low rate of false positives. More recently, Alarifi et al.~\cite{alarifi2016twitter} have collected and manually labelled a dataset of Twitter accounts including bots, human users, and hybrids (i.e., tweets posted by both human and bots). In this work, random forest and Bayesian algorithms reached the best performance at both two-class (bot/human) and three-class classification (bot/human/hybrid). Gilani et al.~\cite{gilani2016stweeler} proposed a framework for collecting, pre-processing, annotating and analysing bots in Twitter. Then, in~\cite{gilani2017bots} they extracted several features such as the number of likes, retweets, user replies and mentions, URLs, and follower friend ratio, among others. They found that humans create more novel contents than bots, which rely more on retweets or URLs sharing. Dickerson et al.~\cite{dickerson2014using} approached instead the bot detection from an emotional perspective. They wondered whether humans were more opinionated than bots or not, showing that sentiment related factors help in identifying bots.  
The Botometer tool\footnote{https://botometer.iuni.iu.edu} (formerly known as BotOrNot) by Onur et al.~\cite{varoletal2017} was employed in several works, for instance by Shao et al. in a work where they analysed the spread of low-credibility content~\cite{shao2018spread}. Botometer was also used by Broniatowski et al.~\cite{broniatowski2018weaponized} to analyse the effect of Twitter bots and Russian trolls in the amplification around the vaccine debate. Recent methods use representation learning based on social and interaction networks~\cite{cresci2020decade} or employ deep learning techniques. For instance, Kudugunta and Ferrara~\cite{kudugunta2018deep} fed an LSTM network with both content and metadata. Similarly, Cai et al.~\cite{cai2017behavior} used an LSTM to analyse temporal text data collected from Twitter.

Influence bots can be defined as those bots whose purpose is to shape opinion on a topic, posing in danger the freedom of expression. Although there are several approaches to bot detection, almost all of them rely on metadata beyond text. Rangel et al.~\cite{rangel2019overview} organised an evaluation task, both in English and Spanish, with the aim to determine whether the author of a Twitter feed is a bot or a human on the basis of her textual information. Participants used different features to address the task (mainly n-grams, stylistics features and word embeddings) that were fed to machine learning algorithms such as SVMs or deep learning techniques such as CNNs, RNNs, and FeedForward neural networks.

Different to bots that are automated, a troll is a real user. Current research in that area focuses on identifying trolls~\cite{addawood2019linguistic} and analysing their political comments~\cite{freelon2020russian} and a few health topics such as vaccines~\cite{broniatowski2018weaponized}, diet~\cite{mitchell2019many}, the health of Hillary Clinton, abortion, and food poisoning~\cite{linvill2020troll}. Recently, Atanasov et al.~\cite{atanasov2019predicting} proposed an approach with the aim to analyse the behaviour patterns of political trolls according to their political leaning (left vs. news feed vs. right) using social media features. They conducted experiments in both supervised and distant supervision scenarios. In the proposed methodology, they leveraged the community structure and the text of the messages from which they extracted different types of representations such as embeddings. Ghanem et al.~\cite{ghanem2019textrolls} proposed a text-based approach to detect the online trolls of the US 2016 presidential elections. Their approach is mainly based on textual features which utilise thematic information, and profiling features to identify the accounts from their way of writing tweets. Alsmadia and O'Brien~\cite{ALSMADI2020102303} addressed the problem both at the account level, considering features related to how often Twitter accounts are tweeting, and at the tweet level. They noticed that bot accounts tend to sound more formal or structured, whereas real user accounts tend to be more informal in that they contain more slang, slurs, cursing, and the like. Finally, Jachim et al.~\cite{jachim2020trollhunter} introduced TrollHunter, an automated reasoning mechanism to hunt for trolls on Twitter during the COVID-19 pandemic in 2020. To counter the COVID-19 infodemic, the TrollHunter leverages a linguistic analysis of a multi-dimensional set of Twitter content features to detect whether or not a tweet was meant to troll.

\subsection{The impact of automatic classification tools in computational social science}
\label{sec:socialbots}
Using machine learning and computational linguistics techniques to identify and categorise messages shared on social media platforms and which are supposedly related to some topic (e.g., debates on immigration, vaccination, mask wearing), has become very common not only in computer science but also in other disciplines, such as political and social sciences. The same is true if the researcher's goal is to identify the entities who control social media accounts, that can be humans, bots, and trolls. In particular, automatically detecting a bot from its behaviour and its writing style is a moving target: if a software learns mimicking humans' behaviour, we can expect that, even in the future, malicious companies and programmers will take advantages of novel tools also to improve the performances of their AI controlled accounts. The problem is twofold: (1) should we be worried about the real impact of bots on social media? (2) given that studying the impact of social bots is worth studying, how much should we trust classification tools? 

The results presented so far suggest that the impact of social bots in social science research is worth studying, and that tools for the automatic classification of the entities controlling social media accounts are of paramount importance to scale down the complexity of analysing huge volumes of data, looking for significant signals of behavioural patterns. As already mentioned, researchers are well aware that bot identification is a moving target, e.g., Cresci et al.~\cite{crescietal2017} noticed that neither Twitter, nor humans, nor tools available until 2017 were capable of accurately detecting a novel family of social spam-bots. As a consequence, state-of-the-art bot detection models still need to deal with generalisation and computational efficiency challenges, that could greatly limit their applications. It is not surprising that Botometer by Varol et al.~\cite{varoletal2017}, one of the most popular bot detection tool in computational social science, is continuously updated: Yang et al.~\cite{Yang_Varol_Hui_Menczer_2020} propose to properly reduce the account meta-data size, to scale up the efficiency of the system when evaluating a stream of public tweets. Also, model accuracy on unseen cases can be improved by strategically selecting datasets from a rich collection used for training and validation. Sayyadiharikandeh et al.~\cite{Sayyadiharikandehetal2020} also observed that an ensemble of classifiers specialised for each class of bots that combine the decisions through the maximum rule, can better generalise for unseen accounts across datasets. They observed that such a methodology leads to an average improvement of 56\% in F1 score. 

The tools' obsolescence problem is still an important issue and it is likely to remain so in the years to come. 
Rauchfleisch and Kaiser point out that tools' unreliability is an important issue in~\cite{Rauchfleischetal2020}, focusing their analysis on Botometer as the most popular tool for social bot identification. In particular, they substantiate their criticism with the observation that many scholars use a threshold of the ``bot score'' to discriminate bots from humans after applying Botometers, although this practice is not recommended by the authors themselves in the FAQ of their easy to access API. Rauchfleisch and Kaiser further discuss that thresholds, even when used cautiously, can lead to a high number of humans classified as bots (false positives), and vice-versa (false negatives). According to their observations, scholars that want to draw general conclusions in computational social science research from the application of identification tools, must keep away from just relying on classifiers results: researchers are encouraged to (1) manually check a significant sample of accounts, calculating the inter-coder reliability between the tool's outcomes and the human's classification; (2) validate results over time, since the scores can naturally change because of modifications to the classification engine or to the status of the entity controlling the account; (3) ponder that some systems are optimised for some languages and not for others (e.g., Botometer is specialised for English); (4) share account IDs (as allowed by the Twitter developer terms) to let other researchers to replicate and further validate the analyses.

\section{Tackling social media manipulation through Networks and Language combined}
\label{sec:network_language}

Although network scientists and NLP experts have been members of quite distant scientific communities for years, network-based and linguistics perspectives have been often adopted together to tackle the problem of disinformation online. From one side, one could investigate the behavioural properties of the news propagation network, to detect malicious content diffusing in an anomalous way. From the other side, one could investigate the stylistic features that differentiate false from true news, or bots and fake news propagators from regular users. For instance, Shu et al.~\cite{Shu2019} is a great survey on methods for detection of fake news from how they spread on networks, but with little mention of the content they deliver. Similarly, Oshikawa et al.~\cite{Oshikawa2020} is a comprehensive survey on detection of fake news using NLP techniques. However, accounting for only the network side of disinformation could result in neglecting the stylistic cues of fake news articles, and the other way around. In 2015 Conroy et al.~\cite{Conroy2015} presented a survey of linguistic-based and network-based methods for fake news detection, but without any mixed contribution, for which they stated to ``see promise in an innovative hybrid approach that combines linguistic cue and machine learning, with network-based behavioural data''. Since then, we are seeing a growing trend of contributions that mix both approaches. 

One of the first pioneering approaches that combined characteristics from both disciplines was attempted by Castillo et al.~\cite{Castillo2011}, that designed features that exploited information from user profiles, tweets and propagation trees. Not surprisingly, this paper corresponds to one of the most influential milestone in the periphery network (see Sec.~\ref{sec:network}), as listed in the Supplementary Material. Similarly, Shu et al.~\cite{Shu2020hierarchical} proposed a model that exploits the structural, temporal and linguistic perspectives of the hierarchical propagation network of fake news to effectively distinguish them from real news. Fed to classical machine learning models, their combination of features outperforms each feature taken alone. Monti et al.~\cite{Monti2019} proposed a CNN fed with an embedding of a high number of features, regarding content (words, including hashtags) and spreading on networks (social connections and cascades trees, among others) for detection of fake news. Jin et al.~\cite{Jin2016} also evaluated news credibility by mining the stance correlations within a graph optimisation framework. In particular, they linked a credibility propagation network of tweets with supporting or opposing relations. The credibility labels were generated by the credibility propagation on the network. Temporal, structural, and linguistic characteristics were also combined by~\cite{kwon2013prominent} for addressing the problem of rumour detection. For the temporal aspect, they considered a periodic time series model that captured daily and external shock cycles. Structural aspect referred to network characteristics such as the number of nodes and links in the friendship network, whereas the linguistic characteristics were based on LIWC and included the number of words that indicated several psycho-linguistic categories. Another interdisciplinary approach was presented by Zhang et al.~\cite{zhang2020fakedetector} who proposed to address the credibility detection problem with a set of explicit and latent features extracted from the textual information of news articles, creators and subjects. In their study, they proposed a deep diffusive network model to incorporate the network structure information into model learning based on the connections among news articles, creators and news subjects. Shu et al.~\cite{shu2019beyond} proposed an approach based on embedding the relationships among news articles, publishers and spreaders on social media. With the embeddings, they represented the news content by using non-negative matrix factorisation, the users on social media, the news-user relationships such as user engagements in news spreading, and the news publisher relationships such as publisher engagement in publishing. Then a semi-supervised machine learning framework is applied for the fake news detection. 

Mixed approaches have been used extensively for the detection of bots and trolls as well. DARPA held a 4-week competition~\cite{subrahmanian2016darpa} with the aim of identifying bots influencing a pro-vaccination discussion on Twitter, providing features that range from the style of texts to the structure of neighbourhood of users. The Botometer tool~\cite{varoletal2017}, largely discussed in Sec.~\ref{sec:socialbots}, computes the probability of a batch of accounts being automated by looking at the content they tweet from a psycho-linguistic point of view, and also by looking at network-based metrics and their ego-network. Consequently, other works that leverage Botometer are inherently using a mixed approach for bots characterisation~\cite{bessi2016social,vilella2021impact,ferrara2016}. Stella et al.~\cite{stella2018bots} paired a deep network analysis of the topology and patterns of human-bots interaction network with an investigation of related concepts in tweets promoted by bots. Dickerson et al.~\cite{dickerson2014using} employed several features to characterise tweets' syntax and semantics, along with users' behaviour and their neighbourhood, to identify bots aiming to influence the 2014 Indian elections. Atanasov et al.~\cite{atanasov2019predicting} leveraged the community structure and the text of the messages from which they extracted different types of representations such as embeddings, in order to distinguish trolls from regular users.

Echo chambers and polarisation are inherently network phenomena, but the textual analysis of the content discussed can better characterise in-group and out-group interactions. For instance, Williams et al.~\cite{WILLIAMS2015126} and Zollo et al.~\cite{Zollo2017Debunking} both found that in-group communications showed a positive sentiment, while out-group interactions were predominantly hostile. The latter found even a negative reaction of conspiracy believers to scientific debunking posts. Bessi~\cite{Bessi2016PersonalityTraits} could link personality traits to the tendency of users to surround themselves with like-minded, by inspecting both the interaction between users and their comments. Lai et al.~\cite{LAI2019101738,Laietal2020} studied polarised debates on constitutional referendums on Twitter observing the dynamics of changing stances and network homophily; also the machine learning models used to detect users' stances over time were trained embedding network measures as features. 

Finally, the recent COVID-19 infodemic phenomenon~\cite{zarocostas2020}, has been studied using both network analysis and text processing tools extensively: Cinelli et al.~\cite{cinelli2020} studied the online discussion on the virus outbreak observing different social platforms, and analysing COVID-19 related topics by means of word embedding techniques. Furthermore, they observed the infodemic diffusion with compartmental epidemic models, estimating the basic reproduction number $R_0$ for each social media platform. Gruzd et al.~\cite{gruzd2021} curated a special theme issue of ``Big Data $\&$ Society'' Journal on Studying the COVID-19 infodemic at scale, mentioning necessary analytical techniques from machine learning, network analysis, agent-based modelling, and text mining. The ``COVID-19 Infodemic Observatory'', a platform for the large-scale analysis of publicly available discourse on Twitter presented by Gallotti et al.~\cite{gallotti2020}, analysed more than 1.6 billion COVID-19 related tweets and estimated that around $40\%$ of them are shared by social bots. Using text mining and network based approaches, they estimates infodemic risks variations across countries, arguing that the level of socio-economic development is not the key discriminant to separate countries with high versus low infodemic risk. For instance, there are G8 countries with remarkable infodemic risk (e.g., Russia and Germany) and developing countries showing  lower risk (e.g., Thailand and the Philippines).

\section{Conclusion}
\label{sec:conclusion}
Information disorders are one of the most studied problems of the recent years, and yet one very hard to see a solution anywhere near. The problem itself is complex, being the outcome of the interplay of different factors: algorithms that promote to users an unfair selection of news, social and personal biases that induce users into reading and sharing manipulated news, deception operated through linguistic shenanigans, social network dynamics of diffusion and peer influence, impersonation and spamming from social bots. Scholars have put a great effort so far on discovering how junk news spread online and why, and how to counter it, but more is needed in order to develop viable solutions. In the next years, research will be crucial to solve many of the unanswered questions we have today. For instance, fragmentation of opinions needs to be better addressed instead of simple polarisation, as it provides a landscape to study how misinformation spreads in non-radicalised debates, and how people get radicalised over time; efficacy of debunking must also be tested under different conditions, with special regard to the early identification of news that need to be debunked immediately from news that have less potential to harm; the role of social bots must be further clarified, in order to design proper defence strategies. Research can also contribute to the development of useful tools against the spread of misinformation online. From one side, research can assist the defence of user online against deceptive news, for instance designing social network interfaces that empower individuals' judgement and promote spontaneous fact checking. From the other side, it can assist the development and the continuous maintenance of tools for a quick detection of fake news online and automated accounts, as both are moving targets, which constantly evolve, and overcome the problems of generalisation and computational efficiency that affect such tools today. The heterogeneity of the problems faced by researchers requires a multidisciplinary approach that must be capable of encompassing different dimensions - networks, linguistics, psychology and others - to develop proper solutions. 

This paper contributes to the research on disinformation and misinformation by offering a multidisciplinary point of view on such a broad topic. It differentiates from other surveys in the attempt to look at the problem through the lenses of Networks and Language together, whilst other literature review usually focus on either one or the other, or a much narrower set of problems (like detection). Additionally, this manuscript is intended as a guide that can help practitioners make sense of a huge literature, providing a data-driven bird-eye overview of the research on mis-/dis-information in the last few years. Finally, this work comes together with a search engine intended to assist researchers in the challenging task of navigating through years of uncoordinated scholarly contributions.

\section*{Acknowledgements}
The work of Paolo Rosso was carried out in the framework of the IBERIFIER hub on Iberian media research and fact-checking funded by the European Digital Media Observatory (INEA/CEF/ICT/A202072381931 n.2020-EU-IA-0252), and the following projects: XAI-Disinfodemics on 
eXplainable AI for disinformation and conspiracy detection during infodemics (PLEC2021-007681) and MARTINI on Malicious Actors pRofiling and deTection In online 
social Networks through artificial
Intelligence (PCI2022-134990-2) of the CHISTERA IV Cofund 2021 
program, funded by MCIN/AEI/10.13039/501100011033 and by the “European Union NextGenerationEU/PRTR”.

\bibliographystyle{abbrv-doi}
\bibliography{appendix,biblio}

\newpage

\textbf{
\large Supplementary Material\\
\\
\large Studying Fake News Spreading, Polarisation Dynamics, and Manipulation by Bots: a Tale of Networks and Language} \\
Ruffo, G., Semeraro, A., Giachanou, A., Rosso, P.\\
\\

Many of the concepts described in this supplementary material are well known to the network scientist or the computational linguistic expert, but we expect that a computer scientist specialised in some other areas, and who is approaching some misinformation related challenges for the first time, could be confused with ideas, methods and tools coming from different disciplines. Hence, the first part of this document focuses on reviewing methods, definitions and concepts that are taken for granted in the survey. The second part focuses explicitly on bibliographic references that emerge as the most influential from the analysis of our citation network. 

\section*{Research methods, definitions and conceptual frameworks}
\label{app:background}
\subsection*{Network analysis}
\label{app:cna}
Network Analysis is a set of quantitative methods that allow the analysts to study the structure and the dynamics of a (physical, social, biological, technological) complex system that could be represented and modelled by means of a network or a graph. Here, we introduce some of the basic measures and concepts of complex networks and graph theory; nevertheless, we recommend the reader to consider some introductory textbooks such as~\cite{menczer2020,barabasi2015,kleinbergeasley2012,newman2010} for a more detailed introduction to the wide field referred as Network Science.

Even if we should remember that not every complex systems can be modelled or represented by means of graphs, networks are pervasive in many domains: a social system is a made of actors (or also individuals, or agents) and the relationships they form are ties that link pairs of actors; information systems (like the Web) are sets of (hyper)links, or citations, from documents (such as files, books, web pages) to other documents, and so on. Such representations help us to easily describe the interconnections between small `units', so that the goal of a network analysis is to characterise the overall systems in terms of structural and dynamical features, that can help us to understand, predict and also manipulate higher level phenomena such as synchronisation, traffic and congestion, spreading processes (such as epidemics and social contagion).

We represent such systems as generic graphs $G=(N,L)$, where $N = \{1, 2, \ldots, n\}$ is the set of nodes' unique identifiers, and $L = \{(i,j) : i, j \in N \} $ the set of links (or edges) between them. Graphs can be directed or undirected, or also weighted or unweighted. For each node $i$ we have a set of \emph{neighbours} $N_i$,  and a \emph{degree} $k_i$, that is the number of links that connect $i$ to its neighbours: $k_i = |N_i|$. In directed networks, neighbours of $i$ can be \emph{predecessors} $P_i$ or \emph{successors} $S_i$, according the direction of the connection that can be, respectively, incoming from or outgoing to the given neighbours. As a consequence, for each node $i$ we have also an \emph{in-degree} $k_i^{in} = |P_i|$ and an \emph{out-degree} $k_i^{out}= |S_i|$, so that $k_i = k_i^{in} + k_i^{out}$. In weighted graphs, we define \emph{strength}, \emph{in-strength}, and \emph{out-strength} as the weighted version of degree, in-degree, and out-degree: $s_i =  \sum_{j \in N_i} w_{ij}, s_i^{in} =  \sum_{j \in P_i} w_{ji}, s_i^{out} =  \sum_{j \in S_i} w_{ij}$.

Traditionally, a graph $G$ is mathematically represented by means of a $n \times n$ \emph{adjacency matrix} $A = \{a_{ij}\}$ , s.t. $a_{ij}= 1 \mbox{ iff } (i,j) \in L$, and $a_{ij}= 
0$ otherwise; if the graph is weighted, then $a_{ij}= 
w_{ij}$. Note that when $G$ is undirected, then $A$ is also symmetrical. This notation is very convenient for mathematical reasoning and stronger formalisation: measures can be defined easily through this notation (e.g., $k_i = \sum_{j}a_{ij})$, and it allows to make explicit the relationship between the graph and the eigenvectors and the eigenvalues of the corresponding adjacency matrix, that is exploited in the so called \emph{spectral graph analysis}. However, it is unpractical for large graphs, likely very sparse, that are translated to memory consuming adjacency matrices full of zeros. For computational purposes, \emph{adjacency lists} or \emph{edge lists} are commonly adopted, as well as other more specific data structures tailored for sparse matrices. 

The main difference between a simple and a complex network is that a simple graph can be represented by very few characteristics: stars, rings, grids, paths are example of simple networks. Without a proper language and framework, it is usually hard to distinguish the topology and the structure of a complex network created by a real dataset (e.g., from relationships declared by social media users) from an artificially generated random graph. First of all, we define a \emph{path} of length $l$ between two nodes $s$ and $t$ as a sequence $p = \{n_1, n_2, \ldots n_l\}$, such that $n_1 = s$ is the source, $n_l = t$ is the target, and $\forall x \in [1, l-1]: (n_x, n_{x+1}) \in L$. When $s=t$ the path is a \emph{cycle}. If we have multiple paths from source to target, we define the \emph{distance} $l_{st}$ between $s$ and $t$ as the shortest path length. The longest shortest path in a network is the \emph{diameter}. When the average distance $\langle l \rangle$ and the diameter are proportional to $\log n$, the graph is said to hold the \emph{small world} property. The graph is \emph{connected} if for every pair of nodes we can always find at least a path connecting them; otherwise, the graph $G$ has multiple connected components, or simply \emph{components}, that are connected subgraphs of $G$. The largest component is usually called \emph{giant component}, because for large scale graphs it happens often that its size is $> 50\%$ of the whole network. even if the graphs are sparse. 

Beyond connectedness, we usually look for classes of nodes with given properties, whose centrality and importance in the system can contribute to assign them different roles. Degree is a measure of ``popularity'' of a node, and high degree nodes are commonly called \emph{hubs}. \emph{Degree centrality} $d_i$ of node i is usually defined as a normalised version of node's degree: $d_i = k_i/n$. A node that is closer to all the other nodes on average has its own centrality in a network: the \emph{closeness} of node $i$ is calculated as $c_i = (\sum_{j\neq i}l_{ij})^{-1}$. Instead of popularity or topological centrality of a node, we may be interested to \emph{bridges}, that are nodes that are crossed by paths more often than other nodes in the networks. If $\rho_{st}$ is the number of shortest paths from $s$ to $t$, and $\rho_{st}(i)$ is the number of shortest paths from $s$ to $t$ passing through node $i$, the \emph{betweenness} of $i$ is $b_i = \sum_{s \ne t \ne i} \frac{\rho_{st}(i)}{\rho_{st}}$. Nodes with higher betweenness are the bridges of our system. Links connected to nodes with high betweenness are said to have high \emph{link betweenness}, and they are also referred as bridges of the network. Observe that calculating degree centrality takes constant computational time; on the contrary, betweenness and closeness are calculated after measuring shortest path lengths of involved pairs of nodes. This can be computationally expensive for large graphs, and algorithms adopting some heuristics are used to reduce such cost in many experimental settings.  Another set of measures is made of variants of the so called \emph{eigenvector centrality}, exploiting the spectral properties of a given graph. \emph{Page Rank} and \emph{Hubs and Authorities scores} (calculated by the \emph{HITS} algorithm) are examples of such measures. Their purpose is to define iteratively the importance of a node according the importance of its neighbours, or only its predecessors in a directed graph in particular. Page Rank, for instance, is widely adopted as a measure of node's influence.

All the centralities described above are just few of the many quantitative measures that can be defined for nodes in a graph, and they can be used to rank nodes accordingly. In small networks, we can ask which nodes are the most important, conversely on large scale networks this approach is often futile. In these cases we use ranks' intervals to identify classes of nodes (or links) with comparable properties, to assume that, under given circumstances, they have similar roles in the network and in its dynamics. Such statistical approach is commonly applied by plotting centrality distributions, that very often show heavy-tailed shapes, corresponding to highly heterogeneous classes of nodes. Focusing on degree centrality, the heterogeneity of the distribution leads to the \emph{emergence of hubs} phenomenon~\cite{barabasi1999}, that means that some nodes' degrees are much higher than the rest of the graph, making statistically meaningless the adoption of the average degree $\langle k \rangle$ as a scale of this measure. In other words, the average degree should not be used as a good estimate of a randomly selected node's degree in many real large scale networks. 

The emergence of hubs and centralities heterogeneity has important implications in many different scenarios. For example, when we want to assess the \emph{robustness} of a networked system, we can simulate what happens when we remove progressively larger fractions of nodes (or links) from the graph. One consequence of this process is that the network, after some iterations, starts to fall apart; in other words, the initial connected graph is split into smaller components and the giant component's size is reduced considerably. When removing nodes (or links) randomly at each step, we are actually simulating nodes' failures, and if the systems is robust, then the giant component size reduces linearly with the number of steps. This happens for many real network, such as the Internet, where random local failures do not usually affect global transmission functions. However, when we target hubs (or also nodes with high betweenness, high page rank, and so on), many real networks are proven to be extremely fragile, and the corresponding giant components are usually disconnected in many smaller components after very few removal steps. This means that some classes of nodes are definitely much more important than others and a targeted attack on a system may lead to the rapid disconnection of the network. However, studying network robustness is a great tool for understanding which is the most promising vaccination or immunisation policy in case of epidemics, for mitigating the diffusion of computer viruses or also controlling information propagation; in fact, finding hubs and bridges in networks is sometimes equivalent to understand how to quickly break down contact and transmission networks.

Targeted attacks are very effective in many real networks because their topologies differ significantly from randomly generated graphs in at least one important aspect: the presence of \emph{clusters} (or \emph{communities}), i.e., subgraphs whose internal nodes are tightly connected to each other (high intra-community cohesion), and that are loosely connected to external nodes (high inter-communities separation). The existence of such communities in real work networks put new lights on the interpretation of the above mentioned weaknesses against targeted attacks; in fact, targeting hubs, bridges and other central nodes in our networks is like removing nodes or links that are responsible for cross-communities connections, as well as internal clusters' cohesion: communities rapidly degenerate into disconnected components. Although exploring the whole space of partitions of a graph is a computationally intractable problem, there are many \emph{community detection} algorithms that use different heuristics to find clusters. Communities in social networks are often due to \emph{homophily}, i.e., the tendency of individuals to link with similar ones~\cite{McPherson2001}. Also information networks can be shaped by homophily or \emph{topicality}: two  documents on similar topics are more likely to have hyperlinks connecting them or to show smaller distances than two other documents on different topics. This has an impact on the design of modern web crawlers, and the biases of datasets collected by way of snow ball sampling processes. 

\emph{Graph drawing} is also an important step in many empirical analyses: properly visualising a graph can help us to easily compare (sub)structures, or spotting interesting network's areas or nodes. Many network analysis packages and tools have layout algorithms that place nodes on a plane to better emphasise some structural features of the graph. \emph{Force-based layouts} are the most widely adopted: they exploit repulsion and attraction laws from physics to place connected nodes near each other, and to move reciprocally away nodes that are topologically distant. This is a way to let clusters be revealed graphically when the network is not too large nor too dense. 

\subsection*{Modelling networks dynamics}
\label{app:dynamics}
Networks' structure has a role in the way nodes' properties change over time; when we focus on aspects such as diffusion of information, transmission of diseases, adoption of innovations or behaviours, then we are studying \emph{networks' dynamics}. Studying networks dynamics may require an additional burden for the beginner, because of the many existing approaches and models proposed over the years in different disciplines such as psycho-sociology, statistical mechanics, graph theory, epidemiology, and so on. 
The spread of misinformation, in particular, is one instance of information diffusion, and the emergence of polarisation and echo chambers is considered one outcome of opinion dynamics: both processes are just two examples of the most studied phenomena that are traditionally observed in networks. If the reader is interested to a comprehensive overview of this field, they can be referred to the complex networks introductory textbooks we already mentioned, and also to~\cite{barrat_barthelemy_vespignani_2008} that is more focused on dynamical processes. For a survey on information diffusion, check also~\cite{guilleetal2013} out. 

There is an important interplay between structure and dynamics, because the first can change, develop or emphasise the second, and vice-versa. This makes some phenomena functions of time, and the observation of their key factors becomes quite subtle in many empirical settings. For example, the emergence of homophily is responsible for high clustering in social networks, and there are two possible mechanisms that can explain this: one is \emph{selection}, the principle according which similar nodes are more likely to be connected, the other is \emph{social influence}, that makes connected nodes more and more similar to each other. However, when we study networks with high clustering, it could be difficult to understand if selection is prevalent w.r.t. social influence, or the opposite, or if both factors are at interplay. Furthermore, we can observe that communities change over time: they can rise and dissolve, merge, split, or even grow, shrink, stabilise. Changes in structure, may have effects on nodes exposed to different information and behaviours, that can lead them to yield to group pressure, or to decide to leave their cluster and join another community. Some of these processes may increase the fragmentation of the network, that is responsible of the formation of \emph{echo-chambers} and \emph{group thinking}, also without considering other factors such as filtering algorithms and recommendation systems, that can accelerate this spontaneous tendency to group conformity. One classical example is given by the Schelling's model showing that urban segregation is likely to be a natural phenomenon that can emerge spontaneously even in very tolerant societies, with the consequence that the causes of ghettoization cannot be solely led back to racism or intolerance. However, segregation may lead to inequalities, that could accelerate ghettoization processes, and then racism and intolerance. Similar dynamics can be found even in online social networks.

Network scientists have many different models to use to study networks dynamics. Although some of these (mainly mathematical) models have been proposed using common design principles, there are important differences to be considered if processes under observation are biological, social, economic, technological, and so on. 

\emph{Compartmental models in epidemiology} are often applied to the mathematical modelling of infectious diseases, and are based on compartments of a population assigned to different states, such as $S$ (Susceptible), $I$ (Infected), and $R$ (Recovered). They have been introduced at the beginning of the 20th century for well mixed populations, and they are often solved with differential equations providing deterministic results. They can also be applied on randomly generated networks, running several simulations on different (stochastic) realisations of the models, in order to fit more realistic scenarios. The basic SIR model is based on the definition of a \emph{transmission probability} $\beta$ between S and I, and a \emph{recovery probability} $\mu$. These parameters allow the definition of a \emph{basic reproduction number} $R_0 = \langle k \rangle \beta / \mu $, where $\langle k \rangle$ is the average degree in a random network. In a well mixed population, or also if the underlying contact network can be assumed to be equivalent to a random graph, it is possible to prove that when $R_0 < 1$, the epidemics is expected to die out naturally very soon. As a consequence, this measure is widely adopted for assessing the danger of an infectious pathogen and to predict the likelihood for an epidemic to scale up as a pandemic, just comparing the empirically calculated value of $R_0$ with the so called \emph{epidemic threshold} (that is equal to 1 for well-mixed networks). However, it has also been proven that the epidemic threshold tends to 0 the more heterogeneous a network is, implying that any epidemics in a real global scale contact network is likely to escalate to a pandemic. Compartmental models are also used to run simulations to compare different immunisation and vaccination strategies: for the sake of simplicity, it is like testing the robustness of the network 'forcing' some given classes of nodes to the R (recovered) state, that is equivalent to remove such nodes from the transmission network.

Many \emph{social contagion} models have been inspired by compartmental models, although epidemiological phenomena are inherently different to the way opinions, ideas, innovations, or ideas spread in a network. The simplest family of frameworks that try to capture how individuals' decisions can be influenced by means of peer pressure is the so called \emph{threshold model}. In this scenario, nodes can be \emph{active} or \emph{inactive}; the state of an inactive node $i$ flips when the number of its active neighbours is above a given threshold $t$. We can also consider a fractional activation condition instead of a linear one, just testing if a ratio of $i$'s active neighbours is above a threshold with values in $[0,1]$. The model can be even more complicated to better fit different and heterogeneous scenarios: for example, we can have a personalised threshold $t_i$ for each node, making some individuals less or more vulnerable to peer pressure. If a spreading phenomenon can be modelled with such a framework, then clusters in a network are important barriers against social contagion: an innovation, for instance, will not penetrate into a densely connected community whose members continue adopting an obsolete technology. 

Alternatively to models based on peer pressure, \emph{independent cascade models} assume that social influence can also work on a one-to-one basis; in fact, a node can be, in principle, convinced by a single, very influential, individual. Another difference with threshold based approaches is that independent cascade models are probabilistic: active node $i$ can convince an inactive node $j$ with a given probability $p_{ij}$. This makes more difficult to control a cascade, and clusters do not necessarily block propagation, but many realisations can be simulated to look for particularly interesting initial configurations. 

Social contagion is also studied by means of many other models, that are variants of the threshold or the independent cascade models. Information diffusion, rumour spreading, opinion dynamics, and other phenomena that need more realistic assumptions, are usually better understood with more complicated models, and also validated with data retrieved from social media platforms. Nevertheless, it must be recalled here that social phenomena are often examples of the so called \emph{complex contagion}, whose dynamics are not fully captured by the basic models described above; in fact, in many realistic settings, there is a fundamental difference between the spreading of rumours or jokes, and the diffusion of a valuable information, the adoption of a new behaviour or innovation, or social mobilisation: individuals usually need \emph{social reinforcement}, i.e., they must be exposed to the same idea or product more than once, before changing their habits, or opinions. Therefore, under complex contagion, the structure of the network or the roles of nodes with high centrality may change substantially w.r.t. simple contagion processes: first, an innovation spreads faster in highly clustered network, even when the diameter is large, in contradiction to what it is observed with threshold models. Second, bridges on big structural holes can be ineffective to push innovation across different communities, contrary to what is expected for high betweenness nodes with infectious diseases. Third, hubs internal to clusters can be super spreaders, equivalently to what happens with epidemics. 

\subsection*{Opinion dynamics}
\label{app:opiniondynamics}

We refer to opinions as states of the users (or the nodes of a social network), and polarisation as mainly the process that leads to an irredeemable divide, i.e., an equilibrium point or a stationary state of the system, between two groups, each representing a one sided view over a topic. 

Social dynamics models have been extensively used to explain how individuals engage with opinions of others (see Lorenz~\cite{Lorenz2007}, Castellano et al.,~\cite{castellanoetal2009}, Dong et al.~\cite{DONG201857} for more detailed surveys on this topic). In some of these models, individuals are entitled of an opinion, but they also are connected with each other, and they regularly update their state depending (to some extent) on the opinions of users they are linked with, interpreting the concept of \emph{peer pressure}, and the way \emph{social influence} may amplify homophily in a social network. This simple yet powerful idea may return good explanations of real social networks dynamics, which may be hard to unveil otherwise. In fact, social networks work for an user as proxies of opinions of their neighbours, and contagion of ideas is a powerful effect on individuals' behaviour. Plus, these models offer a general framework that can be modified to embed many of the biases that social media platforms induce in the process of opinion diffusion: as we noted in Sec. 4 of the survey, online users appraise information filtered by algorithmic biases, and by other widely studied social and cognitive biases that create additional pressures to the individual. Hence, the purpose of this models' review presented here is threefold: (1) to give to scholars studying misinformation a general overview of models well known in some scientific fields (e.g., social science, political science, statistical physics) that may be less popular to researchers active in other disciplines; (2) to propose a general and unique formalising framework to take into account and compare the many existing contributions; (3) to identify under which constraints opinions polarisation may eventually emerge in social networks.


In the following, we will not define opinions as discrete values, because we endorse a view of opinions as a continuum. After an initial configuration where each agent is assigned with a value, the system will eventually reach a steady state, where no nodes will change their opinion anymore. Such final state can represent a \emph{consensus}, where every nodes will share the same opinion, a \emph{polarisation}, where a group of nodes will share one opinion, and the rest will share the opposite opinion. Another typical stationary state is \emph{fragmentation}, when opinions are concentrated around more than two values. 

We start with a very simple model from French–Harary–DeGroot~\cite{French_1956,harary1959criterion,DeGroot}. In this model, each agent in a network of $n$ agents is assigned with an initial opinion, a scalar $x \in[0, 1]$, which represents their degree of support at time step $t$ for the position represented by 1. At each time step $t$ (time is assumed to be discrete), each agent $i$ updates its opinion as a linear combination of its neighbours opinions, as follows:

\begin{equation}
    x_i(t + 1) = \sum_{j=1}^n\alpha_{ij}x_j(t) 
    \label{eq:degroot}
\end{equation}

where $\alpha_{ij}$ is a weight on the link between agents $i$ and $j$. The stronger is the relationship between them, the higher is the weight. Lets observe that if $i$ and $j$ are not connected, then $\alpha_{ij}=0$. By averaging each users' opinion with their neighbour's, i.e., posing $\alpha = \frac{1}{n}$, this model eventually reaches \emph{consensus}. In such a model no polarisation can emerge yet, even with a starting setting with arbitrary homophily, as demonstrated in Dandekar et al.~\cite{Dandekar5791}. However, users do not process information coming from other users equally. It should be noticed that these dynamics can be studied also with the so called ``majority model'', introduced in 1963 in statistical physics by Glauber~\cite{glauber1963} in the context of spin models. 

Weights $\alpha_{ij}$ may be a function of the strength of a social tie, or even a function of similarity of two opinions. In such case, the model would embed a confirmation bias or selective exposure (see Sec. 4.4): by weighting more neighbours' opinion when they are close to their own opinion, agents would consider only ideas they agree with. The tendency to discredit opposite views may be also bounded with a confidence value, as suggested in the theory of social comparison by Festinger~\cite{Festinger1954}: in the \emph{Bounded Confidence Model (BCM)} introduced by Deffuant et al.~\cite{deffuantetal2000}, each individual interacts exclusively with ``close enough'' opinions, i.e., $i$ evaluates $j$'s opinion only if $|x_i(t) - x_j(t)| < \epsilon$. In such scenario, $\epsilon$ is a representation of mind broadness of the agent, or also the quantification of the agent's confirmation bias. Discordant opinions may be neglected, or used as a negative reinforcement. For instance, in Hegselmann et al.~\cite{Hegselmann2009} authors evaluate BCMs where an agent $i$ averages its opinion with concordant agents $j$. We can simply check concordance by means of a Kronecker delta:
\begin{equation}
    \delta_{ij}(t) =
    \begin{cases}
        1, &  \text{if $|x_i(t) - x_j(t)| \leq \epsilon $} \\
        0, & \text{otherwise}
    \end{cases}
\end{equation}
so that opinions are updated again using eq.~\ref{eq:degroot}, 
where $\alpha_{ij} = 0$ if $\delta_{ij} = 0$, and $\alpha_{ij} = (\sum_{z=1}^n \delta_{iz})^{-1}$ if $\delta_{ij} = 1$  (i.e., the weight is set to a value inversely proportional to the total number of agents whose opinion is concordant with both $i$ and $j$). In such model, reaching a consensus is difficult, but convergence towards fragmentation or polarisation depend on $\epsilon$: small confidence intervals mean narrow-minded agents, which rarely accept other ideas, thus more likely to stick to their initial position; larger confidence intervals means that agents adopt ideas from others more often, resulting in a reduced numbers of surviving opinions: polarisation may arise. Authors note, for instance, that with an $\epsilon = 0.01$ up to 38 opinions survived in the end, while with $\epsilon = 0.15$ the system converged on two poles. 

The overconfidence effect, i.e., the tendency of individuals to undervalue other people's ideas, and to update their opinion with hesitancy even when confronted with new evidences, is properly considered by Friedkin and Johnsen~\cite{FriedkinJohnsen1990,FriedkinJohnsen1999} as a generalisation of the basic DeGroot model:  it incorporates a \emph{resistance} in each agent's update rule. At each time step $t$, an user $i$ modifies their opinion as follows:

\begin{equation}
    x_i(t) = g_ix_i(0) + (1-g_i)\left(\sum_{j=1}^n\alpha_{ij}x_j(t)\right) 
\end{equation}

where $g_i$ is a \emph{susceptibility factor}: the higher $g$ is, the more an agent sticks to their initial opinion. Let's observe that the original French–Harary–DeGroot model in eq.\ref{eq:degroot}  is a case of Friedkin and Johnsen's model where $g_i = 0$. This model also leads to disagreement among agents. In a similar fashion, Dandekar et al.~\cite{Dandekar5791} extend the classic DeGroot model as follows:

\begin{quote}
    individual $i$ weights each neighbour $j$’s opinion $x_j(t)$ by a factor $(x_i(t))^{b_i}$ and weights the opposing view $(1 - x_j(t))$ by a factor $(1 - x_i(t))^{b_i}$, where $b_i \geq 0$ is a bias parameter. Informally, $b_i$ represents the bias with which $i$ assimilates his neighbours opinions. When $b_i$ = 0, our model reduces to DeGroot’s, and corresponds to unbiased assimilation.
\end{quote}
Authors confirmed with their simulations that basic French-Harary-DeGroot model is not polarising, but this model is, due to the biased assimilation of other's ideas.

\subsection*{Computational linguistics}
\label{app:linguistics}
Computational linguistics is the automatic processing of natural language. When the aim is to analyse large amounts of raw data and find relevant insights, text mining is applied. Combined with machine learning, it can help in developing text analysis models that learn to classify or extract specific information based on previous observations. A thorough and comprehensive explanation of the different aspects of the topic are presented among other, by Allen~\cite{allen1988natural} and Manning and Schutze~\cite{manning1999foundations}.

The majority of the studies that have explored fake news detection from the perspective of computational linguistics are addressing the task as a text classification task. The aim of a classification task is to assign specific labels to observations based on a model that has been previously trained on known data. The first step of a text classification task is the collection of the data that need to be labelled. After this initial step, the preprocessing follows in order to transform the data in a form that systems can understand. Text mining systems use several Natural Language Processing (NLP) techniques, like tokenization, parsing, lemmatization, stemming and stop removal, to build the input for the classification model.

In particular, the data have to be transformed into vectors since this is what the algorithm can take as input. Vectors represent different features of the existing data. One of the most common approaches for vectorization the text is called bag of words, and consists on counting how many times a word appears in a text. Another way is to use the term frequency–inverse document frequency (TF-IDF) that also reflects how important a word is to a document in a specific collection. Word embeddings is a more sophisticated way to transform text into vectors. Word embeddings represent a word in a lower-dimensional space allowing words with similar meaning to have a similar representation. There are different approaches that can be applied to get word embeddings such as word2vec and Continuous Bowl of Words (CBOW). Another way is to use pre-trained word embeddings such as fastText and SpaCy.

Once there is a vectorized representation of a document, the vector can be fed into a machine learning (e.g., SVM, NB, Decision Tree) or deep learning algorithm (e.g., LSTM, CNN) algorithm which learns to make predictions. Machine learning algorithms build a model based on sample data, known as training data, in order to make predictions or decisions on unseen data. Deep learning is a subset of machine learning and structures algorithms in layers to create an artificial neural network. Recently, researchers also consider the Bidirectional Encoder Representations from Transformers (BERT) to address text classification tasks. BERT that is developed by Google, makes use of Transformer, an attention mechanism that learns contextual relations between words in a text.

The algorithms are trained on a set that is called training set. A validation set is also usually used to optimise the parameters of the model. The performance of the classification model is evaluated on test documents and is measured with accuracy, recall, precision and F1-metric.

\section*{Papers with high centrality in the ``Fake News Reserch'' library network}
\label{app:centralities}
In this appendix, we follow up the analysis described in Sec. 3 of the survey to list all the top-10 papers with high centrality values (global citations, in-degree, betweenness, and page rank). For each measure, we ranked the papers corresponding to nodes in the ``Fake News Reserch'' library network, and in the following tables we show the top-10 for every rank. Assuming that the papers in the core library are more relevant to the topic that is the subject of our review, we filtered out the papers from the so called periphery from almost every table included in this section. Of course, rankings can change while new related papers are published, varying citation dynamics. As a consequence, we recommend the reader to use the search engine at \url{http://fakenewsresearch.net} for updated information and ranks.

We start with the rank for global citations (see Table~\ref{tab:globalcitations}). According to~\cite{DONTHU2021285}, it is important to distinguish between ``global citations'' and ``local citations'', respectively the citations that an article receives as is, and the in-degree, i.e., the citations that an article receives from other articles in our corpus only. It is worth mentioning that every paper in this and in the following lists is associated to a \emph{discipline}, that is deduced from the first level-0 Concept that is returned by the OpenAlex API when queried with the given paper.

\begin{table*}[]
    \centering 
    \begin{adjustbox}{width=1\textwidth}
    \begin{tabular}{|c|l|l|c|c|c|}
    \hline
    \multicolumn{5}{|c|}{\textbf{Paper information}} & \textbf{Ref.} \\
    \cline{1-5}
    \textbf{id} &\textbf{title} &\textbf{discipline} &\textbf{year} & \textbf{cluster} & \mbox{ } \\
    \hline
    \hline
    1 & The spread of true and false news online & psychology & 2018 & {\squareone} 1 & \cite{Vosoughi1146} \\
    \hline
    2 & Social Media and Fake News in the 2016 Election & political science & 2017 & {\squareone} 1 & \cite{Allcott2017} \\
    \hline
    3 & The science of fake news & computer science & 2018 & {\squareone} 1 & \cite{Lazer1094}\\
    \hline
    4 & Fake News Detection on Social Media: A Data Mining Perspective & computer science & 2017 & {\squaretwo} 2 &  \cite{Shuetal2017} \\
    \hline
    5 & Misinformation and Its Correction: Continued Influence and & \mbox{} & \mbox{} & \mbox{} & \mbox{} \\
    \mbox{} & Successful Debiasing & psychology & 2012 & {\squarethree} 3  & \cite{lewandowskyetal2012}\\
    \hline
    6 & The rise of social bots & computer science & 2016 & {\squarefive} 5 & \cite{ferrara2016}\\
    \hline
    7 & How to fight an infodemic & political science & 2020 & {\squarefour} 4 & \cite{zarocostas2020}\\
    \hline
    8 & The spreading of misinformation online & computer science & 2016 & {\squarenine} 9 & \cite{DelVicario554}\\
    \hline
    9 & Management misinformation systems & computer science & 1967 & - & \cite{russell1967}\\
    \hline
    10 & Oops They Did It Again! Carbon Nanotubes Hoax Scientists & \mbox{} & \mbox{} & \mbox{} & \mbox{} \\
    \mbox{} & in Viability Assays & chemistry & 2006 & - & \cite{Worle-Knirsch06}\\
    \hline
   
    \end{tabular}
    \end{adjustbox}
    \caption{Top-10 \textbf{global citations} papers in the \textbf{core} library. Clusters identifiers, as defined in Sec. 3, are shown as well. The ninth and the tenth papers in this rank belong to smaller clusters and are not relevant to the ``fake news'' topic.}
    \label{tab:globalcitations}
\end{table*}

Counting global citations helps us to understand which papers are objectively recognised at a wider level, not necessarily limiting our analysis to papers we collected to create the library, that are more focused on the ``fake news'' topic. 
Quite interestingly, in the top-10 rank in Table~\ref{tab:globalcitations} we find papers that have been covered in this survey, and that we also recognise as fundamental contributions to the development of the given topic, but also two papers that have been added to our library just because they contain the words ``misinformation'' and ``hoax'': paper n. 9 is a classic computer science paper on information system and data base management~\footnote{Quite ironically, in this paper the word ``misinformation'' looks like a mistake in the title, since MIS is used as an acronym for Management Information Systems, and the meaning of ``misinformation'' in this domain is never defined in the article.}, and paper n. 10 describes an anomaly, and how to deal with it, in a chemical empirical experiment. Quite interestingly, in our focused citation graph, both papers do not belong to any of the major nine clusters and their in-degree is very low, suggesting that despite their very high global citations, their impact to the ``fake news'' topic is neglectable. This also suggests that in-degree ranking shown in Table~\ref{tab:indegreecore} is more informative to our purposes than the global citation list.

\begin{table*}[]
    \centering 
    \begin{adjustbox}{width=1\textwidth}
    \begin{tabular}{|c|l|l|c|c|c|}
    \hline
    \multicolumn{5}{|c|}{\textbf{Paper information}} & \textbf{Ref.} \\
    \cline{1-5}
    \textbf{id} &\textbf{title} &\textbf{discipline} &\textbf{year} & \textbf{cluster} & \mbox{ } \\
    \hline 
    1* & The spread of true and false news online & psychology & 2018 & {\squareone} 1 & \cite{Vosoughi1146} \\
    \hline
    2* & Social Media and Fake News in the 2016 Election & political science & 2017 & {\squareone} 1 & \cite{Allcott2017} \\
    \hline
    3* & The science of fake news & computer science & 2018 & {\squareone} 1 & \cite{Lazer1094}\\
    \hline
    4* & Fake News Detection on Social Media: A Data Mining Perspective & computer science & 2017 & {\squaretwo} 2 &  \cite{Shuetal2017} \\
    \hline
    5* & Misinformation and Its Correction: Continued Influence and & \mbox{} & \mbox{} & \mbox{} & \mbox{} \\
    \mbox{} & Successful Debiasing & psychology & 2012 & {\squarethree} 3  & \cite{lewandowskyetal2012}\\
    \hline
    6* & The rise of social bots & computer science & 2016 & {\squarefive} 5 & \cite{ferrara2016}\\
    \hline
    7 & Defining ``Fake News'' & political science & 2018 & {\squareone} 1 & \cite{Tandoc2018}\\
    \hline
    8* & The spreading of misinformation online & computer science & 2016 & {\squarenine} 9 & \cite{DelVicario554}\\
    \hline
    9 & ``Liar, Liar Pants on Fire'': A New Benchmark Dataset for & \mbox{} & \mbox{} & \mbox{} & \mbox{} \\
    \mbox{} & Fake News Detection & computer science & 2017 & {\squaretwo} 2  & \cite{william2017} \\
    \hline
    10* & How to fight an infodemic & political science & 2020 & {\squarefour} 4 & \cite{zarocostas2020}\\
    \hline
    \end{tabular}
    \end{adjustbox}
    \caption{Top-10 \textbf{highest in-degree} papers in the \textbf{core} library. Clusters identifiers, as defined in Sec. 3, are shown as well. Papers that appear also among the top-10 global citations rank are marked with a (*).}
    \label{tab:indegreecore}
\end{table*}

Like global citations, in-degree is a network measure of ``popularity''; however, the basic assumption is that if a paper receives many local citations, as opposed to global ones, it is likely that the community is endorsing that contribution as relevant to the field. In Table~\ref{tab:indegreecore} we show the top 10 papers ranked according in-degree in the core library. As expected, 8 out of 10 papers that are in the first list are also in the top-10 global citations ranking, and the off topics papers have been flushed out because of their very low internal in-degree). Also, papers in this rank have been covered in our survey as important milestones in the area. To be noticed that papers 1, 2, 3, and 7 belong to the first cluster (see Sec. 3). Also, all of them represent highly multidisciplinary contributions that exploit advanced computational techniques to perform empirical analysis on data from social media. In particular, paper 7 contributes to the problem of defining the different variants of ``fake news'' (a terminology that we adopted as well in Sec. 2). The difficulties of debunking and correcting misinformation are usually connected to the misinformation effect, that is described in paper 5, and that is grounded in cognitive psychology. Fake news and social bots detection is one of the most faced challenges by computer scientist, and papers 4, 6 and 9 in the rank are commonly referred as milestones, as well as paper 8 that represents one of the most widely recognised contribution to study misinformation spreading in terms of the role played by echo-chambers in social media. Finally, paper 10 is a one page report published on Feb. 2020 on ``The Lancet'' calling for counter measures against the COVID-19 infodemic to come: an important signal that related sub-topic emerged sharply in the last two years. To be noticed that the majority of these contributions exploit computational linguistics and network science models and tools, confirming once again that a solid background on both scientific areas is necessary to make sense of the overall contributions. 

\begin{table*}[]
    \centering 
    \begin{adjustbox}{width=1\textwidth}
    \begin{tabular}{|c|l|l|c|c|c|}
    \hline
    \multicolumn{5}{|c|}{\textbf{Paper information}} & \textbf{Ref.} \\
    \cline{1-5}
    \textbf{id} &\textbf{title} &\textbf{discipline} &\textbf{year} & \textbf{cluster} & \mbox{ } \\
    \hline 
    1* & Misinformation and Its Correction: Continued Influence and & \mbox{} & \mbox{} & \mbox{} & \mbox{} \\
    \mbox{} & Successful Debiasing & psychology & 2012 & {\squarethree} 3  & \cite{lewandowskyetal2012}\\
    \hline
    2* & The spread of true and false news online & psychology & 2018 & {\squareone} 1 & \cite{Vosoughi1146} \\
    \hline
    3* & Social Media and Fake News in the 2016 Election & political science & 2017 & {\squareone} 1 & \cite{Allcott2017} \\
    \hline
    4* & The spreading of misinformation online & computer science & 2016 & {\squarenine} 9 & \cite{DelVicario554}\\
    \hline
    5 & Systematic Literature Review on the Spread of & \mbox{} & \mbox{} & \mbox{} & \mbox{} \\
    \mbox{} & Health-related Misinformation on Social Media & psychology & 2019 & {\squarethree} 3 & \cite{WANG2019}\\
    \hline
    6* & Fake News Detection on Social Media: A Data Mining Perspective & computer science & 2017 & {\squaretwo} 2 & \cite{Shuetal2017} \\
    \hline
    7 & Social Media, Political Polarization, and Political Disinformation: & \mbox{} & \mbox{} & \mbox{} & \mbox{} \\
    \mbox{} & A Review of the Scientific Literature & political science & 2018 & {\squarefive} 5 & \cite{tucker18} \\
    \hline
    8 & Beyond misinformation: Understanding and coping & \mbox{} & \mbox{} & \mbox{} & \mbox{} \\
    \mbox{} & with the ``post-truth'' era & psychology & 2017 & {\squarethree} 3 & \cite{BeyondMisinformation} \\
    \hline
    9 & The COVID-19 social media infodemic & computer science & 2020 & {\squarefour} 4 & \cite{cinelli2020} \\
    \hline
    10* & The science of fake news & computer science & 2018 & {\squareone} 1 & \cite{Lazer1094}\\
    \hline
    \end{tabular}
    \end{adjustbox}
    \caption{Top-10 \textbf{higher betweenness} papers in the \textbf{core} library. Clusters identifiers, as defined in Sec. 3, are shown as well. Papers that appear also among the top-10 in-degree rank are marked with a (*).}
    \label{tab:betweennesscore}
\end{table*}

A node with high betweenness is likely to be a ``bridge'' in a graph. Given the clustered structure of our citation graph, a bridge node could corresponds to highly cited papers, or to studies that are relevant for different disciplines, or both; in fact, 6 out of 10 papers in Table~\ref{tab:betweennesscore} are also in the in-degree top-10. Among the ``newly'' discovered contributions, paper 5 is a systematic literature review that describes a bibliometric analysis tailored on health-related misinformation; paper 7 is another survey that overviews political science most relevant contributions; paper 8 focuses on the psychological misinformation bias and its implication on the post-truth era (as paper 1); finally, paper 9 is one of the pioneering work describing data driven observations on the COVID-19 social media infodemic. It is probably not surprising that all of these studies could be of interest for a wide range of different disciplines.

\begin{table*}[]
    \centering 
    \begin{adjustbox}{width=1\textwidth}
    \begin{tabular}{|c|l|l|c|c|c|}
    \hline
    \multicolumn{5}{|c|}{\textbf{Paper information}} & \textbf{Ref.} \\
    \cline{1-5}
    \textbf{id} &\textbf{title} &\textbf{discipline} &\textbf{year} & \textbf{cluster} & \mbox{ } \\
    \hline 
    1* & Social Media and Fake News in the 2016 Election & political science & 2017 & {\squareone} 1 & \cite{Allcott2017} \\
    \hline
    2* & The spread of true and false news online & psychology & 2018 & {\squareone} 1 & \cite{Vosoughi1146} \\
    \hline
    3* & Fake News Detection on Social Media: A Data Mining Perspective & computer science & 2017 & {\squaretwo} 2 &  \cite{Shuetal2017} \\
    \hline
    4* & The science of fake news & computer science & 2018 & {\squareone} 1 & \cite{Lazer1094}\\
    \hline
    5* & Misinformation and Its Correction: Continued Influence and & \mbox{} & \mbox{} & \mbox{} & \mbox{} \\
    \mbox{} & Successful Debiasing & psychology & 2012 & {\squarethree} 3  & \cite{lewandowskyetal2012}\\
    \hline
    6* & The rise of social bots & computer science & 2016 & {\squarefive} 5 & \cite{ferrara2016}\\
    \hline
    7* & How to fight an infodemic & political science & 2020 & {\squarefour} 4 & \cite{zarocostas2020}\\
    \hline
    8* & ``Liar, Liar Pants on Fire'': A New Benchmark Dataset for & \mbox{} & \mbox{} & \mbox{} & \mbox{} \\
    \mbox{} & Fake News Detection & computer science & 2017 & {\squaretwo} 2  & \cite{william2017} \\
    \hline
    9* & The spreading of misinformation online & computer science & 2016 & {\squarenine} 9 & \cite{DelVicario554}\\
    \hline
    10 & Automatic deception detection: methods for finding fake news & computer science & 2015 & {\squaretwo} 2 & \cite{conroyetal2015}\\
    \hline
\end{tabular}
    \end{adjustbox}
    \caption{Top-10 higher \textbf{page rank} papers in the \textbf{core} library. Clusters identifiers, as defined in Sec. 3, are shown as well. Papers that appear also among the top-10 in-degree and betweenness rankings are marked with a (*).}
    \label{tab:pagerankcore}
\end{table*}

Page rank has already been used for finding ``hidden gems'' in citation networks (e.g., in~\cite{chen2007}): although some papers do not receive a huge number of citations such as other hubs, they can be referred as very influential by other very important contributions.
In Table~\ref{tab:pagerankcore}, we show the top-10 papers with highest page rank in our core library.  Nevertheless, here we find that all the papers in this list have been already mentioned because of their high in-degree or betweenness scores, so that apparently page rank does not help to let emerge some other highly influential paper. However, we must consider that this network measure works well in the medium-long term, because we need to have different layers of ``important'' papers that cite other papers. This scientific field is probably too young to give page rank more insights abilities than other simpler measures.

Looking at the periphery (see Table~\ref{tab:indegreeperiphery}) allows us to go back to pioneering studies that preceded the boost of interest aroused after 2016. For the sake of brevity, we show here only the top-20 papers by in-degree only. 

\begin{table*}[]
    \centering 
    \begin{adjustbox}{width=1\textwidth}
    \begin{tabular}{|c|l|l|c|c|}
    \hline
    \multicolumn{4}{|c|}{\textbf{Paper information}} & \textbf{Ref.} \\
    \cline{1-4}
    \textbf{id} &\textbf{title} &\textbf{discipline} &\textbf{year} & \mbox{ } \\
    \hline
    1 & Long Short-Term Memory &  computer science & 1997 & \cite{Hochreiter1997} \\
    \hline
    2 & Glove: Global Vectors for Word Representation &  computer science & 2014 & \cite{pennington2014} \\
    \hline
    3 & Information credibility on twitter & computer science & 2011 & \cite{Castillo2011}\\
    \hline
    4 & The case for motivated reasoning & psychology & 1990 & \cite{Kunda1990} \\
    \hline
    5 & When Corrections Fail: The Persistence of Political Misperceptions & political science & 2010 & \cite{NyhanReifler2010}\\
    \hline
    6 & The moderator–mediator variable distinction in social psychological & \mbox{} & \mbox{} & \mbox{} \\
    \mbox{} & research: Conceptual, strategic, and statistical considerations & psychology & 2014 & \cite{Baron1986} \\
    \hline
    7 & The theory of planned behavior & psychology & 1991 & \cite{AJZEN1991}\\
    \hline
    8 & The Nature and Origins of Mass Opinion & political science & 1992 & \cite{zaller_1992} \\
    \hline
    9 & Source monitoring & psychology & 1993 &  \cite{Johnson1993}\\
    \hline
    10 & Collective dynamics of ‘small-world’ networks & computer science & 1998 & \cite{watts1998}\\
    \hline
    11 & Cutoff criteria for fit indexes in covariance structure analysis:  & \mbox{} & \mbox{} & \mbox{} \\
    \mbox{} & Conventional criteria versus new alternatives & mathematics & 1999 & \cite{Hu1999}\\
    \hline
    12 & Emergence of Scaling in Random Networks & computer science & 1999 & \cite{barabasi1999}\\
    \hline
    13 & What is Twitter, a social network or a news media? &  &  & \cite{Haewoon2010}\\
    \hline
    14 & Motivated Skepticism in the Evaluation of Political Beliefs & political science & 2006 & \cite{taber2006}\\
    \hline
    15 & The Strength of Weak Ties & business & 1973 & \cite{granovetter1973}\\
    \hline
    16 & Birds of a Feather: Homophily in Social Networks & sociology & 2001 & \cite{McPherson2001}\\
    \hline
    17 & Social Network Sites: Definition, History, and Scholarship & sociology & 2007 & \cite{boyd2007}\\
    \hline
    18 & Judgment under Uncertainty: Heuristics and Biases & computer science & 1974 & \cite{Tversky1974}\\
    \hline
    19 & Users of the world, unite! The challenges and opportunities of Social Media & business & 2010 & \cite{KAPLAN2010}\\
    \hline
    20 & The Benefits of Facebook ``Friends:'' Social Capital & \mbox{} & \mbox{} & \mbox{} \\
    \mbox{} & and College Students’ Use of Online Social Network Sites & sociology & 2007 & \cite{Ellison2007}\\
    \hline
    \end{tabular}
    \end{adjustbox}
    \caption{Top-20 \textbf{highest in-degree} papers in the library's \textbf{periphery}.}
    \label{tab:indegreeperiphery}
\end{table*}

The list is quite impressive to the scholar's eye: very few papers are directly connected to the recent ``fake news'' hype, but for the pioneering work of Castillo et. al (paper 3) that is probably the first study that addressed the problem of assessing information credibility on Twitter, and that proposed an early automatic detection model made of textual and network based features. However, we have a strong signal that the current knowledge on how misinformation spreads is based on network science and computational linguistics, besides cognitive psychology, political science and sociology. In fact, some papers present artificial intelligence tools very popular in the computational linguistic community, such as LSTM (paper 1) and Glove (paper 2), classic statistical methods (papers 11 and 18), milestones in network science (papers 10, 12, 15, and 16), and early studies on emerging social media providing plenty of data to scientists and analysts (papers 13, 17, 19, and 20). To be noticed that current works on opinion dynamics, political partisanship and cognitive motivations for misinformation often refer to classic political/social/psychological science studies (book 8, papers 4, 5, 6, 7, 9, and 14). This could be very likely the list of the essential recommended readings for every advanced ``social media manipulation'' course.

\section*{Online resources and tools}
\label{app:online}

In the last runs of our bibliometric analysis, we used OpenAlex~\cite{OpenAlex} to download papers and follow their references to build the ``Fake News Research'' library, as described in Section 3. Our repository is accessible and searchable via the platform available at \url{http://fakenewsresearch.net}.

Network analysis was performed with most common tailored Python packages (e.g., pandas, networkx, igraph); graph plots (as in Fig. 2) were created with Gephi~\footnote{\url{https://gephi.org}}, and with R/matplotlib.

\end{document}